\documentclass[aip,reprint,showkeys,nofootinbib]{revtex4-1}

\draft 
\usepackage{graphicx}
\usepackage{dcolumn}
\usepackage{bm}

\usepackage[utf8]{inputenc}
\usepackage[T1]{fontenc}
\usepackage{mathptmx}
\usepackage{graphicx}
\usepackage{epstopdf, epsfig}
\usepackage{color,soul}
\usepackage{amssymb}
\usepackage{amsmath}
\usepackage{bm}
\usepackage{multirow}
\usepackage{xcolor}
\usepackage{tikz}
\usepackage{subfig}
\usepackage{subfloat}
\usepackage{multirow}
\usepackage{rotating}
\usepackage{booktabs}
\usepackage{relsize}
\usepackage{nomencl}
\usepackage{etoolbox}
\usepackage{float}
\usepackage{url}
\usepackage{hyperref}
\usepackage{algorithm2e}
\usepackage[export]{adjustbox}
\usepackage{upgreek}

\makeatletter
\newcommand{\labeltext}[3][]{%
    \@bsphack%
    \csname phantomsection\endcsname
    \def\tst{#1}%
    \def\labelmarkup{}
    \def\refmarkup{}%
    \ifx\tst\empty\def\@currentlabel{\refmarkup{#2}}{\label{#3}}%
    \else\def\@currentlabel{\refmarkup{#1}}{\label{#3}}\fi%
    \@esphack%
    \labelmarkup{#2}
}
\makeatother

\newlength{\offsetpage}
{\end{adjustwidth}}

\newcommand{\buf}{\textbf{u}^\mathrm{f}}

\newcommand{\bbf}{\textbf{b}^\mathrm{f}}

\newcommand{\stf}{\bm{\sigma}^{\mathrm{f}}}
\newcommand{\stnf}{\bm{\sigma}}
\newcommand{\bxf}{\textbf{x}^\mathrm{f}}
\newcommand{\df}{\Omega^\mathrm{f}(t)}

\newcommand{\ifsnt}{\Gamma^\mathrm{fs}}
\newcommand{\vf}{\mu^\mathrm{f}}

\usepackage{amsthm}

\begin{document}

\title{Three-dimensional deep learning-based reduced order model for unsteady flow dynamics with variable Reynolds number} 



\affiliation{Department of Mechanical Engineering, University of British Columbia, V6T 1Z4, Canada}
\author{Rachit Gupta}
\email{rachit.gupta@ubc.ca}

\author{Rajeev Jaiman}
\email{rjaiman@mech.ubc.ca}


\date{\today}

\begin{abstract}
In this article, we present a deep learning-based reduced order model (DL-ROM) for predicting the fluid forces and unsteady vortex shedding patterns. We consider the flow past a sphere to examine the accuracy of our DL-ROM predictions. 
The proposed DL-ROM methodology relies on a three-dimensional convolutional recurrent autoencoder network (3D CRAN) to extract the low-dimensional flow features from the full-order snapshots in an unsupervised manner.  
The low-dimensional features are evolved in time using a long short-term memory-based recurrent neural network (LSTM-RNN) and reconstructed back to the full-order as flow voxels. 
These flow voxels are introduced as static and uniform query probes in the point cloud domain to reduce the unstructured mesh complexity while providing convenience in the 3D CRAN training. We analyze a novel procedure to recover the interface description and the instantaneous force quantities from these 3D flow voxels. 
The 3D CRAN-based DL-ROM methodology is first applied to an external flow past a static sphere at single Reynolds number ($Re$) of $ Re = 300$ to test the 3D flow reconstruction and inference. We provide an assessment of the computing requirements in terms of the memory usage, training costs and testing times associated with the 3D CRAN framework.  
Subsequently, variable $Re$-based flow information is infused in one 3D CRAN to learn a complicated symmetry-breaking flow regime (280 $\leq$ $Re$ $\leq$ 460) for the flow past a sphere.
Effects of transfer learning are analyzed for training this complicated 3D flow regime on a relatively smaller time series dataset.
The 3D CRAN framework learns the complicated flow regime nearly $20$ times faster than the parallel full-order model and predicts this flow regime in time with an excellent to good accuracy. Based on the predicted flow fields, the network demonstrates an $\mathrm{R}^2$ accuracy of $98.58 \%$ for the drag and $76.43 \%$ for the lift over the sphere in this flow regime. 
The proposed framework aligns with the development of a digital twin for 3D unsteady flow field and instantaneous force predictions with variable $Re$ effects. 
\end{abstract}
\maketitle 

\section{Introduction} \label{Intro}
Unsteady flows described by the Navier-Stokes partial differential equation (PDE) possess highly nonlinear and multi-scale characteristics.
In particular, the underlying flow phenomenon due to the features (e.g., flow separation, shear layer, vortex shedding, near wake) exhibits complex spatial-temporal dynamics as functions of geometry and physical parameters \cite{miyanawala2019decomposition}. 
Examples of such spatial-temporal behaviors include the flow past fluttering flags and thin foils \cite{gurugubelli2015self}, oscillating hydrofoils with cavitation \cite{kashyap2021}, two-phase flows with fluid-structure interaction in offshore and marine applications \cite{joshi2019hybrid} and among others.
An accurate understanding of the unsteady flow features and their physical interactions is essential for operational decisions, structural designs, and the development of control strategies. 
The predictions of the unsteady flows are widely investigated and accurately modeled via reliable numerical methods based on PDEs that describe physical laws. 
Using state-of-the-art discretizations such as the finite element method, accurate solutions have been possible by solving millions of flow variables using full-order methods on high-dimensional PDEs. 
These techniques primarily involve solving the unsteady Navier-Stokes equations in a 3D computational domain for various geometries and boundary conditions which can involve moving interfaces and fluid-structure interaction \cite{jaimancomputational}. 

While the full-order modeling techniques provide high fidelity data, it is well known that such techniques are computationally expensive.
Furthermore, the analysis of flow involving three-dimensional geometries involves high resolutions near the interface. 
This results in the generation of a large number of unknown variables for accuracy gains and thereby scales the model's fidelity to millions of variables. 
Running such high-resolution and multi-scale simulations with regular PDE discretization requires a large computational time even in the supercomputing environment. 
As a result, the forward and high-dimensional problems become less attractive for multiquery analysis, control and structural design optimization.

\subsection{Deep leaning-based reduced order modeling}
For addressing issues of high-dimensionality, reduced-order models (ROMs) are instead constructed for low-dimensional analysis and have been widely investigated to identify dominant flow patterns and make dynamical predictions. One of the principal tools is to project a high-dimensional dataset onto an optimal low-dimensional subspace either linearly or nonlinearly to reduce spatial dimension and extract flow features. These low-dimensional analyses can provide essential flow dynamics for operational decision and efficiency improvement.  
Various projection techniques such as proper orthogonal decomposition (POD) \cite{sirovich1987turbulence}, dynamic mode decomposition (DMD) \cite{schmid2010dynamic}, balanced-POD \cite{singler2012balanced} and Koopman operators \cite{peitz2019koopman} have been extensively studied for the dimensionality reduction, control and mode decomposition of field dataset into relevant features. The mode decomposition can be considered as a mathematically optimal linear representation of the flow field and can provide interpretable analysis on flow features. For instance, Miyanawala and Jaiman \cite{miyanawala2019decomposition} showcased that for flows involving low $Re$, the POD modes represent one of the large-scale flow features such as vortex shedding, shear layer or near-wake bubble. 

However, projection-based ROMs can pose difficulty in the dimensionality reduction for complex flow patterns and hyperbolic PDEs as the number of required modes increases significantly. Instead, neural network-based autoencoders\cite{baldi1989neural, plaut2018principal} are explored as an alternative for nonlinear approximation because of their ability to automatically encode flow datasets and address some of the limitations of linear projection techniques.
Using encoder-decoder networks and activation functions, 
autoencoders allow to learn nonlinear relations between the input and the output dataset.
In contrast to  the projection-based ROMs, autoencoders provide larger compression and a greater flexibility for the dimensionality reduction of the data arising from the Navier-Stokes equations.
Autoencoders have been employed in a variety of fields such as object detection \cite{park2018multimodal}, sensor data-analysis \cite{ma2018deep} and biometric recognition \cite{yu2017multitask} due to their ease of implementation and low computational cost. For autoencoder and its variants, one can refer to a review work of Dong \emph{et. al} \cite{dong2018review}. To achieve data-driven prediction of dynamical problems using projecting methods, many researchers combine ROM spaces with deep learning to enhance predictive abilities which can be termed hybrid DL-ROMs. Such hybrid architectures consider spatio-temporal domain knowledge and achieve data-driven time series predictions. Recently proposed POD-based DL-ROMs are the POD-CNN by Miyanawala and Jaiman \cite{miyanawala2019hybrid}, the POD-RNN by Bukka \emph{et al.}\cite{bukka2021assessment}, the POD-enhanced autoencoders by Fresca and Manzoni\cite{fresca2021real}. These hybrid architectures have been demonstrated for 2D bluff body flows with and without fluid-structure interaction.   

\subsection{Review of physics-based deep leaning}
Deep learning in physical simulation has been boosting from the past decade owing to its effectiveness to automatically classify functional relations and make inference from a set of training data. Deep neural networks rely on the universal approximation of functions \cite{cybenko1989, chen1995universal, hornik1990universal}.
Despite the fact that deep neural networks are heavily overparametrized, they have an inherent bias to induce and make inferences from unseen data, which is known as inductive bias \cite{bronstein2021geometric}. 
Convolutional neural nets, for example, have an implicit inductive bias due to shared weight convolutional filters (i.e., translational symmetry) and pooling to exploit scale separation \cite{bronstein2021geometric}.
However, these black-box deep learning techniques ignore prior domain knowledge, which is crucial for interpretability, data efficiency, and generalization.
%

Many promising approaches have been established in the research community for a synergistic coupling of deep learning and physics-based models \cite{bukka2019data,bukka2021assessment}.
These models are trained to represent a full or partial parametrization of a forward process to reduce computational costs while emulating physical laws.
For instance, the trained parameters can be used to achieve {the state-to-state time advancements} \cite{ham2019deep, brown2008neural,san2018machine} and {inverse modeling} \cite{chen2017low, lunz2018adversarial,parish2016paradigm}. 
The state-to-state time advancement implies inferring dependent physical variables from the previous states. Inverse modeling, on the other hand, identifies physical system parameters from output state data. In order to increase generality, we can divide the physics-based machine learning into three categories: (a) adding a regularizer to the objective or loss function, (b) modifying neural architecture designs, and (c) combining deep learning and projection-based model reduction.

In the first category, physical losses are applied to the objective functions using regularizers in neural networks. These networks are trained to solve supervised learning tasks while adhering to governing equations. For instance Karpatne \emph{et al.} \cite{karpatne2017physics}, Raissi \emph{et al.} \cite{raissi2019physics}, Zhu \emph{et al.} \cite{zhu2019physics}, Erichson \emph{et al.} \cite{erichson2019physics} and among others have applied such regularizers to boost generalizability while training neural networks. Wang \emph{et. al} \cite{wang2017physics} also employed physics-informed machine learning to model the reconstruction of inconsistencies in the Reynolds stresses. As the second category of neural architectural designs,  convolutional neural networks are  utilized for the prediction of steady laminar flows \cite{guo2016convolutional} and the bulk quantities of interest \cite{miyanawala2017efficient} for bluff bodies. Similarly, Lee \emph{et. al} \cite{lee2019data} employed CNNs and generative adversarial networks (GANs) with and without loss function modification to predict recursive unsteady flows. For the prediction of unsteady flow over a cylinder and airfoil, Han \emph{et al.}\cite{han2019novel} constructed a convolutional long-short term memory (LSTM) network. To estimate the drag force, Ogoke \emph{et al.} \cite{ogoke2020graph} developed a graph-based convolutional neural network by depicting the flow field around an airfoil using unstructured grids. Likewise, Snachez \emph{et al.} \cite{sanchez2020learning} and Pfaff \emph{et al.} \cite{pfaff2020learning} utilized graph neural networks to represent the state of a physical system as nodes in a graph and compute the dynamics by learning message-passing signals. The third category in the physics-based deep learning involves the development of the hybrid DL-ROMs that take into account the spatial and temporal domain knowledge in neural networks \cite{miyanawala2019hybrid, bukka2021assessment, gupta2021hybrid, fresca2021real}. We refer to such DL-ROM frameworks as physics-based because they incorporate physical interpretability via proper orthogonal decomposition and its variants.

\subsection{Gaps and our contribution}
Using physics-based deep learning, the above works report substantial speed-ups during the online predictions compared with their full-order counterpart. However, most of these works are centered around two-dimensional geometries \cite{bhatnagar2019prediction, guo2016convolutional, jin2018prediction, eichinger2021stationary, murata2020nonlinear}. 
In this paper, we develop a deep learning-based reduced-order modeling framework for (a) three-dimensional unsteady flow predictions, (b) parametric flow and force predictions with variable Reynolds number, and (c) data-driven computational speed-ups for both offline training and online prediction of complex 3D wake flow. Although there are a few studies that use CNNs and autoencoder for parametric dependent flow problems \cite{lee2019data, fresca2021real}, there is no work that attempts to develop a DL-ROM methodology for the flow past 3D geometries in a way that can provide an effective mean to couple with real-time flow field snapshots \cite{rabault2017performing, lee2017piv}.

The present work builds upon our previous works \cite{bukka2021assessment, gupta2021hybrid} on the development of a convolutional recurrent autoencoder framework for three-dimensional flow past a sphere. 
Specifically, the present work utilizes 3D CNNs to extract low-dimensional features from full-order flow snapshots. The LSTM-RNN is utilized to evolve the low-dimensional features in time. 
Through 3D convolutional architecture, we model the flow variables in a uniform voxel grid. 
Using an unstructured irregular grid for the full-order simulation, we utilize snapshot-field transfer and load recovery (snapshot-FTLR)\cite{gupta2021hybrid} to select the structured grid for the fluid domain via interface force recovery. This simplified procedure reduces the unstructured mesh complexity while providing convenience in the training of the 3D CRAN framework. 
We emphasize the learning and inference capabilities of 3D CRAN on a complicated variable $Re$ flow regime for flow past a sphere. Using transfer learning, we reduce the offline training time and hyperparameter search of 3D CRAN. With the transfer learning process, this 3D CRAN maintains speed-up in training and provides a coarse-grained learning model for 3D unsteady flows compared with the FOM. The proposed DL-ROM provides a near real-time prediction of 3D flow fields and forces with variable Reynolds numbers. 
The end-to-end  DL-ROM framework is 3D and entirely data-driven; hence, it aligns with developing a digital twin involving variable $Re$ information.

The article is organized as follows: Section \ref{fom_vs_rom_v2} describes the full-order governing equations and reduced-order modeling of flow fields. Section \ref{3D-FTLR} introduces the voxel-based data generation process from full-order flow fields. Section \ref{3D-DL-ROM} presents the DL-ROM methodology employing 3D CNNs, convolutional recurrent autoencoder network, and transfer learning. The implementation of the proposed DL-ROM for the flow past a sphere with single and variable $Re$ is demonstrated in section \ref{R_and_d_sphere}. The article ends with the discussion on the variation of predicted forces to $Re$ and concluding remarks in section \ref{conclusions_sph}.

\section{Full-order and reduced-order modeling} \label{fom_vs_rom_v2}
This section starts by describing the full-order governing equations of the incompressible Navier-Stokes, followed by a brief description of reduced-order modeling.  

\subsection{Full-order modeling}
The isothermal viscous fluid in an Eulerian domain $\Omega^{\mathrm{f}}(t)$ can be described using the incompressible Navier-Stokes equations as: 
\begin{align}
\rho^\mathrm{f}\frac{\partial  \buf}{\partial t}+ \rho^\mathrm{f} \buf  \cdot\bm{\nabla}\buf = \bm{\nabla}\cdot\stf+\bbf \quad \text{on} \quad \df, \label{eq1}\\ 
\frac{\partial \rho^\mathrm{f} }{\partial t} + \bm{\nabla}\cdot (\rho^\mathrm{f} \buf) = 0 \quad \text{on}\quad \df, \label{eq2}
\end{align}
where the fluid variables are denoted using the superscripts $\mathrm{f}$. Here, $\buf$, $\bbf$ and $\stf$ represent the fluid velocity, body force and Cauchy stress tensor, respectively. For a Newtonian fluid with density $\rho^\mathrm{f}$ and dynamic viscosity $\vf$, $
\stf = -\textit{p}^{\mathrm{f}}\textbf{I}+\vf\left(\bm{\nabla}\buf+(\bm{\nabla}\buf)^{\mathrm{T}}\right) 
$. Here,  $\textbf{I}$ represents the identity tensor and $p^\mathrm{f}$ denotes the hydrodynamic pressure in the fluid. The partial time derivatives in Eqs. (\ref{eq1})-(\ref{eq2}) is with respect to the Eulerian referential coordinate system $\bxf$ and must satisfy the boundary conditions in the fluid domain
\begin{align}
\buf&= \buf_{D} \quad \forall \quad \bxf \in \Gamma^{\mathrm{f}}_{D}, \label{eq3}\\ 
\stf\cdot\textbf{n}^{\mathrm{f}} &= \textbf{h}^{\mathrm{f}} \quad \forall \quad \bxf \in \Gamma^{\mathrm{f}}_{N}, \label{eq4} \\
\buf&= \buf_{0} \quad \text{on} \quad \Omega^{\mathrm{f}}(0). \label{eq5}
\end{align}
While $\buf_{0}$ is the initial condition, $\buf_{D}$ and $\textbf{h}^{\mathrm{f}}$ represent the Dirichlet and Neumann boundary conditions on $\Gamma^{\mathrm{f}}_{D}$ and $\Gamma^{\mathrm{f}}_{N}$, respectively.  $\textbf{n}^{\mathrm{f}}$ denotes the unit normal on $\Gamma^{\mathrm{f}}_{N}$. The fluid-solid interface $\Gamma^{\mathrm{fs}}$ in the fluid domain is modeled using a no-slip Dirichlet boundary condition via Eq. (\ref{eq3}) with $\buf_{D} = 0$. The fluid force along the fluid-solid boundary is computed by integrating the surface traction, from the Cauchy stress tensor, over the first boundary layer elements on the fluid-solid surface. At a time instant, the force coefficients $\mathrm{C}_{\mathrm{x}}$, $\mathrm{C}_{\mathrm{y}}$ and $\mathrm{C}_{\mathrm{z}}$ in the respective Cartesian directions are given as
\begin{equation}
\begin{aligned}
    \mathrm{C}_{\mathrm{x}} = \frac{1}{\frac{1}{2}\rho^{\mathrm{f}}U_{\infty}^{2}D} {\int_{\ifsnt} (\stf.\mathrm{\mathbf{n}}).\mathrm{\mathbf{n}}_{\mathrm{x}} \mathrm{d}\Gamma },\\
    \mathrm{C}_{\mathrm{y}} = \frac{1}{\frac{1}{2}\rho^{\mathrm{f}}U_{\infty}^{2}D} {\int_{\ifsnt} (\bm{\stf}.\mathrm{\mathbf{n}}).\mathrm{\mathbf{n}}_{\mathrm{y}} \mathrm{d}\Gamma },\\ 
    \mathrm{C}_{\mathrm{z}} = \frac{1}{\frac{1}{2}\rho^{\mathrm{f}}U_{\infty}^{2}D} {\int_{\ifsnt} (\bm{\stf}.\mathrm{\mathbf{n}}).\mathrm{\mathbf{n}}_{\mathrm{z}} \mathrm{d}\Gamma }, 
\end{aligned}
\label{force_eqns}
\end{equation}
where $U_{\infty}$ and $D$ are the reference velocity and reference length, respectively. For the unit normal $\mathbf{n}$ of the surface, $\mathbf{n}_{\mathrm{x}}$, $\mathbf{n}_{\mathrm{y}}$ and $\mathbf{n}_{\mathrm{z}}$ are the x, y and z Cartesian components, respectively. The weak form of the incompressible Navier-Stokes equations is solved in space using an equal-order isoparametric finite elements for the fluid velocity and pressure. A numerical scheme implementing the Petrov–Galerkin finite element and the semi-discrete time stepping is utilized \cite{jaiman2011transient, jaiman2016partitioned} to generate the full-order 3D flow states. 

\subsection{Reduced-order modeling} \label{rom_details}
\begin{figure*}
\centering
\includegraphics[width=1.0\textwidth]{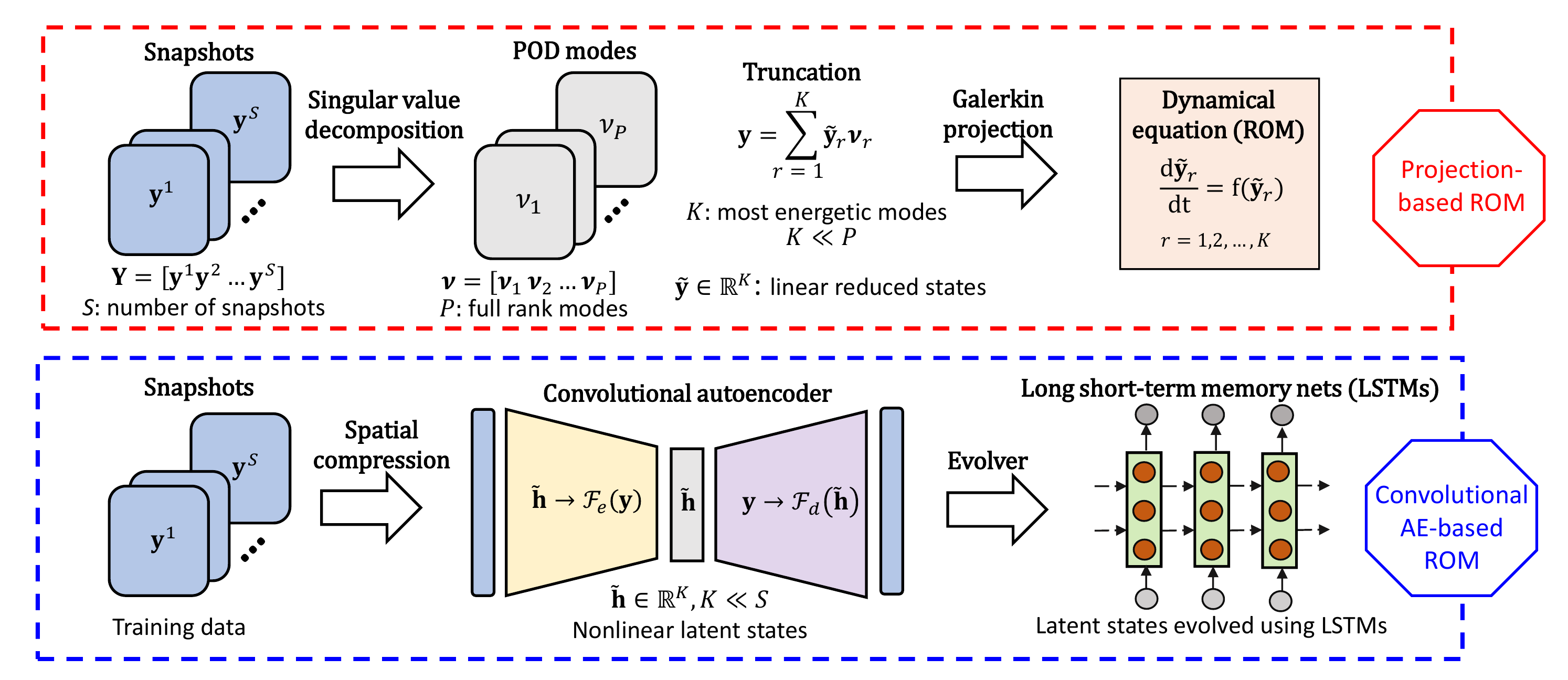}
\caption{Illustration of reduced-order modeling: (i) Projection-based ROM for creating orthonormal basis functions (modes) for the reduced representation and (ii) convolutional autoencoder ROM framework. Refer to the details of the variables in the main text in section \ref{rom_details}.}
\label{rom_simple}
\end{figure*}

In relation to the reduced-order modeling of these state variables, the PDEs described by  Eqs.~(\ref{eq1})-(\ref{eq2}) can be re-written in an abstract state-space form as
\begin{equation} \label{dynamic_eq}
\begin{split}
\frac{d \mathbf{y}}{d t} = \textbf{F} (\textbf{y}), 
\end{split}
\end{equation}
$\mathbf{y} \in \mathbb{R}^{M}$ represents a state vector consisting of $M$ system variables, while $ d \mathbf{y} / d t$ is the dynamics of the state vector. In the present case, the state vector consists of the fluid velocity and the pressure dataset as $\mathbf{y} = \{\mathbf{u}^\mathrm{f}, p^\mathrm{f}  \}$. $\textbf{F}$ denotes a vector-valued differential operator describing the spatially discretized PDEs. The spatial-temporal dynamical equation is useful for creating the reduced-order approximation of a coupled dataset. 
In reduced-order modeling, the differential operator $\textbf{F}$ is constructed by projecting the full-order state variables on a trial subspace. For this purpose, $\textbf{F}(\textbf{y})$ is decomposed into the constant $\mathbf{C}$, linear $\mathbf{B} \mathbf{y}$ and nonlinear $\mathbf{F}^{\prime}(\mathbf{y})$ dynamical components as 
\begin{equation} \label{dynamic_exp}
\mathbf{F}(\mathbf{y})=\mathbf{C}+\mathbf{B} \mathbf{y}+\mathbf{F}^{\prime}(\mathbf{y}).
\end{equation} 

\subsubsection{Projection-based reduced order modeling}
The Galerkin-based ROMs project the state vector $\mathbf{y} \in \mathbb{R}^{M}$ via a subspace spanned  $\mathcal{V} \in \mathbb{R}^{M \times K}$.
The subspace matrix $\mathcal{V}$ represents the reduced basis onto which the full-order dynamics is generally projected with $K<<M$. 
This reduces the full-order state vector $\mathbf{y}$ as $\mathcal{V}\mathbf{\tilde{y}}$ with $\mathbf{\tilde{y}}\in \mathbb{R}^{K}$ and thereby approximating the system dynamics using Eqs. (\ref{dynamic_eq})-(\ref{dynamic_exp}) as  
\begin{equation} \label{low_order_dynamics}
\frac{d \mathbf{\tilde{y}}}{d t}=\mathcal{V}^{T} \mathbf{C}+\mathcal{V}^{T} \mathbf{B} \mathcal{V} \mathbf{\tilde{y}}+\mathcal{V}^{T} \mathbf{F}^{\prime}(\mathcal{V} \mathbf{\tilde{y}}).
\end{equation}
The reduced-order space is defined by a set of modes $\mathcal{V} \in \mathbb{R}^{M \times K}$ using the snapshot matrix $\textbf{Y} = \{\mathbf{y}^{1} \; \mathbf{y}^{2} \; \ldots  \; \mathbf{y}^{S}\} \in \mathbb{R}^{M \times S} $, with $M$ number of variables and $S$ denotes the number of snapshots. 
The projection forms an orthonormal basis of $\mathbf{\tilde{Y}} = \{\mathbf{\tilde{y}}^{1} \; \mathbf{\tilde{y}}^{2} \; \ldots \; \mathbf{\tilde{y}}^{S}\} \in \mathbb{R}^{K \times S}$, a $K$ dimensional subspace of $\mathbb{R}^{M}$.
For example, the POD method first creates the orthonormal basis functions for the reduced representation, which is followed by the Galerkin projection onto the subspace spanned by a set of truncated basis functions \cite{taira2017modal}.
While these projection-based models are effective in constructing a low-dimensional subspace of low Kolmogorov n-width problems, they may not provide efficient reconstruction for general nonlinear systems.
As a further approximation of these nonlinear systems for efficiency, hyperreduction techniques such as the discrete empirical interpolation method  \cite{chaturantabut2010} and  energy-conserving sampling and weighting \cite{An2008ECSW} method can be employed.
However, these empirical projection-based reduced-order models can come at the cost of large subspace dimensions for convection-dominated or turbulence problems characterized by
large Kolmogorov n-width. 

\subsubsection{Autoencoder-based reduced order modeling}

\begin{figure*} 
\centering
\includegraphics[width=1.0\textwidth]{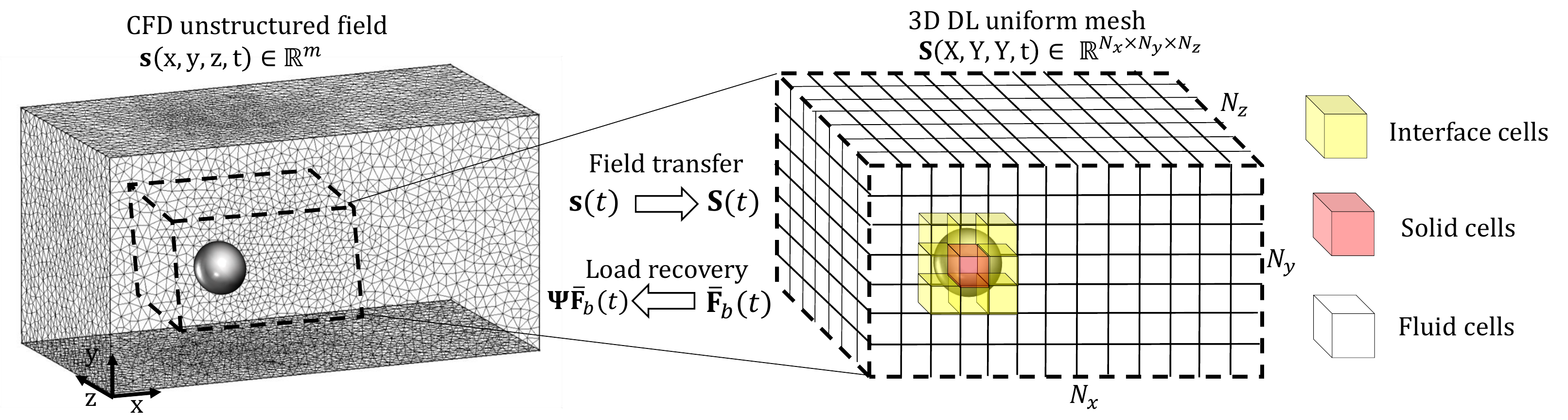}
\caption{Schematic of a 3D mesh-to-mesh field transfer and load recovery process to determine the 3D CRAN snapshot grid. Refer to the details of all variables in section~\ref{3D-FTLR}.}
\label{unstruct_to_voxel}
\end{figure*}

The drawbacks of the projection-based ROMs can be addressed by the use of autoencoders that provide a promising alternative for constructing the reduced-order space of the snapshot matrix $\textbf{Y}$. An autoencoder comprises an encoder and decoder network and  can be interpreted as a nonlinear and flexible generalization of POD for a dimensionality reduction of flow field\cite{bukka2020deep}. 
An autoencoder, $\mathcal{F}$, is trained to output the same input data $\textbf{Y}$ so that $\textbf{Y} \approx \mathcal{F}(\textbf{Y} ; w)$, where $w$ are the parameters of the end-to-end autoencoder model. Using an iterative minimization of an error function $E$, the parameters $w$ are trained as
\begin{equation}
    \boldsymbol{w}=\operatorname{argmin}_{w}[E(\textbf{Y}, \mathcal{F}(\boldsymbol{\textbf{Y}} ; w))].
\end{equation}
The dimension of the low-order space $\tilde{\textbf{H}}$ in an autoencoder is smaller than that of the input or output data $\textbf{Y}$ and is referred to as a latent space. When the output $\mathcal{F}(\textbf{Y})$ is comparable to the input such that $\textbf{Y} \approx \mathcal{F}(\textbf{Y})$, the latent space is a low-dimensional representation of its input which provides a low-rank embedding. 
The dimension compressor in an autoencoder is called the encoder $\mathcal{F}_e$, and the counterpart is the decoder $\mathcal{F}_d$. The internal mechanism of the autoencoder can be stated as follows using the encoder-decoder architecture:
\begin{equation}
\tilde{\textbf{H}}=\mathcal{F}_{e}(\textbf{Y}), \quad \textbf{Y}=\mathcal{F}_{d}(\tilde{\textbf{H}}).
\end{equation}

The subspace projection in an autoencoder is achieved using an unsupervised training of the snapshot matrix $\textbf{Y}$.
Throughout the paper, we utilize the convolutional autoencoder process for general reduced-order modeling.  In particular, a 3D convolutional mapping in the autoencoder can provide an efficient latent space to characterize the reduced dynamics of 3D unsteady flows. 
After convergence,  linear autoencoders with a latent dimension of $K$ span the same subspace as POD using $K$ modes as illustrated in \cite{baldi1989neural, plaut2018principal}.
In the next section, we turn our attention toward an optimal procedure for selecting the full-order dataset for a 3D convolutional autoencoder. 

\section{Field transfer and coarse graining} \label{3D-FTLR}

Flow simulations involving bluff body flows are generally modeled in a non-uniform and unstructured body conformal mesh for computational fluid dynamics (CFD) applications. %
Primarily, an unstructured mesh offers the advantage of allocating a greater node resolution in the region of importance in the flow domain, for example, the boundary layer mesh along the fluid-solid interface. 
Although creating an unstructured mesh for CFD problems can provide an accurate interface modeling, the number of nodes can scale to millions of variables for the fluid domain. 
This generally results in a point cloud domain with complex spatial connectivity information for every bluff body geometry. 
The complex mesh connectivity information can be directly difficult to incorporate in a neural network especially involving the use of CNN filters. 
As a result, the time series flow snapshots are point cloud information that may not contain spatial connectivity information in the dataset. 

To retain the spatial connectivity information in the dataset, we project the unstructured field dataset on a uniform and structured voxel grid to coarse-grain the field information as shown in Fig.~\ref{unstruct_to_voxel}. This simple process brings uniformity in the field information together with convenience in training the CRAN driver. The projection and interpolation of information are achieved via the snapshot-field transfer and load recovery process introduced in the following subsections. Snapshot-FTLR is an iterative data processing step that allows recoverable interface information by preserving the forces. Once this loss is observed in the training forces, they are corrected by reconstructing to a higher-order CFD force.

\subsection{Field transfer}

We extend the snapshot-FTLR to 3D flow fields throughout this paper. Let $ {\textbf{s}} = \{ {\textbf{s}}^{1} \; {\textbf{s}}^{2} \; ... \; {\textbf{s}}^{n} \} \in \mathbb{R}^{m \times n}$ be the full-order flow fields generated from the Navier-Stokes solver and $ {\overline{\textbf{F}}}_{\ifsnt} =  \{ {\overline{\textbf{F}}}_{\ifsnt}^{1} \; {\overline{\textbf{F}}}_{\ifsnt}^{2} \; ... \; {\overline{\textbf{F}}}_{\ifsnt}^{n} \} \in \mathbb{R}^{3 \times n} $ be the corresponding interface forces over the fluid-solid body along the three Cartesian directions. Here, $m$ represent the number of full-order nodes and $n$ denotes the number of snapshots. The point cloud dataset at every time step is interpolated and projected to a uniform voxel grid as shown in Fig.~\ref{unstruct_to_voxel}. The size of this grid can be chosen ($N_{x} \times N_{y} \times N_{z}$).  $N_{x}$, $N_{y}$ and $N_{z}$ are the data probes in the respective Cartesian axes. These probes are structured as spatially uniform 3D space for coarse-graining the field information as well as retaining a uniform mesh connectivity. We employ Scipy's \emph{griddata} function \cite{SciPy} to interpolate the scattered CFD data and fit a surface to generate the 3D flow snapshots. The generated information $\textbf{S}=\{\textbf{S}^1 \; \textbf{S}^2 \; ... \; \textbf{S}^n \} \in \mathbb{R}^{N_{x}\times\ N_{y} \times N_{z} \times n}$ denotes the 3D snapshots of a field dataset (for instance,  pressure or velocity), where $\textbf{S}^i \in \mathbb{R}^{N_{x}\times N_{y} \times N_{z}}$.

\subsection{Load recovery}
The presence of the interface in the DL voxel grid is ensured by masking the entire 3D spatial region covered by the solid domain with a mandatory function which zeroes out the field values inside the fluid-solid interface. This function identifies the coarse interface voxels (shown as yellow cells in Fig.~\ref{unstruct_to_voxel}) that contain the solid description. We now integrate the total voxel force $\overline{\textbf{F}}_{b} = \{ \overline{\textbf{F}}_{b}^{1} \; \overline{\textbf{F}}_{b}^{2} \; ... \; \overline{\textbf{F}}_{b}^{n} \}$ exerted by these interface cells over the solid using
\begin{equation}
\begin{aligned} \label{eq12}
\overline{\textbf{F}}_{{b}}^{i} = \sum_{\mathrm{k}=1}^{N_{F}} \int_{\Gamma^{\mathrm{k}}} \stnf_{\mathrm{k}}^{i}.\mathrm{\mathbf{n}}\mathrm{d} \Gamma, \qquad i=1,2,...,n, 
\end{aligned}
\end{equation}
where $N_{F}$ are the number of interface voxels. For every interface cell ${\mathrm{k}}$, $\stnf_{\mathrm{k}}^{i}$ is the Cauchy stress tensor at a time step $t^{i}$. We calculate this tensor using the finite difference approximation and integrate over the faces of the voxel $\Gamma^{\mathrm{k}}$. For the present case, we only consider the pressure component while calculating the voxel forces. 

The coarsening effect of the full-order data onto the voxel grid brings a spatial uniformity in the input state.
However, this coarsening may also lead to a loss of accurate forces on the physical interface, even in the training data. 
This is accounted due to a considerable loss of interface resolution in the voxel grid as compared with the FOM mesh. 
Herein, we recover the data loss in the voxel forces $\overline{\textbf{F}}_{b} = \{ \overline{\textbf{F}}_{b}^{1} \; \overline{\textbf{F}}_{b}^{2} \; ... \; \overline{\textbf{F}}_{b}^{n} \}$ by mapping it to full-order $ {\overline{\textbf{F}}}_{\ifsnt} =  \{ {\overline{\textbf{F}}}_{\ifsnt}^{1} \; {\overline{\textbf{F}}}_{\ifsnt}^{2} \; ... \; {\overline{\textbf{F}}}_{\ifsnt}^{n} \} $  and still maintaining a lower DL grid resolution. 
This is achieved by constructing the functional recovery mapping $\Psi$. 
We select the coarse grid $N_{x} \times N_{y} \times N_{z}$ that recovers these bulk force quantities with $\Psi$ mapping by avoiding the need of super-resolution. 
The process of constructing the $\Psi$ mapping is summarised as:
\begin{itemize}
\item {Get the voxel forces} $\overline{\textbf{F}}_{b} = \{ \overline{\textbf{F}}_{b}^{1} \; \overline{\textbf{F}}_{b}^{2} \; ... \; \overline{\textbf{F}}_{b}^{n} \}$ {and full-order forces} $ {\overline{\textbf{F}}}_{\ifsnt} =  \{ {\overline{\textbf{F}}}_{\ifsnt}^{1} \; {\overline{\textbf{F}}}_{\ifsnt}^{2} \; ... \; {\overline{\textbf{F}}}_{\ifsnt}^{n} \} $ {for} $n$ {training time steps,}

\item {Define the the mean and fluctuating force components}
\begin{equation} \label{eq13}
\begin{aligned}
\overline{\textbf{F}}_{b}^{'} &= \overline{\textbf{F}}_{b} - \mathrm{mean}(\overline{\textbf{F}}_{b}),\\ 
\overline{\textbf{F}}_{\ifsnt}^{'} &= \overline{\textbf{F}}_{\ifsnt} - \mathrm{mean}(\overline{\textbf{F}}_{\ifsnt}),
\end{aligned}
\end{equation}
  
\item {Calculate the time-dependent derivative error} $\overline{E}_{c}$
\begin{equation} \label{eq14}
\begin{aligned}
\overline{E}_{c} &= (\overline{\textbf{F}}_{\ifsnt}^{'} - \overline{\textbf{F}}_{b}^{'}) ./(\overline{\textbf{F}}_{b}^{'}), \;\; \mathrm{with} \;\; \overline{\textbf{F}}_{b}^{'} \neq 0, 
\end{aligned}
\end{equation}

\item {Reconstruct the voxel forces to full-order with mean and derivative corrections}
\begin{equation} \label{eq15}
\begin{aligned}
\overline{\textbf{F}}_{b} &= \overline{\textbf{F}}_{b}^{'} + \mathrm{mean}(\overline{\textbf{F}}_{\ifsnt}) + \mathrm{mean}(\overline{E}_c) \overline{\textbf{F}}_{b}^{'}, \\
&= \Psi \overline{\textbf{F}}_{b}.
\end{aligned}
\end{equation}
\end{itemize} 

This mesh-to-mesh field transfer and load recovery process brings uniformity in the field information to coarse-grain the full-order information. It also recovers the interface information together with convenience in training the CRAN driver. 

\section{Deep learning-based reduced order modeling framework} \label{3D-DL-ROM}
\begin{figure*}
\centering
\includegraphics[width=1.0\textwidth]{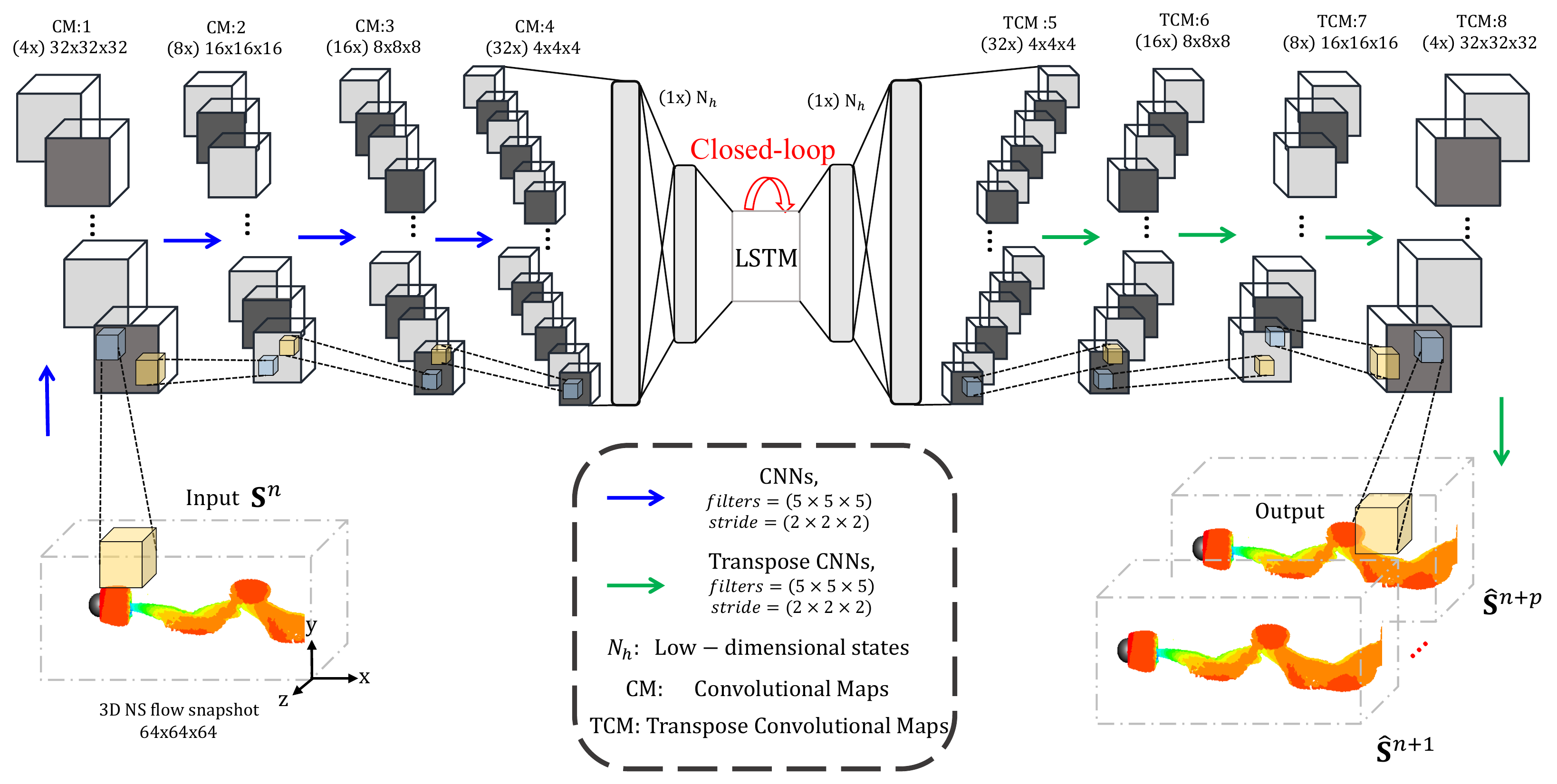}
\caption{Schematic of the 3D CRAN framework for unsteady flow predictions. The encoding is constructed by utilising 3D CNNs to reduce the input dimension, and the decoding is achieved by using transpose 3D CNNs. The LSTM-RNN evolves the low-dimensional states between the encoder and decoder networks.}  
\label{3D_CRAN}
\end{figure*}
We present a hybrid DL-ROM framework that can be trained with variable $Re$-based full-order unsteady information on a range of 3D flow patterns. 
In conjunction with a 3D convolutional autoencoder for the dimensionality reduction, we use the well-known LSTM-RNNs for evolving the low-dimensional states in time.
The LSTM-RNN represents a mathematical framework that depicts a nonlinear state-space form, making it suitable for nonlinear dynamical systems.
When the low-dimensional states are obtained using a 3D convolutional autoencoder and evolved in time with the LSTM-RNN, the hybrid framework is called the 3D CRAN. 

We use an approach known as transfer learning to efficiently lean and predict variable $Re$-based unsteady flow for a 3D CRAN.
Transfer learning is the process of improving learning in a new task by transferring knowledge from a previously learned related task.
The CRAN architecture learns the task of time series prediction for single $Re$-based flow information with a remarkable accuracy. 
The goal of transfer learning is to improve learning in the variable $Re$-based flow by
leveraging knowledge from the single $Re$-based flow task. 
We next elaborate a brief description of 3D CNNs, formulation of our 3D hybrid DL-ROM framework and it's integration with transfer learning for predicting variable $Re$-based unsteady flow in time. 

\subsection{3D convolutional neural networks}
%
In this study, we consider 3D convolutional neural networks for our hybrid DL-ROM technique. 
They are utilized to extract relevant features from the 3D unsteady flow data to construct the reduced-order state.
The application of 2D CNNs as a reduced-order model has been explored by Miyanawala and Jaiman \cite{miyanawala2017efficient} to predict bluff body forces.
For the sake of explanation, we briefly describe the feature extraction process of a 3D CNN that is useful for constructing the encoder network of the convolutional autoencoder. 
Analogous to a $2 \mathrm{D}$ operation, a $3 \mathrm{D}$ CNN layer takes a set of vectors as input and applies the discrete convolutional operation. 
This is achieved via a number of $3 \mathrm{D}$ kernels as shown using the blue arrows in Fig.~\ref{3D_CRAN}. 

For the first 3D convolution operation, the input is the flow field snapshot: $\textbf{S}^{n} \in \mathbb{R}^{N_{x} \times N_{y} \times N_{z}} $. 
For simplicity, let us denote this input using ${D(\mathrm{x}, \mathrm{y}, \mathrm{z})}$. 
For any $L^{t h}$ convolutional layer, let $k_{L}$ denote the number of feature kernels. We group these kernels of size $f_x \times f_y \times f_z$ into a 4D tensor $K^{L} \in \mathbb{R}^{f_x \times f_y \times f_z \times k_{L}}$.
When we apply the first operation of $3 \mathrm{D}$ convolution on the input matrix, i.e. $L=1$, it generates a $4 \mathrm{D}$ tensor $Y^{C 1}=\left\{Y_{i j k l}^{C 1}\right\}$
\begin{equation} \label{3d_cnn_opr}
\begin{aligned}
Y_{i j k l}^{C 1}&=D_{i j k} \star K_{i j k l}^{1},\\
&= \sum_{c=1}^{N_z} \sum_{b=1}^{N_y} \sum_{a=1}^{N_x} D_{a b c} K_{(i-a+1)(j-b+1)(k-c+1)l}^{1},
\end{aligned}
\end{equation}
where $l=1,2,\dots, k_L$. The $\star$ sign represents the convolutional process, which allows local features to be extracted from a 3D space. Eq.~(\ref{3d_cnn_opr}) is modified slightly if the convolutional blocks are skipped on more than one element of the input function along any cartesian direction. 

The skipping lengths along the three directions of the input is termed as the stride $s_{L}=\left[\begin{array}{ll}s_{x} \;\;  s_{y} \;\; s_{z} \end{array}\right]$ and is an important hyperparameter for the dimensionality reduction. 
For the 3D CNN operation of the CRAN architecture, we utilize a filter length $(f_x \times f_y \times f_z) = (5 \times 5 \times 5)$ and stride $s_{L}=\left[\begin{array}{ll}2 \;\;  2 \;\; 2 \end{array}\right]$. 
The input to the convolutional layer is a 4D tensor, except for the input layer.
The $L^{t h}$ layer takes each 3D slice of the tensor $Y^{C(L-1)}$ and convolutes them with all the kernels which create a 4D tensor $Y^{C L}$. 
Because convolution is a linear operation, a nonlinearity must be introduced in this process to capture complicated nonlinear flow features such as vortex patterns.
We employ the sigmoid activation $\sigma(z) = \left(1+e^{- z }\right)^{-1}$  function for this purpose. We next briefly review the 3D CRAN framework which utilizes these 3D CNNs to construct the low-dimensional states and infer 3D fields in time. 
\subsection{3D convolutional recurrent autoencoder network}\label{3D_CRAN_details}
This projection and propagation technique extracts the low-dimensional encoding of the flow variables by using 3D CNNs. 
The obtained low-dimensional states are propagated in time using the LSTM-RNN to evolve the encoding.
We build a decoding space of 3D transpose convolutions that up-samples the low-dimensional encoding back to the high-dimensional space to decode the evolving features. 
It is assumed that the 3D unsteady flow solutions draw a low-dimensional subspace, allowing an embedding in the high-dimensional space. 
The end-to-end three-dimensional convolutional recurrent autoencoder architecture is illustrated in Fig.~\ref{3D_CRAN}.

The 3D CRAN framework relies on the high-fidelity 
snapshots of flow field acquired from the full-order simulation. 
Let $\textbf{S}=\{\textbf{S}^1 \; \textbf{S}^2 \; ... \; \textbf{S}^n \} \in \mathbb{R}^{N_{x}\times\ N_{y}\times N_{z}\times n}$ denote the 3D snapshots of any field dataset. $\textbf{S}^i \in \mathbb{R}^{N_{x}\times N_{y} \times N_{z} }$ indicates a field snapshot and $n$ represents the number of such snapshots.
As described in section \ref{3D-FTLR}, $N_{x}$, $N_{y}$ and $N_{z}$ represent the number of data probes in the respective Cartesian axes for the uniform voxel grid. For the present case, $N_{x} = N_{y} = N_{z} = 64$. The target of the CRAN-based data-driven prediction is to encode-evolve-decode to the future values at the field probes: $\hat{\textbf{S}}^{n+1}, \hat{\textbf{S}}^{n+2}, ... \hat{\textbf{S}}^{n+p}$ for some finite time steps.  
For the sake of clarity, let us assume that $\textbf{S}$ is a dataset matrix consisting of time series information for a single $Re$ flow field obtained from the Navier-Stokes solver. 
The preparation of the dataset matrix for variable $Re$ is straightforward and is described later in section~\ref{Tansfer_learning_VAR_RE}. 

\subsubsection{Proposed architecture} 
\begin{figure*}
\centering
\includegraphics[width=0.9\textwidth]{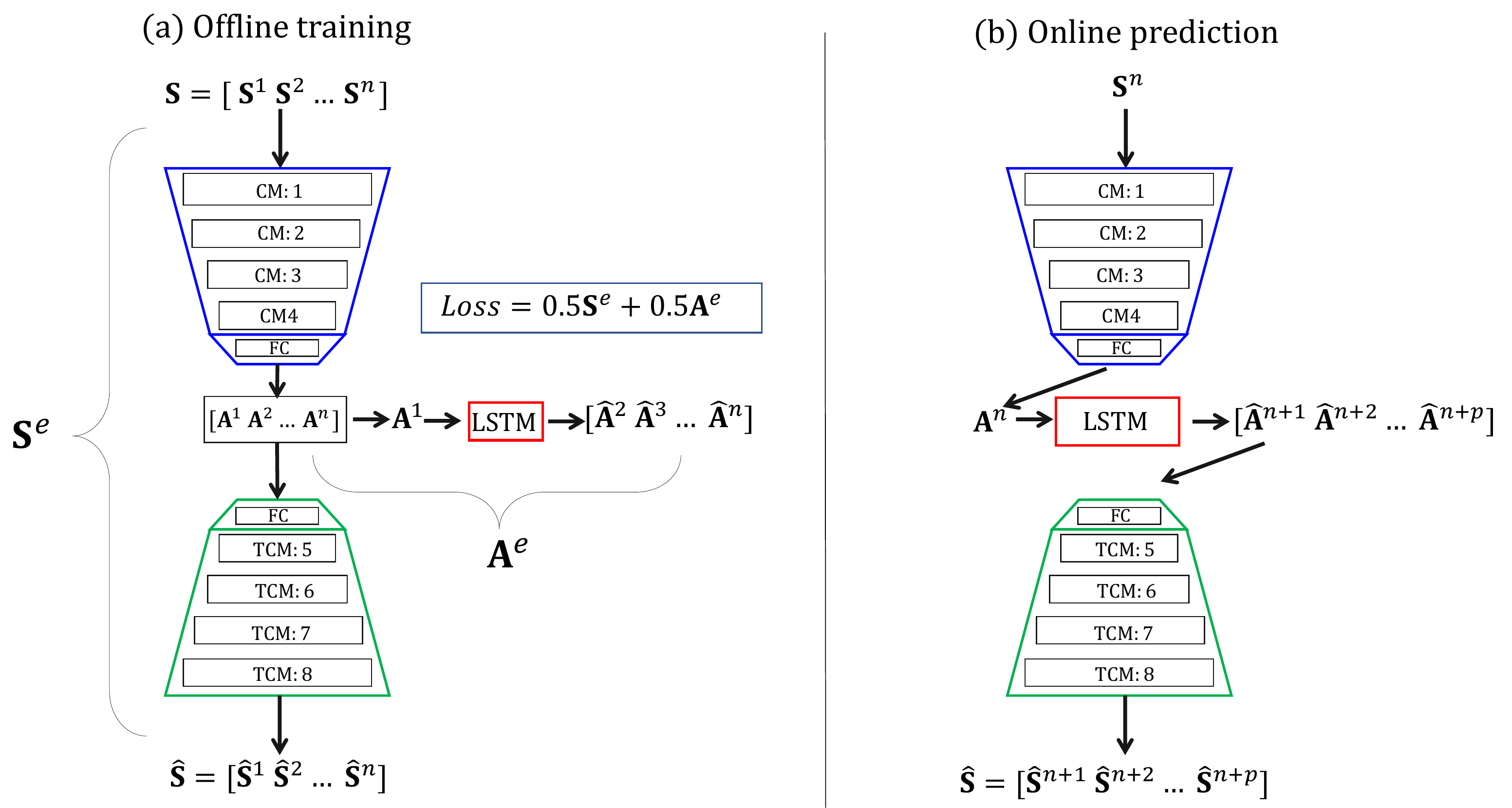}
\caption{Illustration of the training and prediction process of the CRAN architecture. (a) An offline training strategy for the 3D CRAN. (b)  Iterative process of online predictions using the 3D CRAN. Refer to all the variable details in section~\ref{3D_CRAN_details}.}
\label{3D_CRAN_training_prediction}
\end{figure*}

Using the 3D encoder-decoder architecture of the convolutional autoencoder, the low-dimensional states $\textbf{A}_{c} =\{\textbf{A}_{c}^1\; \textbf{A}_{c}^2\; ...\; \textbf{A}_{c}^n\} \in \mathbb{R}^{h \times n}$ of the full-order dataset $\textbf{S}$ are determined in an unsupervised manner. The nonlinear encoding feature space can be expressed as
\begin{equation}
\textbf{A}_{c}^i = f_{enc}(\textbf{\textbf{S}}^{i};\theta_{enc}),  \qquad i=1,2,...,n,
\end{equation}
where $f_{enc}$ is a compositional function consisting of a trainable encoder space that is parametrized by $\theta_{enc}$. As shown in Fig.~\ref{3D_CRAN}, $f_{enc}$ consists of four blocks of 3D convolutional layers followed by plain vanilla feed forward networks to return a vectorized feature map $\textbf{A}_{c}^i \in \mathbb{R}^{h} $.  Unlike the energy or orthogonality constraint of the linear POD-Galerkin, the convolutional autoencoder automatically reduces the order of high-dimensional field from $O(N_{x}\times N_{y} \times N_{z})$ to $O(h)$ via a nonlinear manifold projection. $h < n \ll N_{x} \times N_{y} \times N_{z}$ and $h$ is usually an unknown hyperparameter for the optimal feature extraction based on the input dataset $\textbf{S}$.

LSTM networks are employed to model the evolution of the low-dimensional states. Using a one-to-one dynamic transformation $g_{lstm}$, the LSTM evolver maps the low-dimensional states $\{\textbf{A}_{c}^1\;\textbf{A}_{c}^2\;...\;\textbf{A}_{c}^{n-1}\}$ to the time advanced low-dimensional states $\{\hat{\textbf{A}}_{c}^2\; \hat{\textbf{A}}_{c}^3\;...\;\hat{\textbf{A}}_{c}^n\}$ in a closed-loop recurrent manner. 
\begin{equation} \label{evolver}
\hat{\textbf{A}}_{c}^i = g_{lstm}(\textbf{A}_{c}^{i-1};\theta_{lstm}), \qquad i=2,...,n,
\end{equation}
where $\theta_{lstm}$ represents the weight operators of the LSTM cell. 
Notably, in the current work, a single layer LSTM network is found to be sufficient for evolving the feature vectors $\textbf{A}_{c}$ obtained from the 3D snapshots. The process of 3D transpose convolutions is carried out via the
last four layers of the 3D CRAN and is depicted using the green arrows in Fig.~\ref{3D_CRAN}. The 3D transpose convolution
can be interpreted as the mirror of 3D convolution that upsamples/decodes the low-dimensional representations across four layers to reconstruct the high-dimensional state. For an evolved low-dimensional state $\hat{\textbf{A}}_{c}^{n+1}$, the decoder compositional function of the 3D CRAN can be expressed using 
\begin{equation}
\hat{\textbf{S}}^{n+1} = f_{dec}(\hat{\textbf{A}}_{c}^{n+1};\theta_{dec}),  
\end{equation}
where $f_{dec}$ is the trainable decoder space that is parametrized by $\theta_{dec}$ and $\hat{\textbf{S}}^{n+1}$ is a predicted high-dimensional 3D field. The end-to-end 3D CRAN on flow variables is unsupervised and is capable of predicting flow variables at fixed 3D probes as long as the prediction error is within an acceptable range of accuracy.

\subsubsection{Training and prediction:} \label{train_predict_CRAN}
\begin{figure*}
\centering
\includegraphics[width=0.9\textwidth]{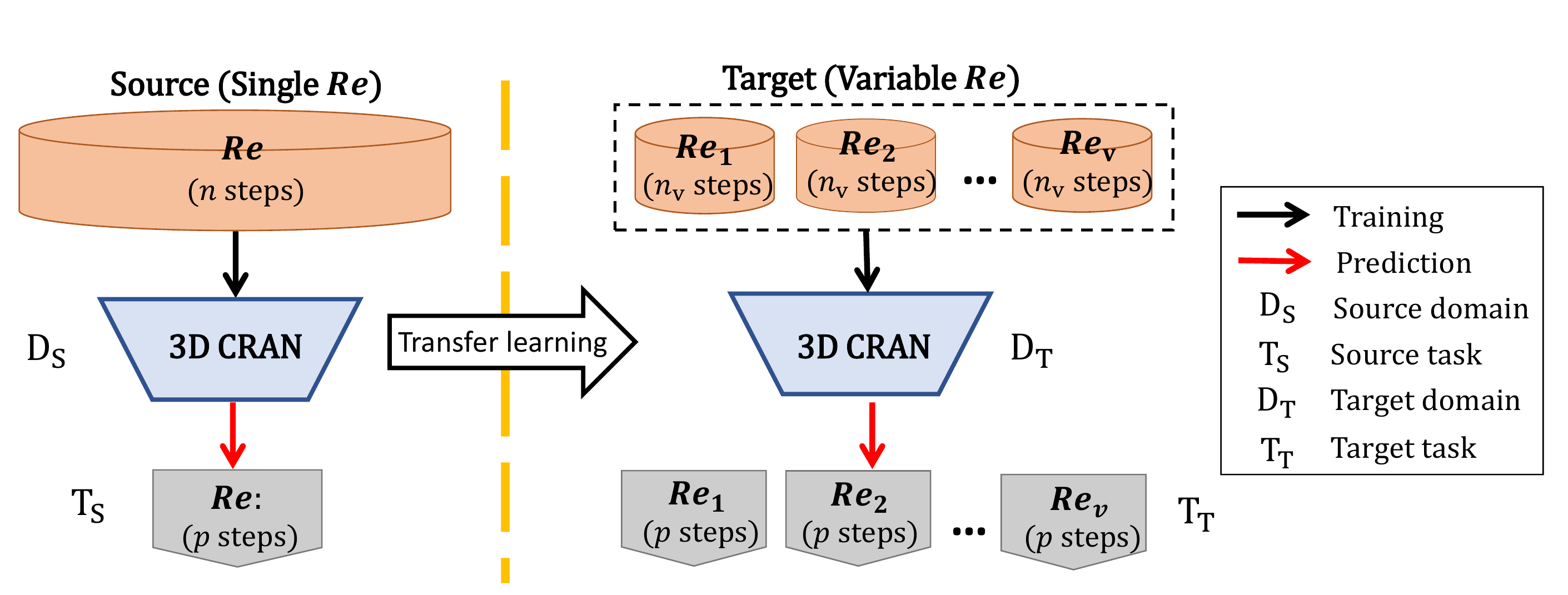}
\caption{Illustration of the transfer learning process for training variable $Re$ flows. Source refers to learning and extrapolating single $Re$ flows in time. Target refers to learning and extrapolating multiple $Re$ flows in time. Note that for training steps $n_{\mathrm{v}} < n$. }
\label{transfer_learning}
\end{figure*}
The snapshot dataset matrix $\textbf{S}$ is normalised and batch-wise arranged  for ease in training.
By subtracting the temporal mean from the dataset matrix and scaling the fluctuations, the dataset matrix is first normalized.  
The resultant dataset is re-arranged into a set of ${N}_{s}$ finite time training sequences, where each sequence consists of $N_{t}$ snapshots. This data processing converts the snapshot matrix $\textbf{S}$ into a 5D tuple of the following form: 
\begin{equation} \label{dn_cran_3}
\mathcal{S}=\left\{\mathcal{S}_{s}^{\prime 1}\; \mathcal{S}_{s}^{\prime 2}\; \ldots\; \mathcal{S}_{s}^{\prime N_{s} } \right\} \in[0,1]^{N_{x} \times N_{y} \times N_{z} \times N_{t} \times N_{s}}, 
\end{equation}
where each training sample $\mathcal{S}_{s}^{\prime j}=\left[\mathbf{S}_{s, j}^{\prime 1} \; \mathbf{S}_{s, j}^{\prime 2} \ldots \mathbf{S}_{s, j}^{N_{t}}\right]$ is a matrix consisting of the scaled database and each $\mathbf{S}_{s, j}^{\prime i} \in [0,1]^{N_{x} \times N_{y} \times N_z}$ is a normalized 3D flow snapshot. 

We utilize an unsupervised-supervised training strategy to train the CRAN architecture as there are two types of losses obtained in a CRAN. One is the unsupervised loss accounted for the 3D convolutional autoencoder reconstruction $\textbf{S}^{e}$ and the other is the supervised loss $\textbf{A}^{e}$ obtained from the evolution of the low-dimensional representations using the LSTM. 
This can be clarified from Fig.~\ref{3D_CRAN_training_prediction}(a). 
A hybrid loss function $E_h$ is constructed that equally weights the error in the unsupervised  and supervised losses of the CRAN. The target of the training is to find the CRAN parameters $\theta = \{ \theta_{enc}, \theta_{dec}, \theta_{lstm}\}$ such that for any sequence $\mathcal{S}_{s}^{\prime j}=\left[\mathbf{S}_{s, j}^{\prime 1} \; \mathbf{S}_{s, j}^{\prime 2} \ldots \mathbf{S}_{s, j}^{N_{t}}\right]$ and its corresponding low-dimensional representation $ \left[\mathbf{A}_{c,j}^{1} \; \mathbf{A}_{c,j}^{2} \ldots \mathbf{A}_{c,j}^{N_{t}}\right] $ the hybrid loss $E_h$ is minimized: 
\begin{equation}  \label{CRAN_loss_1}
\begin{aligned}
\theta &= \mathrm{argmin}_{\theta} (E_h), \\ 
&= \mathrm{argmin}_{\theta} (0.5\textbf{S}^{e} + 0.5 \textbf{A}^{e}).   
\end{aligned}
\end{equation}
Here, $\textbf{S}^{e}$ and $\textbf{A}^{e}$ are given using the following expressions: 
\begin{equation}  \label{CRAN_loss_2}
\begin{aligned}
\textbf{S}^{e} = \frac{1}{N_{t}} \sum_{i=1}^{N_{t}} \frac{\left\|\mathbf{S}_{s,j}^{\prime i}- f_{dec}(f_{enc}(\mathbf{S}_{s,j}^{\prime i}; \theta_{enc}); \theta_{dec})\right\|_{2}^{2} }{\left\|\mathbf{S}_{s,j}^{\prime i}\right\|_{2}^{2}}, \\
\textbf{A}^{e} = \frac{1}{N_{t}-1} \sum_{i=2}^{N_{t}} \frac{\left\|\mathbf{A}^{i}_{c,j}- g_{lstm}(\mathbf{A}^{i}_{c,j}; \theta_{lstm})\right\|_{2}^{2}}{\left\|\mathbf{A}^{i}_{c,j}\right\|_{2}^{2}}.
\end{aligned}
\end{equation}
The hybrid loss function given by Eqs.~(\ref{CRAN_loss_1})-(\ref{CRAN_loss_2}) is minimized over all the $j = 1,2,\dots,N_{s}$ training sequences using the adaptive moment optimization \cite{kingma2014adam}. 

The online prediction is straight forward and is depicted in Fig. \ref{3D_CRAN_training_prediction}(b) for easy comprehension. The low-dimensional representation 
$\textbf{A}_{c}^{n}\in \mathbb{R}^{h}$ is constructed using the encoder network for a given initial 3D flow snapshot $\textbf{S}^{n}\in [0,1]^{N_{x}\times N_{y} \times N_{z}}$ and the trained 3D CRAN parameters $\theta$. Eq. (\ref{evolver}) is applied iteratively for $p$ steps with  $\textbf{A}_{c}^{n}\in \mathbb{R}^{h}$ as the initial solution to generate predictions of the low-dimensional representations $ \hat{\textbf{A}}_{c} =  \{\hat{\textbf{A}}_{c}^{n+1}\; \hat{\textbf{A}}_{c}^{n+2}\;...\;\hat{\textbf{A}}_{c}^{n+p}\}$. Finally, the 3D state is reconstructed  from the low-dimensional representations at every time steps using the decoder network. We next detail the formulation of the training strategy for the 3D CRAN for variable $Re$-based flows using transfer learning. 

\subsection{Variable $Re$-based flows} \label{Tansfer_learning_VAR_RE}
As discussed in the previous section, scaling the CRAN to 3D is algorithmically straightforward as the framework largely relies on 3D CNNs to extract the flow features. 
However, this simple algorithmic extension can considerably increase the memory requirement, hyperparameter space, and training costs for the end-to-end learning model.
In some cases, the offline training time for learning a simple flow regime can take a matter of days or weeks if the CRAN framework starts learning from scratch. This often results in:  
(a) enormous computing power for training than the full-order model itself and, 
(b) large training data requirement for the neural network.
These challenges can complicate the process of training a 3D CRAN-based framework, especially for learning complex flow patterns involving variable $Re$-based flows. 

To overcome these challenges, transfer learning is beneficial. 
Transfer learning, employed in machine learning, refers to the use of a previously trained network to learn a new task. In transfer learning, a machine uses a previously learned task to increase generalisation about another. 
The neural parameters and task for a pre-trained network are called as the source domain and source task, respectively.  
Whereas, the neural parameters and task for a new network are called as the target domain and target  task, respectively.  
This is further elaborated in Fig.~\ref{transfer_learning}. 
For a source domain $\mathrm{D}_{\mathrm{S}}$ with a corresponding source task $\mathrm{T}_{\mathrm{S}}$ and a target domain $\mathrm{D}_{\mathrm{T}}$ with a corresponding task $\mathrm{T}_{\mathrm{T}}$, transfer learning is the process of improving the target CRAN predictive function by using the related information from $\mathrm{D}_{\mathrm{S}}$ and $\mathrm{T}_{\mathrm{S}}$, where $\mathrm{D}_{\mathrm{S}} \neq \mathrm{D}_{\mathrm{T}}$ or $\mathrm{T}_{\mathrm{S}} \neq \mathrm{T}_{\mathrm{T}}.$ The single source domain defined here can be extended to multiple target domains or tasks. 
%
This study employs transfer learning to train the 3D CRAN for a variable $Re$ flow regime on a limited data and training time. For this purpose, we load a pre-trained model of a single $Re$ case and optimize for variable $Re$ flows. The source domain and task are to learn and predict single $Re$ flows in time using 3D CRAN. On the flip side, the target domain and task become learning and prediction of multi-$Re$ flows in time using one 3D CRAN.

\subsubsection{Training and prediction}

We use the same 3D CRAN architecture and, training and prediction algorithms for variable $Re$ flows as described in section \ref{3D_CRAN_details}. The only difference is the preparation of the dataset matrix. 
For variable $Re$ flows, the time series data are 
snapshots of flow fields for different $Re$ values acquired from the full-order simulation. Consider that  $Re_{m}  = [Re_1 \; Re_2 \; ... \; Re_{\mathrm{v}}] $ be the range of Reynolds number for generating the full-order fields. The training dataset matrix $\textbf{S}(Re_{m})$ consists of time series data for different $Re$ fields (pressure or velocity) that are stacked as a matrix
\begin{equation}
   \textbf{S}(Re_{m}) = 
   \begin{Bmatrix} 
   [ \textbf{S}^1(Re_1) \;\; \textbf{S}^2(Re_1) \; \cdots \;  \textbf{S}^{n_{\mathrm{v}}}(Re_1) ], \\
   [ \textbf{S}^1(Re_2) \;\; \textbf{S}^2(Re_2) \; \cdots \; \textbf{S}^{n_{\mathrm{v}}}(Re_2) ], \\
     \vdots  \\
   [ \textbf{S}^1(Re_{\mathrm{v}}) \;\; \textbf{S}^2(Re_{\mathrm{v}}) \; \cdots \;  \textbf{S}^{n_{\mathrm{v}}}(Re_{\mathrm{v}}) ]
\end{Bmatrix}.  
\end{equation}
Note that $\textbf{S}(Re_{m}) \in \mathbb{R}^{N_x \times N_y \times N_z \times ({\mathrm{v}} n_{\mathrm{v}})}$ consists of $\mathrm{v}$ Reynolds numbers and each consisting $n_{\mathrm{v}}$ snapshots. $N_{x} = N_{y} = N_{z} = 64$ represents the number of data probes for the uniform voxel grid.  
The training dataset matrix is normalised and batch-wise arranged for ease in training. The matrix re-arranges in the following form  
\begin{equation}
\mathcal{S}(Re_m)=\left\{\mathcal{S}_{s}^{\prime 1}\; \mathcal{S}_{s}^{\prime 2}\; \ldots\; \mathcal{S}_{s}^{\prime N_{s} } \right\} \in[0,1]^{N_{x} \times N_{y} \times N_{z} \times N_{t} \times N_{s}}, 
\end{equation}
where each training sample $\mathcal{S}_{s}^{\prime j}=\left[\mathbf{S}_{s, j}^{\prime 1} \; \mathbf{S}_{s, j}^{\prime 2} \ldots \mathbf{S}_{s, j}^{N_{t}}\right]$ is a time series data at a particular $Re$ value.  $N_t$ are the evolver steps while $N_s$ are the number of batches. While training, sequence of different $Re$ field is selected randomly $\mathcal{S}_{s}^{\prime j} \subset \mathcal{S}(Re_m)$ to optimize the loss function given by Eqs.~(\ref{CRAN_loss_1})-(\ref{CRAN_loss_2}).

The prediction algorithm is the same as illustrated in Fig. \ref{3D_CRAN_training_prediction}(b). 
For a given initial 3D flow snapshot  $\textbf{S}^{n_{\mathrm{v}}}(Re_i)\in [0,1]^{N_{x}\times N_{y} \times N_{z}}$ at a Reynolds number say $Re_i \subset Re_{m}$, the trained 3D CRAN parameters and Eq. (\ref{evolver}) is applied iteratively for $p$ steps with  $\textbf{A}_{c}^{n_{\mathrm{v}}}(Re_i)\in \mathbb{R}^{h}$ as the initial solution. This helps to infer multiple-$Re$ fields in time for a chosen value of $Re_i$. 
With transfer learning, a 3D CRAN model can be built with comparatively less training data and time because the model is already pre-trained. This can be valuable in tasks where the data can be limited and unlabeled, for instance variable $Re$ flows. 

\section{Results and discussion} \label{R_and_d_sphere}
In this section, we test our proposed 3D snapshot-FTLR and DL-ROM methodologies for data-driven prediction of flow past a sphere. We are interested in integrating an end-to-end 3D spatial encoding-decoding and temporal evolution for a realistic CFD problem with usual boundary conditions. Of particular interest is to forecast flow fields for single and variable $Re$ flow information in the DL space using an optimized CRAN framework, while preserving the interface description from the voxel grid. 

\begin{figure}
\centering
\subfloat[]{\includegraphics[width = 0.5\textwidth]{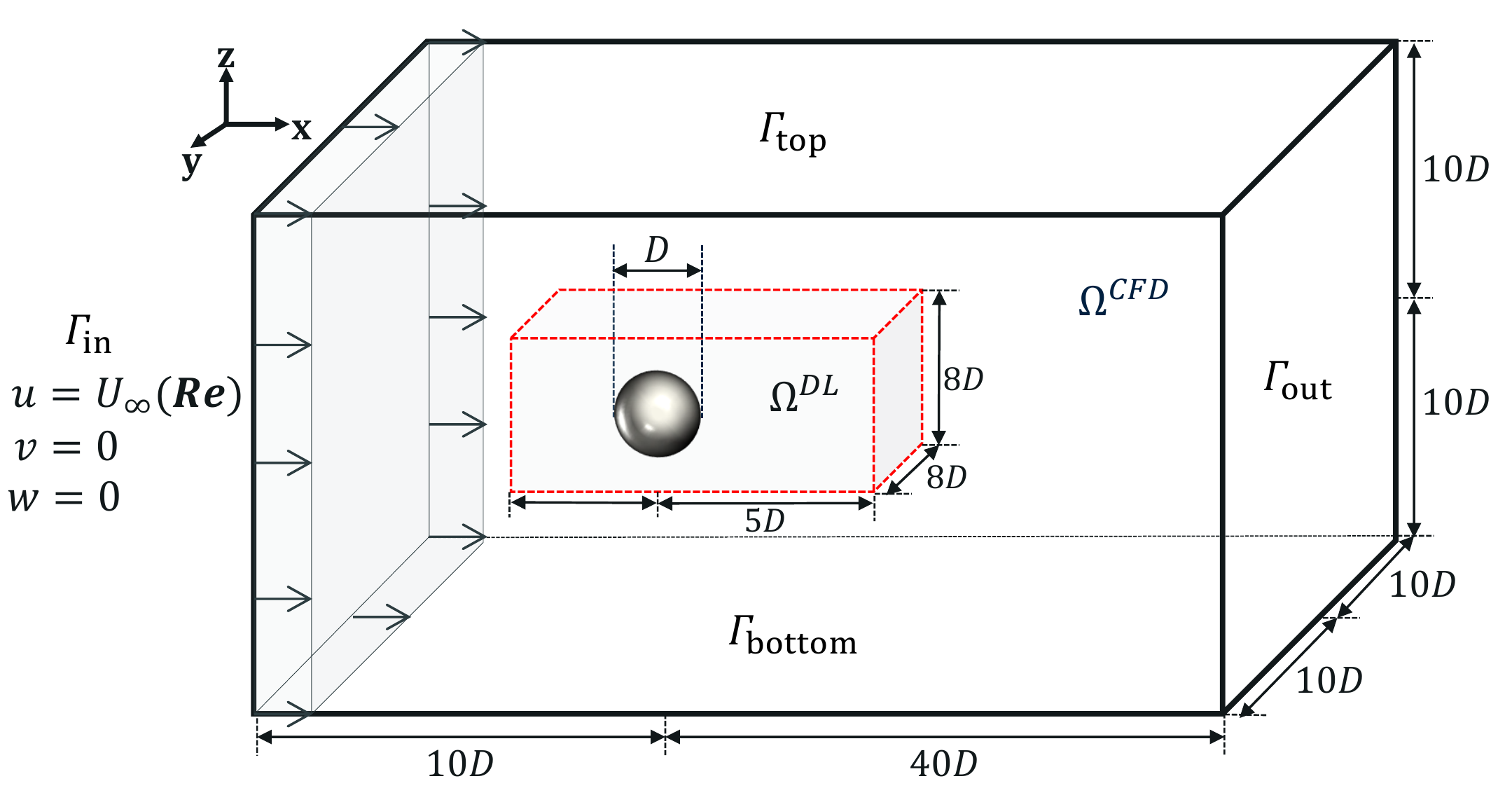}}\\ 
\subfloat[]{\includegraphics[width = 0.5\textwidth]{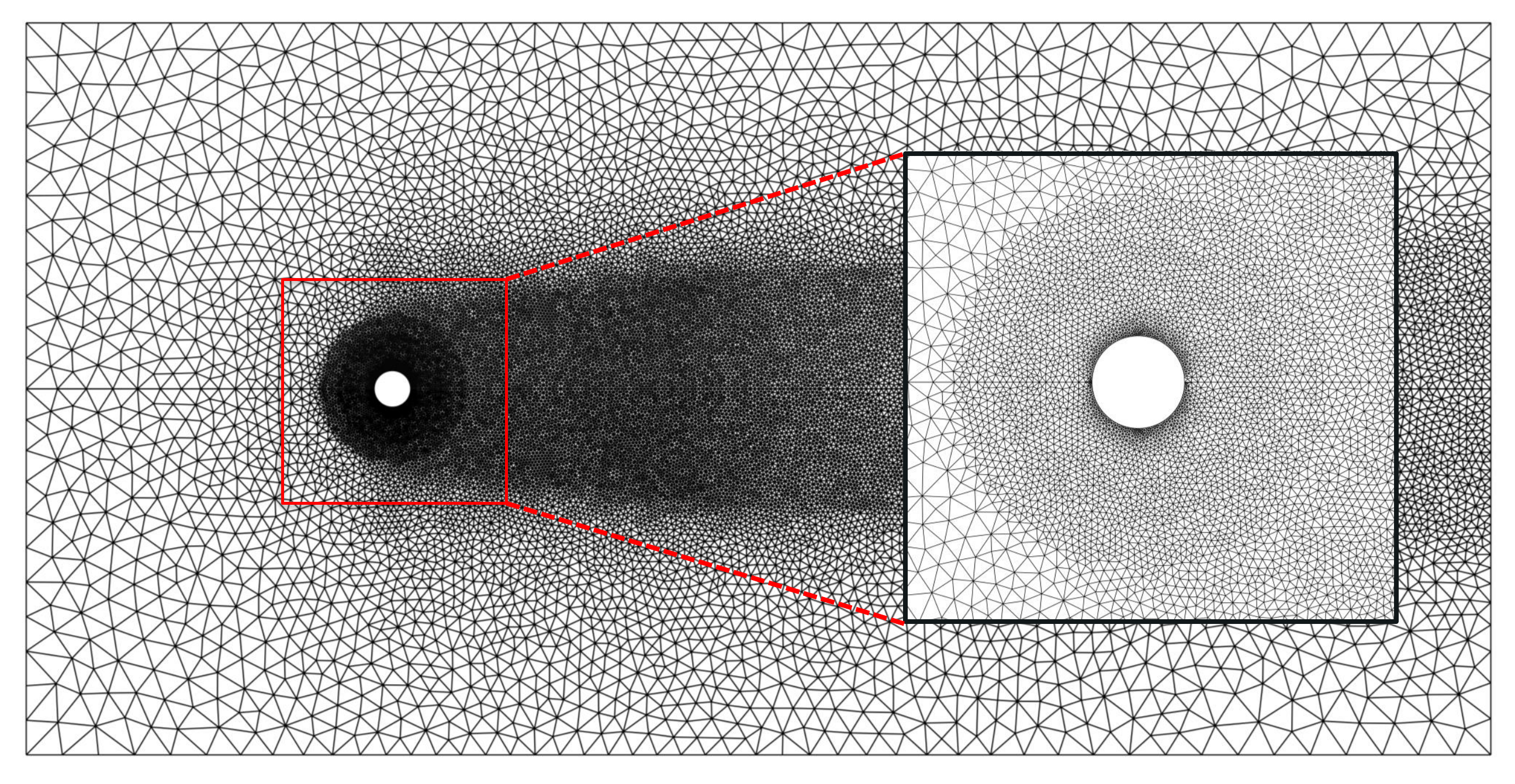}}
\caption{(a) Schematic and associated boundary conditions of flow past a sphere and deep learning domain of interest. (b) Representative CFD mesh for the entire domain sliced in $Z/D=10$ plane.}
\label{setup_single}
\end{figure}

\begin{figure*}
\centering
\subfloat[]{\includegraphics[width = 0.65\textwidth]{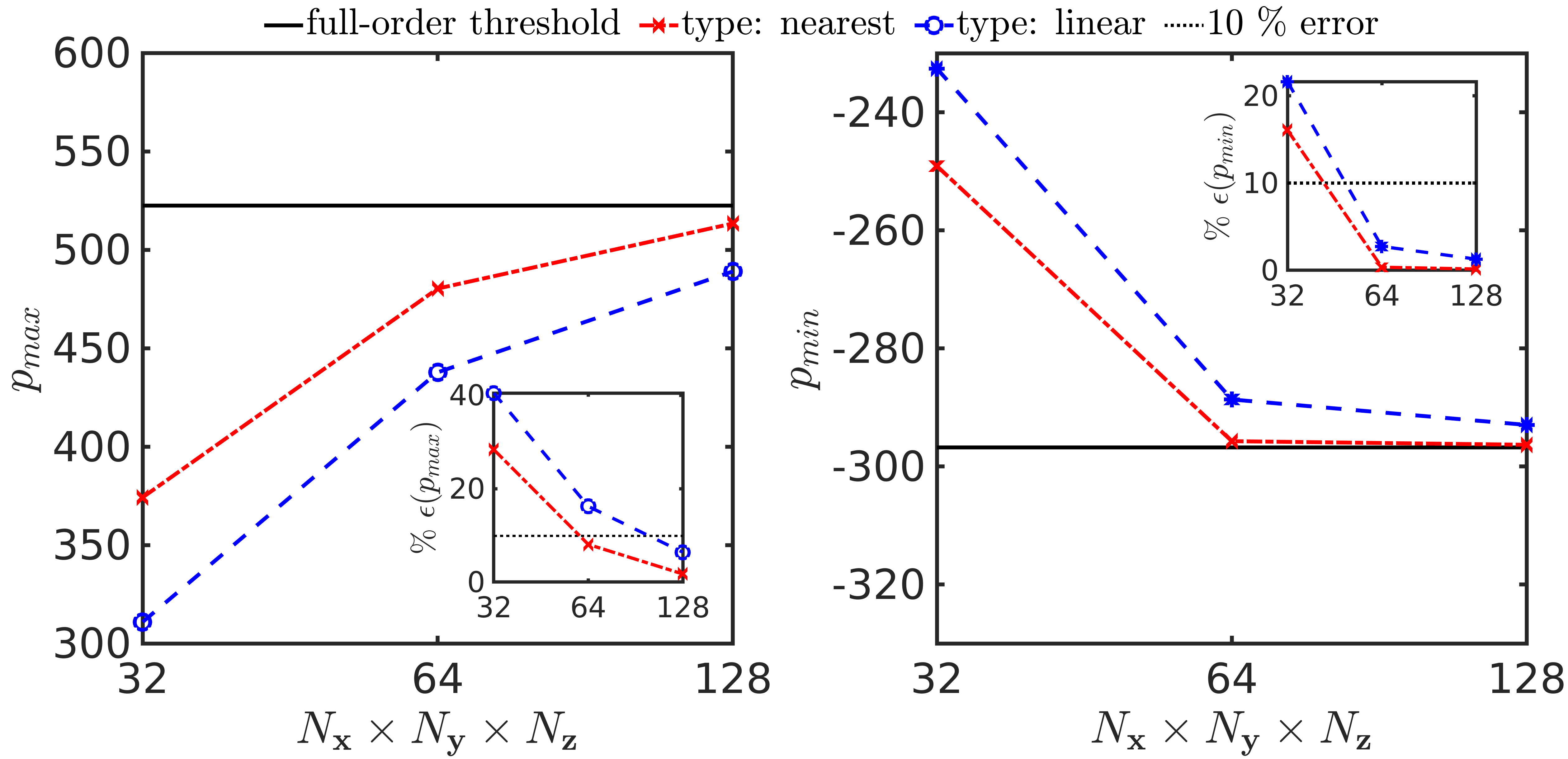}} 
\\
\subfloat[]{\includegraphics[width = 0.33\textwidth]{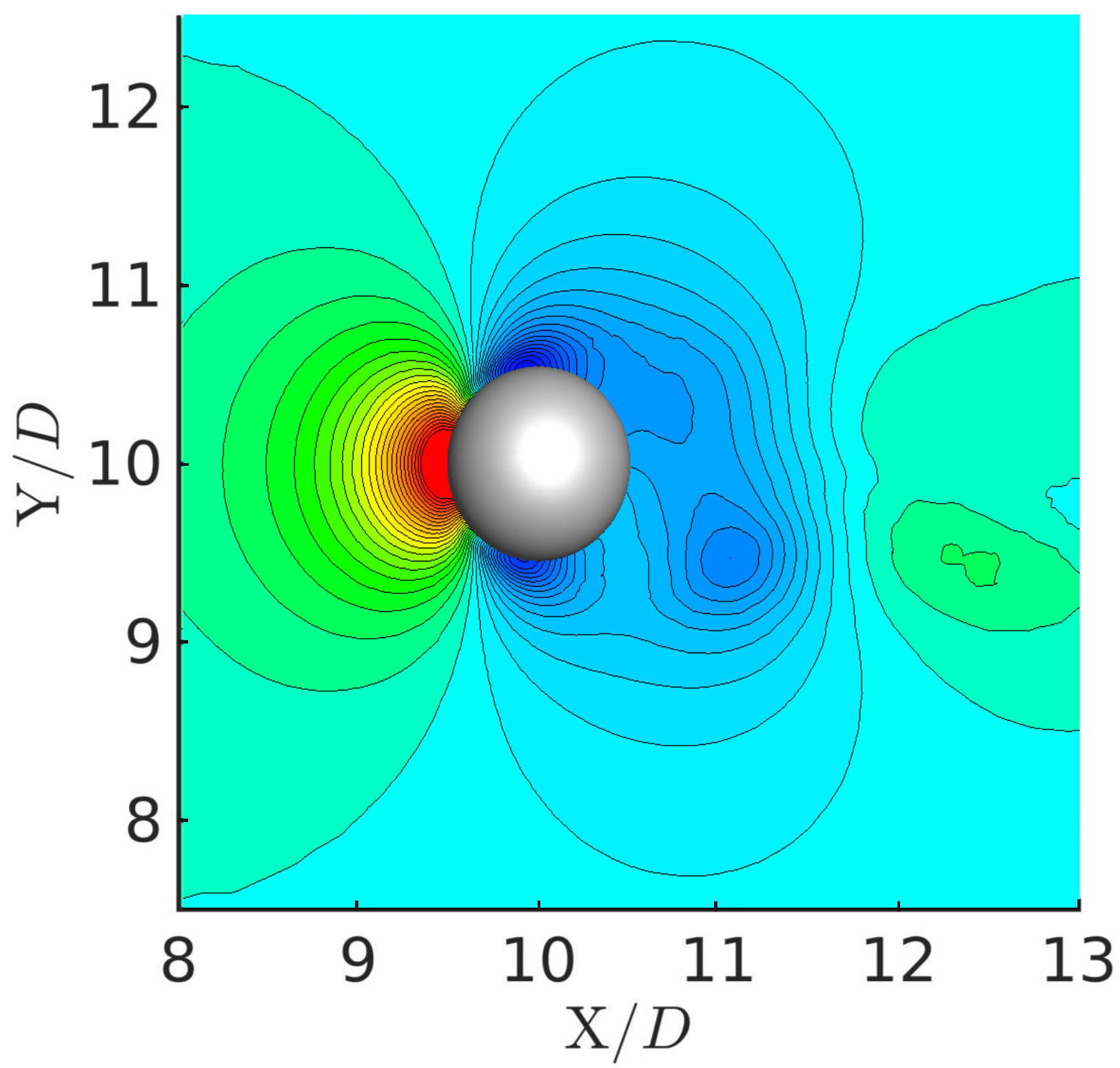} 
\includegraphics[width = 0.295\textwidth]{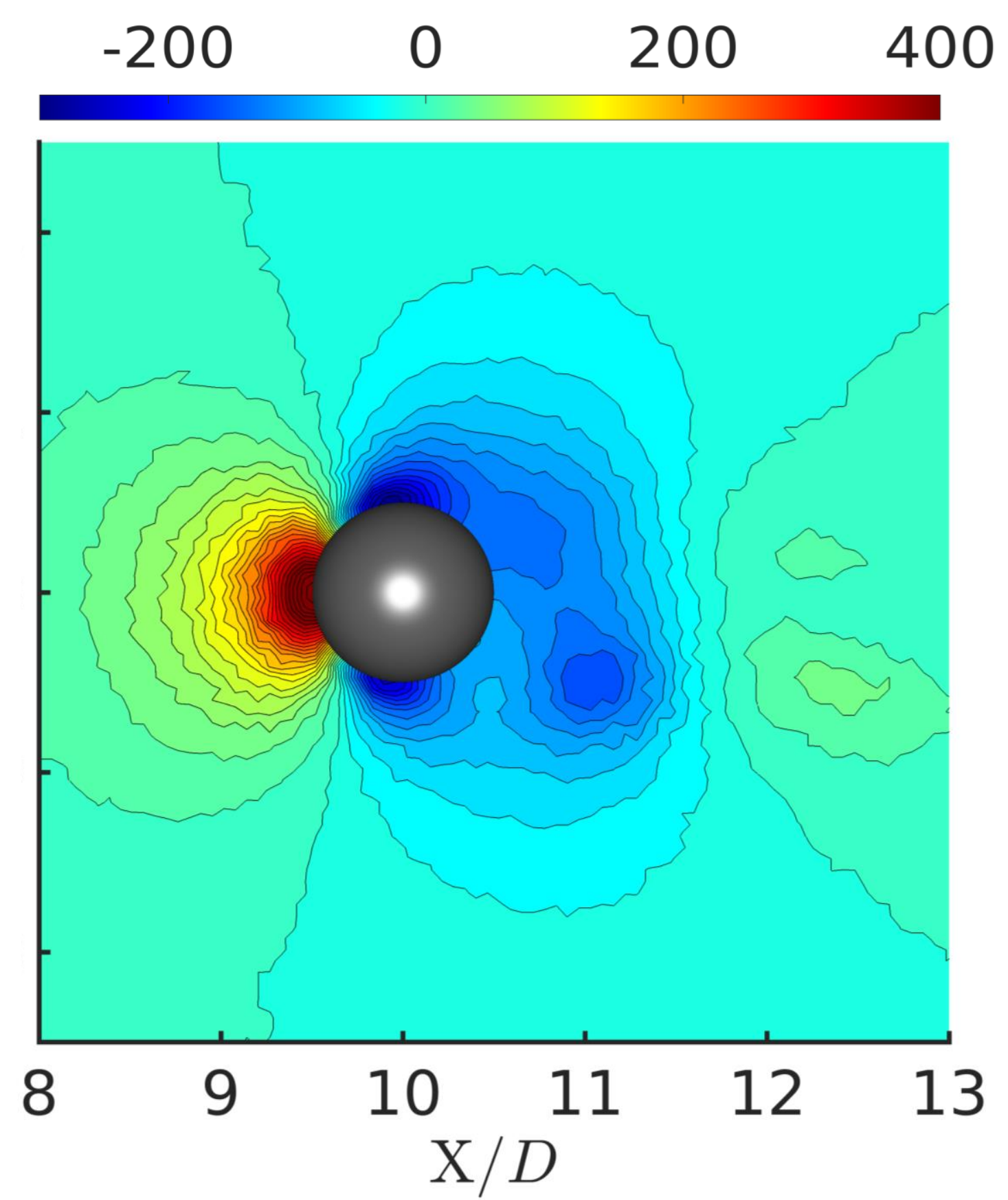}
\includegraphics[width = 0.28\textwidth]{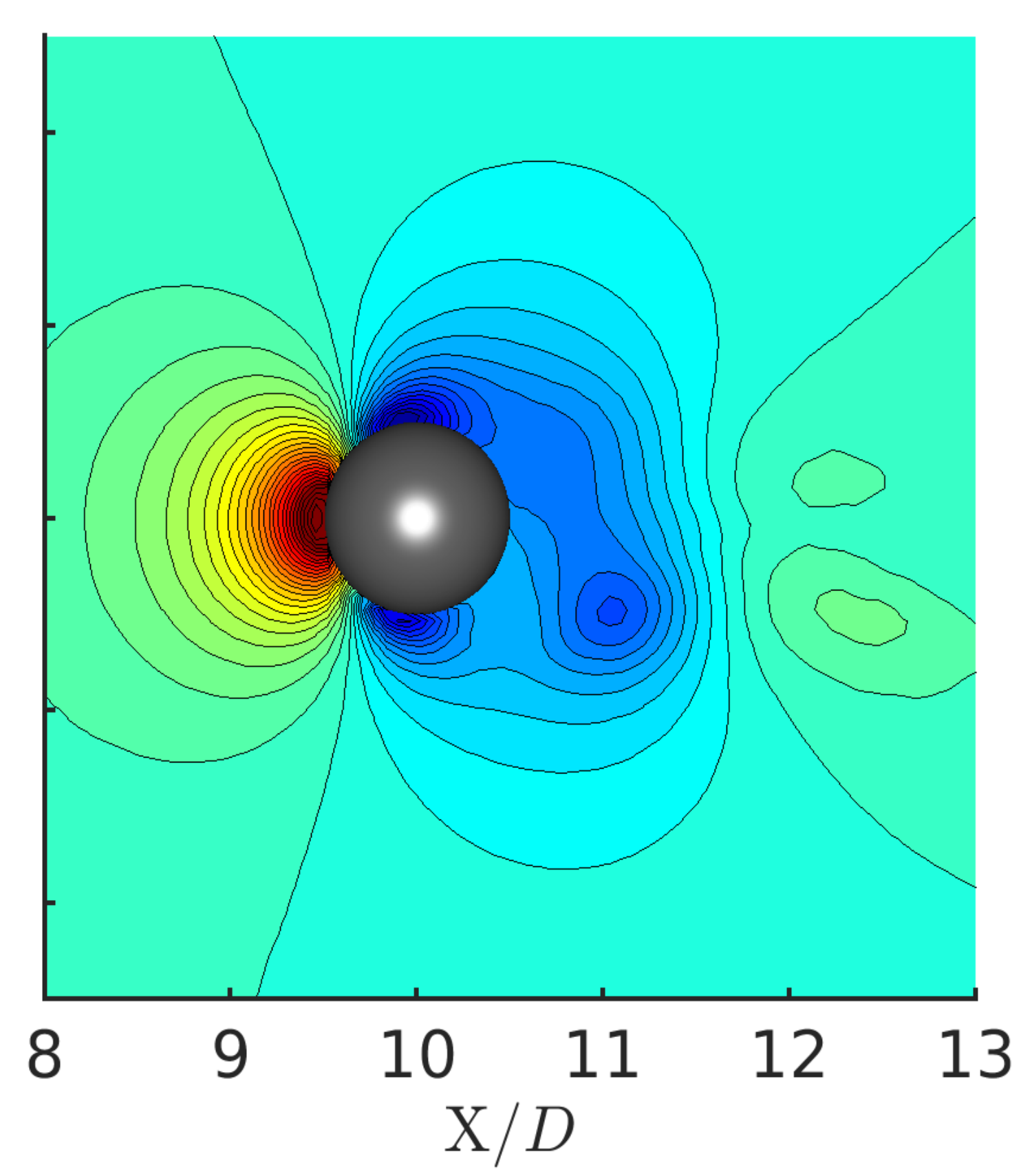}} 
\caption{The flow past a sphere: (a) Pressure field convergence with number of flow voxels for interpolation scheme variants. $\epsilon(.)$ is the respective relative error. (b) Descriptive behaviour of nearest-neighbour (middle) and linear interpolation (right) techniques for pressure field in the DL space  $128 \times 128 \times 128$ with respect to CFD space (left) sliced at $\mathrm{Z}/D=10$. Plots correspond to $t U_{\infty} / D = 200$.}
\label{pressure_convergence}
\end{figure*}

A schematic of the problem configuration employed for the full-order data generation $\mathrm{\Omega}^{CFD}$, of a stationary mounted sphere, is shown in Fig~\ref{setup_single} (a). The sphere system of diameter $D$ is installed in the 3D computational domain of size $50D \; \times \; 20D \; \times \; 20D$, with center at sufficient distances from the far-field boundaries to capture the downstream sphere wake. ${u}$, ${v}$ and ${w}$ depict the streamwise, transverse and vertical flow velocities in the $\mathrm{x}$, $\mathrm{y}$ and $\mathrm{z}$ directions, respectively. A uniform free-stream velocity profile $\{u, v, w\} = \{U_{\infty},0,0\}$ is maintained at the inlet boundary (${\Gamma_{\mathrm{in}}}$). Here, the free stream velocity is adjusted by defining the $Re$ of the problem using $Re= \rho^{\mathrm{f}} U_{\infty} D/ \mu^{\mathrm{f}} $, with $\rho^{\mathrm{f}}$ and $\mu^{\mathrm{f}}$ being the fluid density and viscosity, respectively. Along the top ${\Gamma_{\mathrm{top}}}$, bottom ${\Gamma_{\mathrm{bottom}}}$, and side surfaces, a slip-wall boundary condition is implemented while a traction-free Neumann boundary is maintained on the outlet ${\Gamma_{out}}$. The streamwise $\mathrm{C}_\mathrm{x}$, the transverse $\mathrm{C}_\mathrm{y}$ and  the vertical force coefficients $\mathrm{C}_\mathrm{z}$ on the submerged sphere are calculated by integrating the Cauchy stress tensor $\stf$ on the sphere $\ifsnt$ using Eq.~(\ref{force_eqns}).

In the following sub-sections, we apply the 3D snapshot-FTLR and CRAN methodologies for synchronously predicting the flow fields and the pressure force coefficients, by selecting a  DL domain of interest $\mathrm{\Omega}^{DL}$ as shown in Fig.~\ref{setup_single} (a).

\subsection{Flow past sphere at constant $Re$} \label{Re300}
\begin{figure*}
\centering
{\includegraphics[width = 0.33\textwidth]{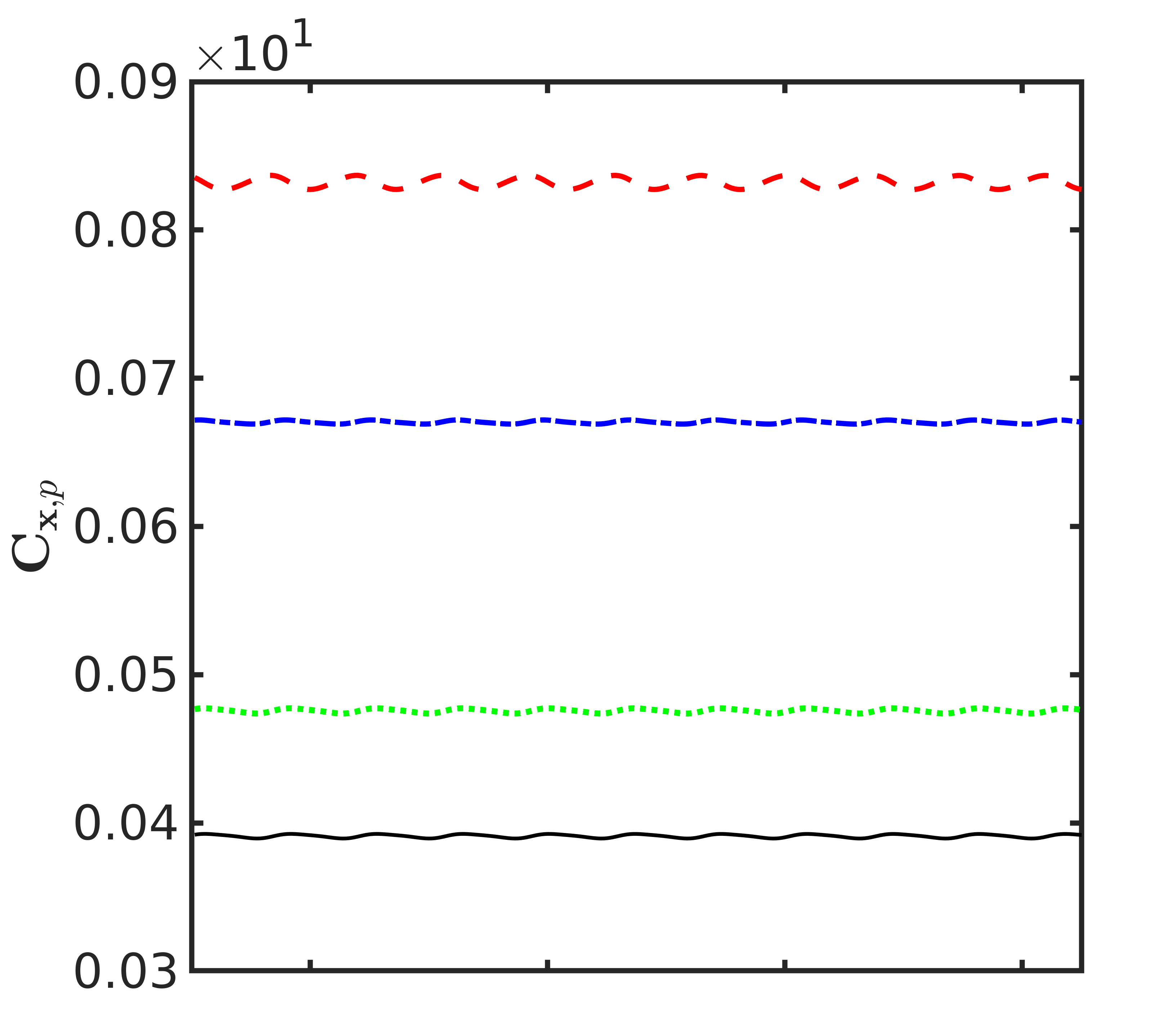}} 
{\includegraphics[width = 0.33\textwidth]{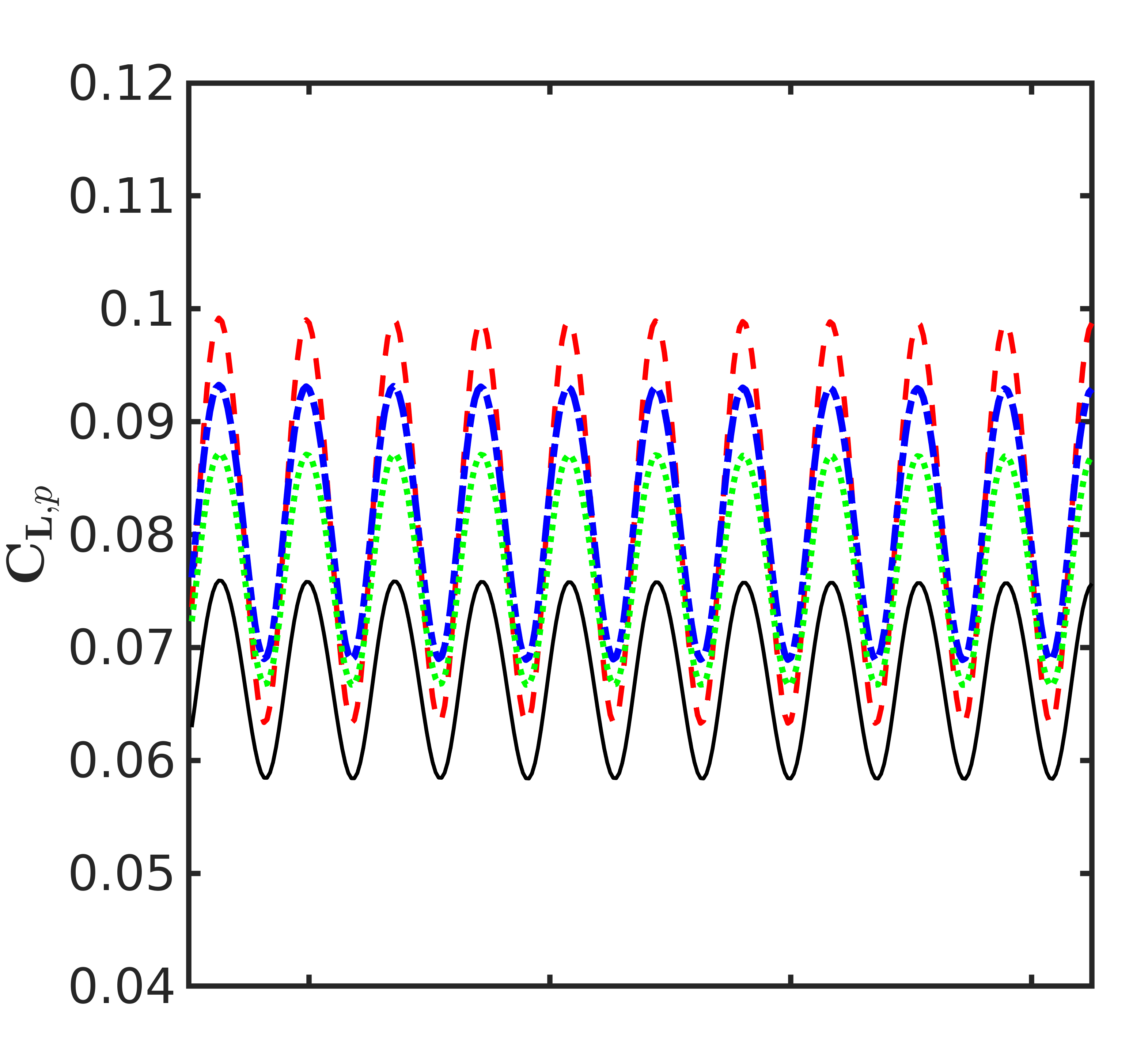}} 
\\ 
{\includegraphics[width = 0.33\textwidth]{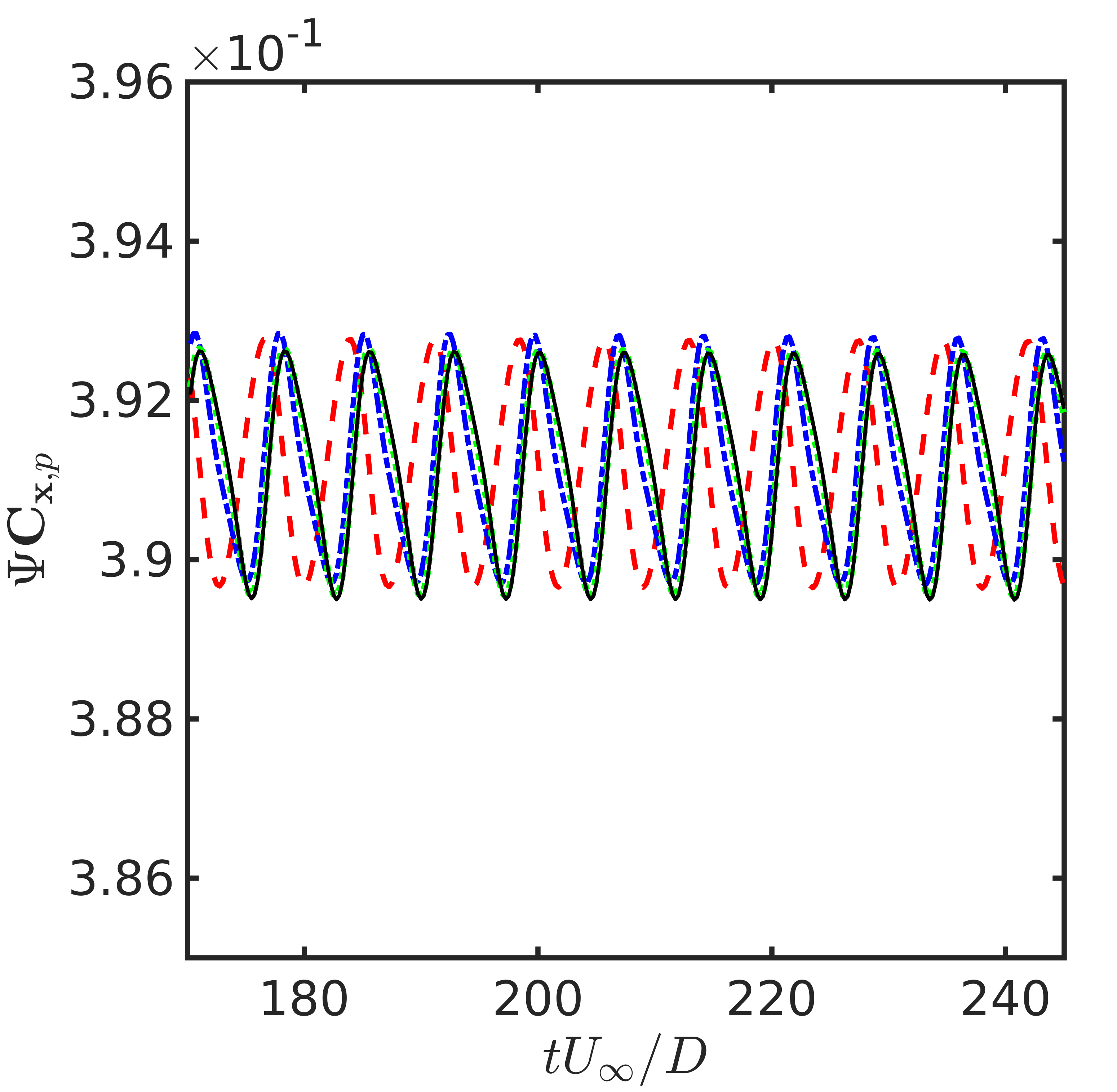}} 
{\includegraphics[width = 0.33\textwidth]{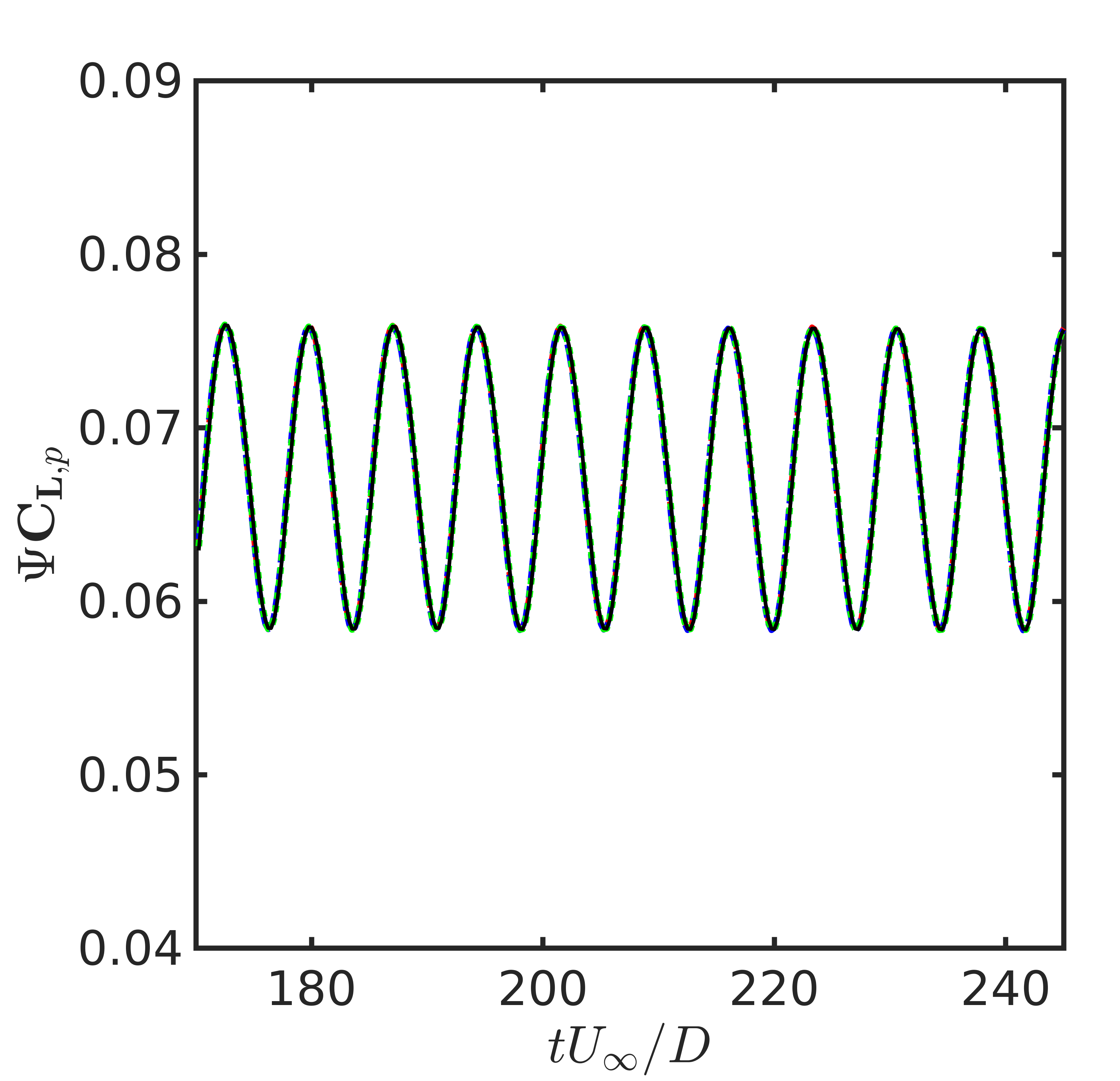}} 
\\
\caption{The flow past a sphere: Voxel interface force propagation and load recovery effects on various snapshot 3D DL grids (shown from 170-245 $t U_{\infty}/D$). Voxel drag and lift components (Row 1).  Corresponding voxel force recovery (Row 2). Red, blue and green dashed lines represent the grid $32 \times 32 \times 32$, $64 \times 64 \times 64$ and $128 \times 128 \times 128$, respectively. The black line depicts the full-order force.}
\label{forces_behavior_voxel}
\end{figure*}

To assess the 3D flow reconstruction and coarse-grain field predictions, we first examine our 3D data-driven DL-ROM framework on an unsteady fully submerged sphere problem in external flow at a single $Re$. The objective, herein, is to learn the strength and shedding orientation of unsteady planar-symmetric flow at $Re=300$, where the downstream hair-pinned shaped vortices shed strongly periodic in the near sphere wake. %
The drag $\mathrm{C}_{\mathrm{x}}$ and total lift $\mathrm{C}_{\mathrm{L}} = \sqrt{\mathrm{C}_{\mathrm{y}}^{2} + \mathrm{C}_{\mathrm{z}}^2}$ coefficients  demonstrate periodic behavior pattern from $ t U_{\infty}/D \approx 200$ onward. 
The full-order unsteady flow simulation is first carried out in the unstructured CFD domain to generate the full-order data. 
We use a time step of $\mathrm{\Delta} t = 0.05\;tU_{\infty}/ D$ for a total of $400\;tU_{\infty}/ D$ at $Re=300$. The final mesh is obtained using standard principles of mesh convergence and the full-order output details are tabulated in Table~\ref{tab:validation} to validate the FEM solver. 

\begin{table}
\centering
\caption{The flow past a sphere: The present study's full-order force values compared to benchmark data. $\overline{\mathrm{C}}_{\mathrm{D}}$ and $\overline{\mathrm{C}}_{\mathrm{L}}$ represent the mean drag and lift force coefficients, respectively. $St$ is the Strouhal number. }
\begin{tabular}{p{2.8cm}|p{1.5cm}|p{1.5cm}|p{1.5cm}}
\toprule
\toprule
Study &   $\overline{\mathrm{C}}_{\mathrm{D}}$ & $\overline{\mathrm{C}}_{\mathrm{L}}$ & $St$ \\
\bottomrule
\bottomrule
Present & $0.669$ & $0.082$ & $0.137$\\
Johnson and Patel\cite{johnson1999flow} & $0.656$ & $0.069$ & $0.137$ \\
\bottomrule
\end{tabular} 
\label{tab:validation}
\end{table}

A total of $1600$ time snapshots of point cloud data are saved at every $0.25\;tU_{\infty}/D$ for the pressure and $\mathrm{x}$-velocity field. From these full-order data, $n=1000$ snapshots (from 95-345 $tU_{\infty}/D$) are kept for training and $n_{ts}=100$ (from 345-370 $tU_{\infty}/D$) are reserved for testing. Thus, the total time steps in this analysis are $N=1100$. We further organize the test data in groups of every $p=20$ time steps to assess the compounding error effects in the multi-step predictions from the 3D CRAN solver. 
%
After generating the full-order point cloud dataset, we apply the 3D CRAN framework to forecast the flow fields past a sphere in a DL-based voxel grid for $Re=300$ while preserving the exact interface description.

We employ the snapshot-FTLR to bring field uniformity in the DL space, while recovering forces as described in section \ref{3D-FTLR}. A DL space $\mathrm{\Omega}^{DL}$ of dimension $8D \times 8D \times 8D$ is selected with $ \approx 5D$ length kept for downstream sphere wake as shown in Fig.~\ref{setup_single}(a). The point cloud CFD training/testing data (for instance pressure field) ${\textbf{s}} = \{ {\textbf{s}}^{1} \; {\textbf{s}}^{2} \; ... \; {\textbf{s}}^{N} \} \in \mathbb{R}^{m \times N}$ is interpolated and projected as spatially uniform 3D snapshots ${\textbf{S}} = \{ {\textbf{S}}^{1} \; {\textbf{S}}^{2} \; ... \; {\textbf{S}}^{N} \} \in \mathbb{R}^{N_{x} \times N_{y} \times \times N_{z} \times N}$ in the chosen DL space. $N_{x}$, $N_{y}$ and $N_{z}$ are number of flow voxels in the $\mathrm{x}$, $\mathrm{y}$ and $\mathrm{z}$ directions, respectively. The field uniformity reduces the model-order fidelity and unstructured mesh complexity by mapping the $m$-dimensional unstructured dataset on a 3D reference grid.
We compare the field interpolation methods provided by $griddata$\cite{SciPy}: nearest and linear methods, with respect to the number of flow voxels. This is performed by sampling the field's maximum and minimum values for various 3D DL grid resolution at an instant $t U_{\infty} / D = 200$. On DL grid refinement, the nearest method levels-off to the true $p_{max}$ and $p_{min}$ for a pressure instant as illustrated in Fig.~\ref{pressure_convergence}(a). Because this method assigns the value of the nearest neighbour in the unstructured information, this effect is expected. The linear interpolation approach, however, linearly converges $p_{max}$ and $p_{min}$ to the true full-order values on grid refinement. 

Fig.~\ref{pressure_convergence}(b) compares the field interpolation methods for pressure field with respect to the full-order on a $128 \times 128 \times 128$ DL voxel grid. The presence of the sphere boundary is ensured by masking the exact interface description in the 3D DL grid. It can be interpreted that, because of a discontinuous assignment of fields at the specified probes, the nearest method contains oscillations compared to the full-order description. With the linear interpolation and projection, a nearly perfect match is obtained in terms of the descriptive near wake snapshot ($\epsilon({p_{max}}) \approx 7\%,\;\epsilon({p_{min}}) \approx 2\%$) devoid of noises. The qualitative description is further substantiated by the convergence behavior in Fig.~\ref{pressure_convergence}(a). Hence, we rely on the linear interpolation technique for 3D coarse-grain field assignment.

\begin{figure*}
\centering
\includegraphics[width = 0.75\textwidth]{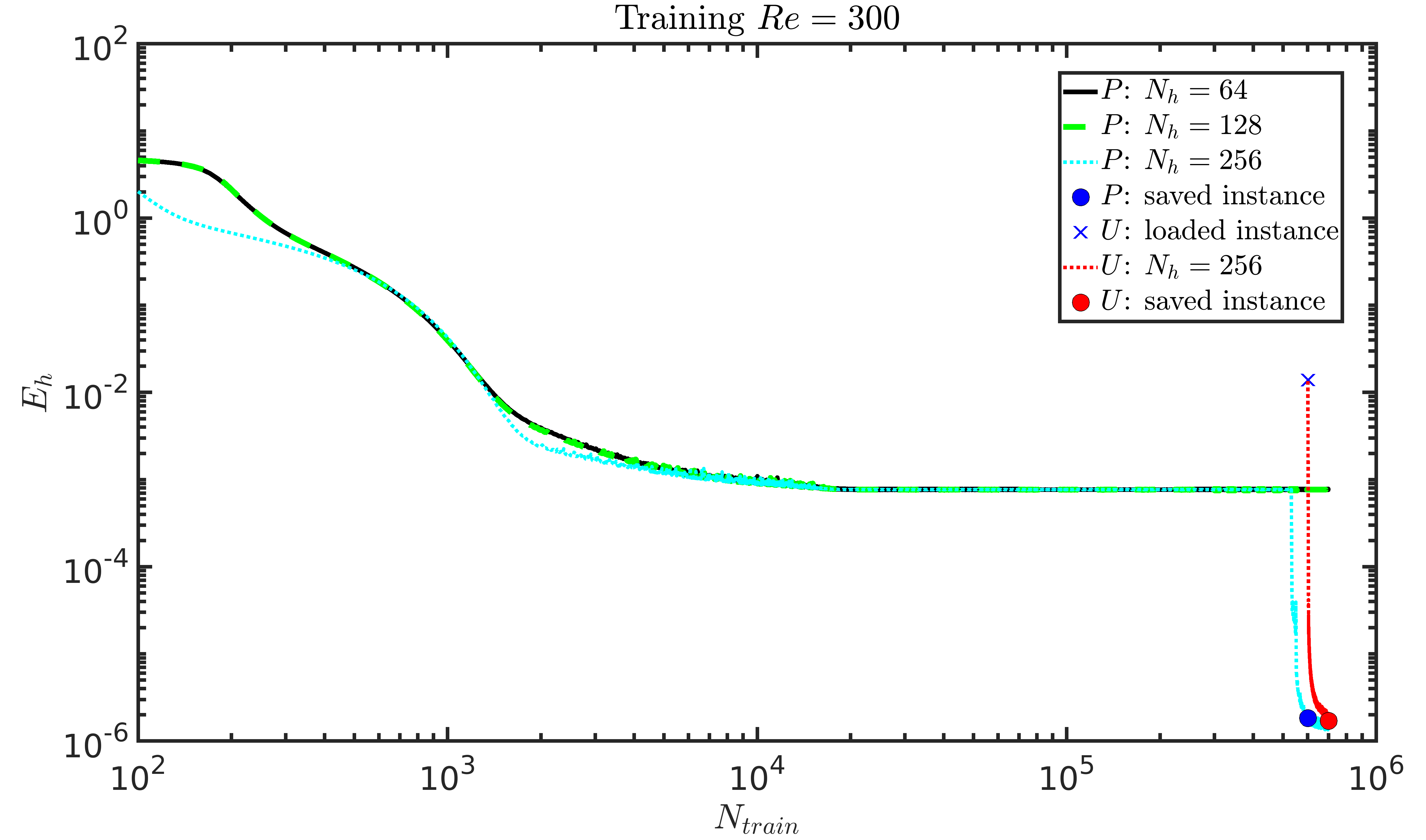}
\caption{The flow past a sphere: Evolution of the loss function $E_{h}$ with training iterations for different evolver cell sizes $N_h$. $P$ and $U$ denote the 3D CRAN trained with pressure and x-velocity datasets, respectively. Blue and red dots depict the saved instances for testing the pressure and x-velocity fields, respectively. The blue cross represents the initialization of velocity training from saved pressure parameters. }
\label{Re300_tr_loss}
\end{figure*}

The total voxel force propagation (drag and lift force components) $\overline{\textbf{F}}_{b} = \{ \overline{\textbf{F}}_{b}^{1}, \overline{\textbf{F}}_{b}^{2}, ..., \overline{\textbf{F}}_{b}^{N}  \}$ are obtained from the interface voxels on various 3D DL Cartesian grids using Eq.~(\ref{eq12}). The primary idea is to recover this bulk quantity by a functional corrective mapping $\Psi$ on the chosen DL grid using the exact fluid-solid interface location.  Fig.~\ref{forces_behavior_voxel} demonstrates the DL grid dependence of normalized pressure voxel forces $\overline{\textbf{F}}_{b} / 0.5 \rho^{f}U_{\infty}^{2}D $ and their data recovery effects using the mapping $\mathrm{\Psi}$. It can be interpreted that low-resolution leads to mean and derivative errors in the voxel forces compared with the full-order CFD forces. The primary reasons are the considerable loss of boundary fidelity in the DL grid and a linear force calculation using finite difference. 

Fig.~\ref{forces_behavior_voxel} also depicts the force correction by observing the $\Psi$ mapping and correcting on the training forces. For the flow field predictions using 3D CRAN, we rely on the DL grid $64 \times 64 \times 64$ because it accounts for a reasonable force recovery across all components  ($\epsilon(\mathrm{C}_{\mathrm{x},p}) = 0.0019,\;\epsilon(\mathrm{C}_{\mathrm{L},p}) = 0.0026$) without requiring the need of super-resolution. With super-resolution, the voxel force errors are indeed decreased. However, we want to refine the DL grid to the point where the force coefficients can be transformed to the full-order with mean and derivative error quantification using Eqs.~(\ref{eq13})-(\ref{eq15}). This process facilitates an optimal uniform grid to carry the neural prediction. The snapshot-FTLR procedure is scalable for voxel grid selection so that 3D CNNs can be conveniently integrated with the point cloud full-order dataset. 

With the chosen DL grid $N_{x}=N_{y}=N_z=64$, the 3D CRAN is employed for the coarse-grain flow field predictions. The coarse-grain pressure information $\textbf{S} = \left\lbrace\textbf{S}^{1}\;\textbf{S}^{2}\dots\;\textbf{S}^{N}\right\rbrace \in \mathbb{R}^{N_{x} \times N_{y} \times N_{z} \times N}$  is decomposed into $n=1000$ learning data (95-345 $tU_{\infty}/D$) and $n_{ts}=100$ analysis data (345-370 $tU_{\infty}/D$). 
Following this, standard principles of data normalization and batch-wise arrangement are adopted to generate the 5D scaled featured input 
$
\mathcal{S}=\left\{\mathcal{S}_{s}^{\prime 1} \; \mathcal{S}_{s}^{\prime 2}  \; \ldots \; \mathcal{S}_{s}^{\prime N_{s} } \right\} \in[0,1]^{N_{x} \times N_{y} \times N_{z} \times N_{t} \times N_{s}}
$
with $N_s$ = 40 and $N_t = 25$. 
The encoding space of the 3D CRAN encodes the $64 \times 64 \times 64$ flow voxel-based input dimension via four layers of 3D CNN operation with a constant kernel size of $5 \times 5 \times 5$  and stride $2 \times 2 \times 2$.
Every CNN operation reduces the input size by half, with number filters increasing by twice in every layer. Three feed-forward networks further map the feature vectors until a finite low-dimensional encoding $\textbf{A}_{c} \in \mathbb{R}^{N_{h}}$ is achieved with $N_h << N_x \times N_y \times N_z$. The architecture is detailed in section \ref{3D_CRAN_details}. Since 3D CNNs can considerably increase the trainable variables, the cost of hyperparameter tuning and training is very high. This can result in an increase in computing power for training the framework. Table~\ref{tab:cran_training_resources} depicts a comparison of the computational resources used for the 3D CRAN training and the 2D CRAN\cite{bukka2021assessment}. We note that scaling the CRAN architecture to three-dimension increases the trainable parameters by an order magnitude with the increase in random access memory (RAM) and training time.

\begin{table}
\centering
\caption{Comparison of computational resources used for 3D CRAN and 2D CRAN training.}
\begin{tabular}{p{3.0cm}|p{2.5cm}|p{2.5cm}}
\toprule
\toprule
 &  3D CRAN &  2D CRAN\cite{bukka2021assessment} \\
  &  (flow past sphere) & (flow past cylinder)   \\
\bottomrule
\bottomrule
DL grid & $64 \times 64 \times 64 $ & $64 \times 64  $\\
Training snapshots & $n=1000$  &  $n=1000$ \\
RAM (gB) & 32 & 16\\
Processor number  & Single GPU node & Single CPU node\\
Processor type  & NVDIA v100 & Intel E5 2690 \\
Trainable parameters   & $\theta \approx 6 \times 10^6$ & $\theta \approx 3 \times 10^5$ \\
Mini-batch size   & $n_s= 1-2$ & $n_s= 2-5$ \\
Training time  & 64 h & 16 h \\
\bottomrule
\end{tabular} 
\label{tab:cran_training_resources}
\end{table}

\begin{figure*}
\centering
{\includegraphics[width = 0.332\textwidth]{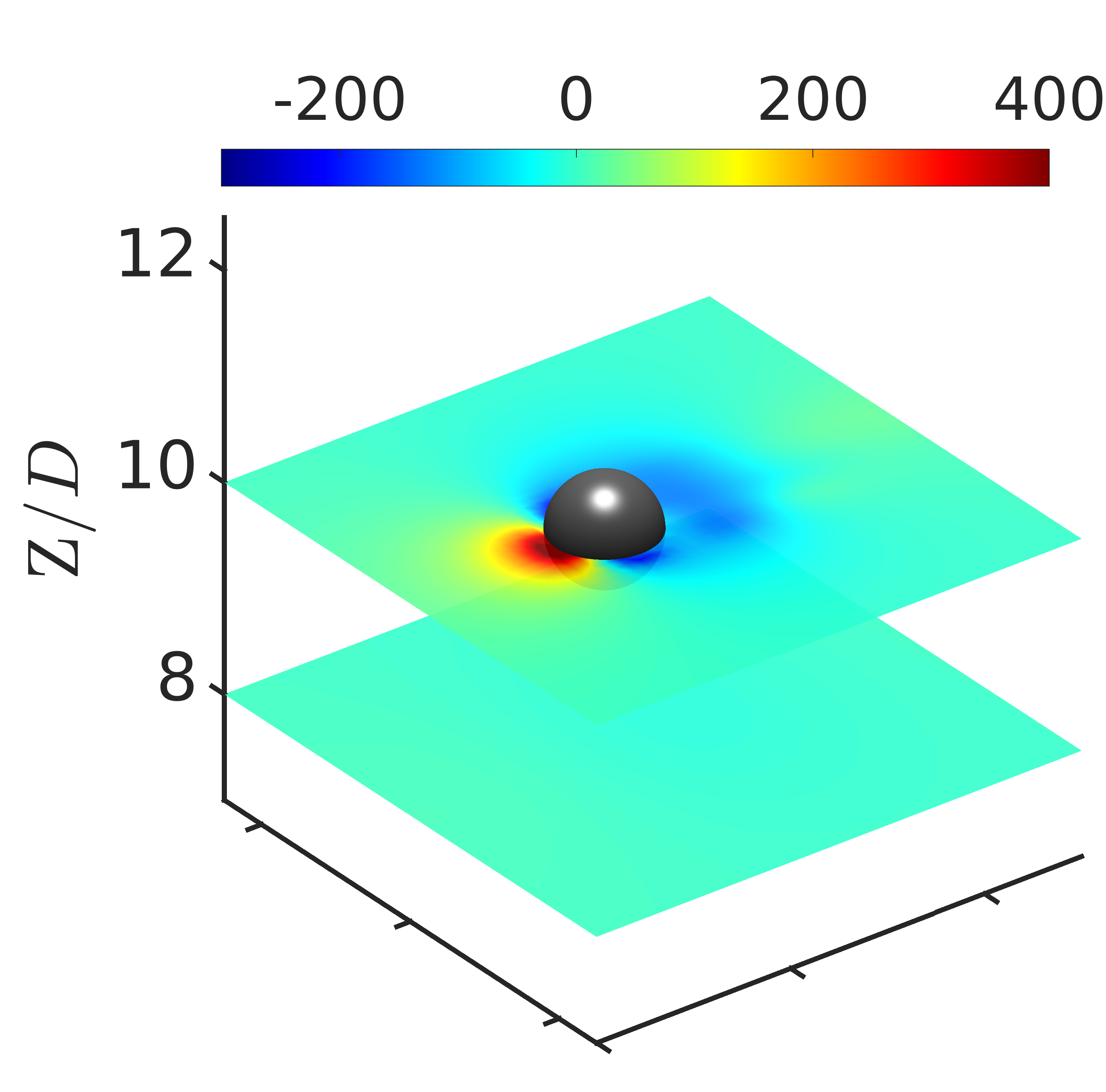}
\hspace{0.005\textwidth}
\includegraphics[width = 0.3\textwidth]{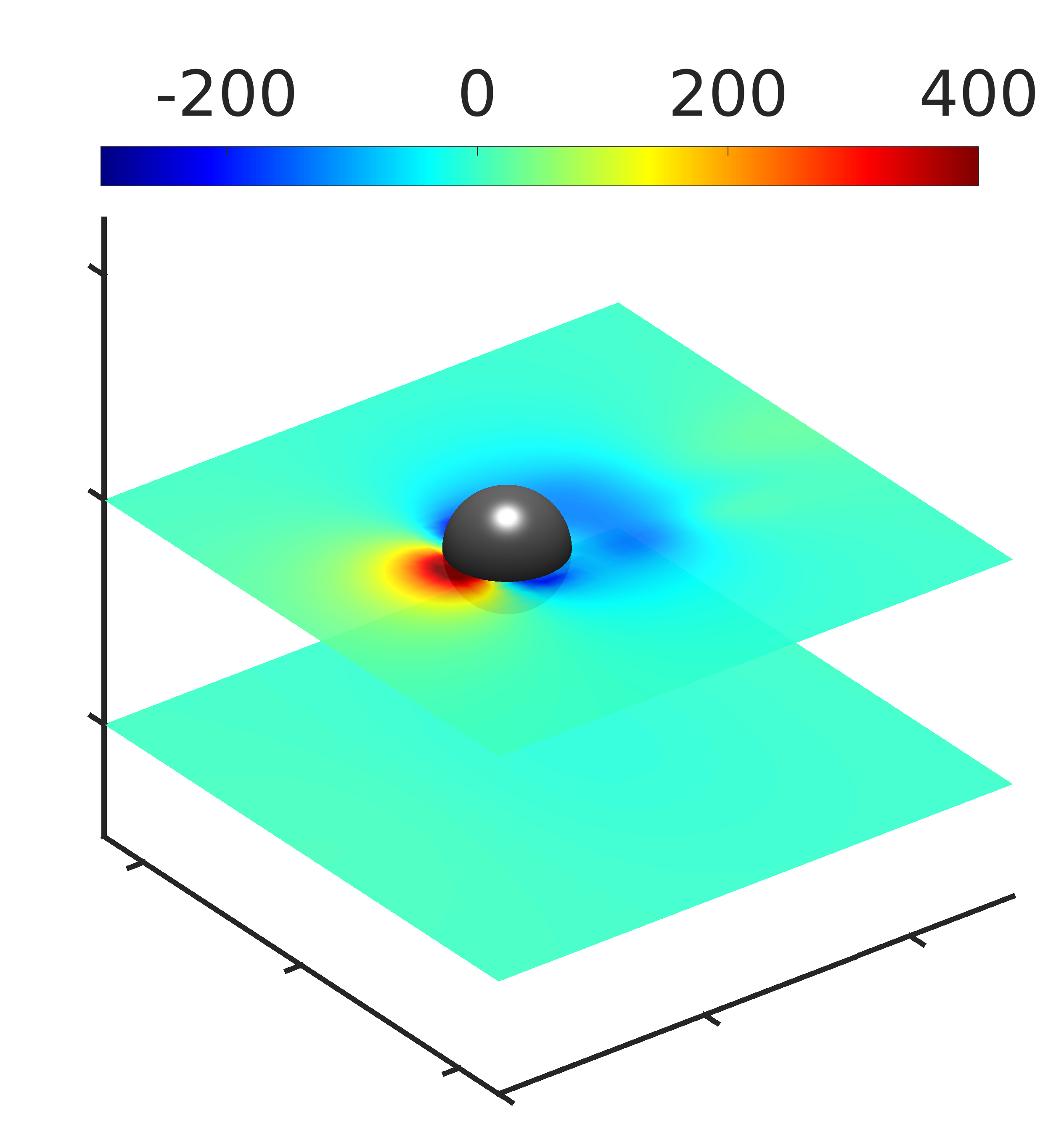}
\hspace{0.005\textwidth}
\includegraphics[width = 0.3\textwidth]{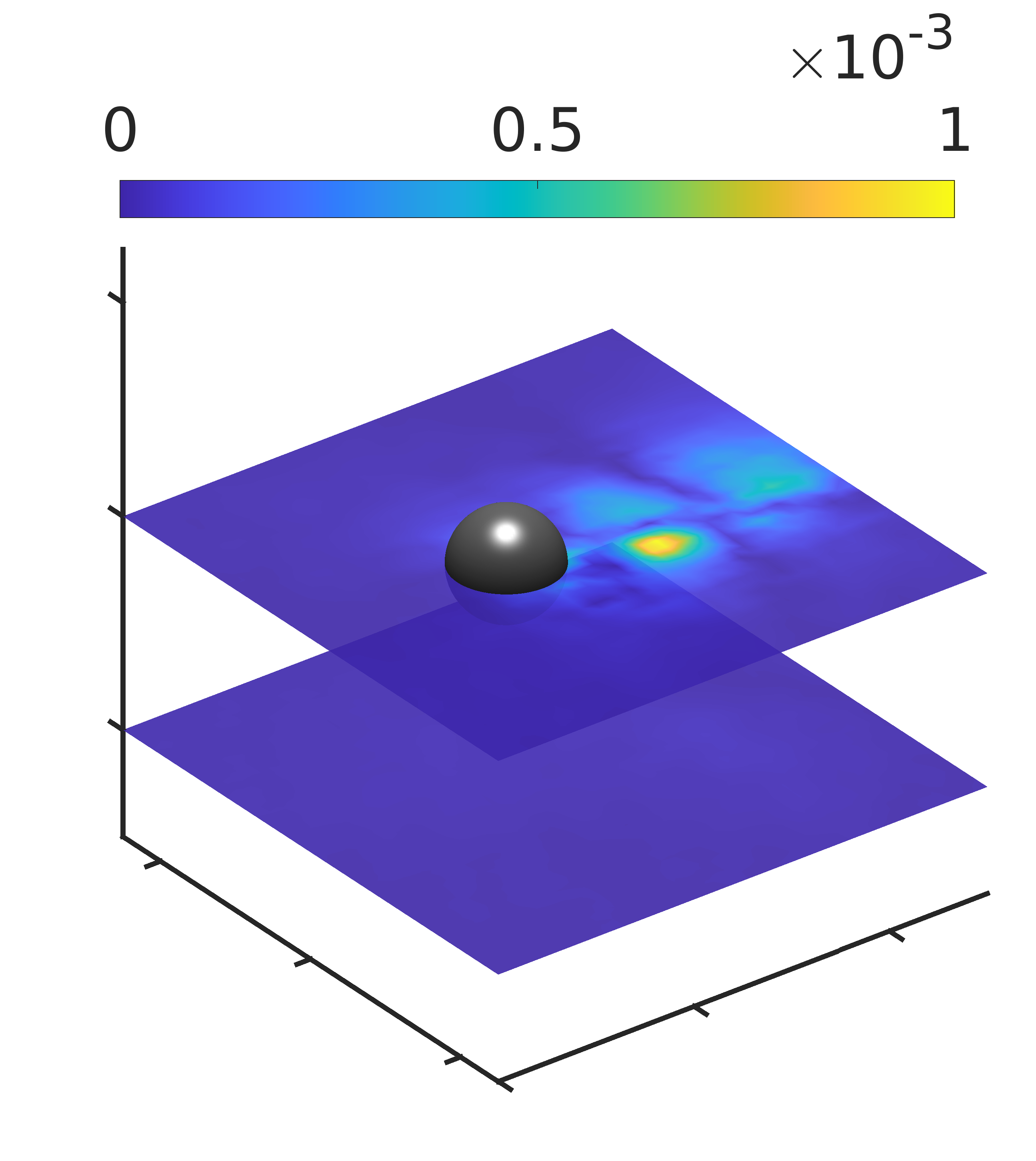}}
\\
{\includegraphics[width = 0.332\textwidth]{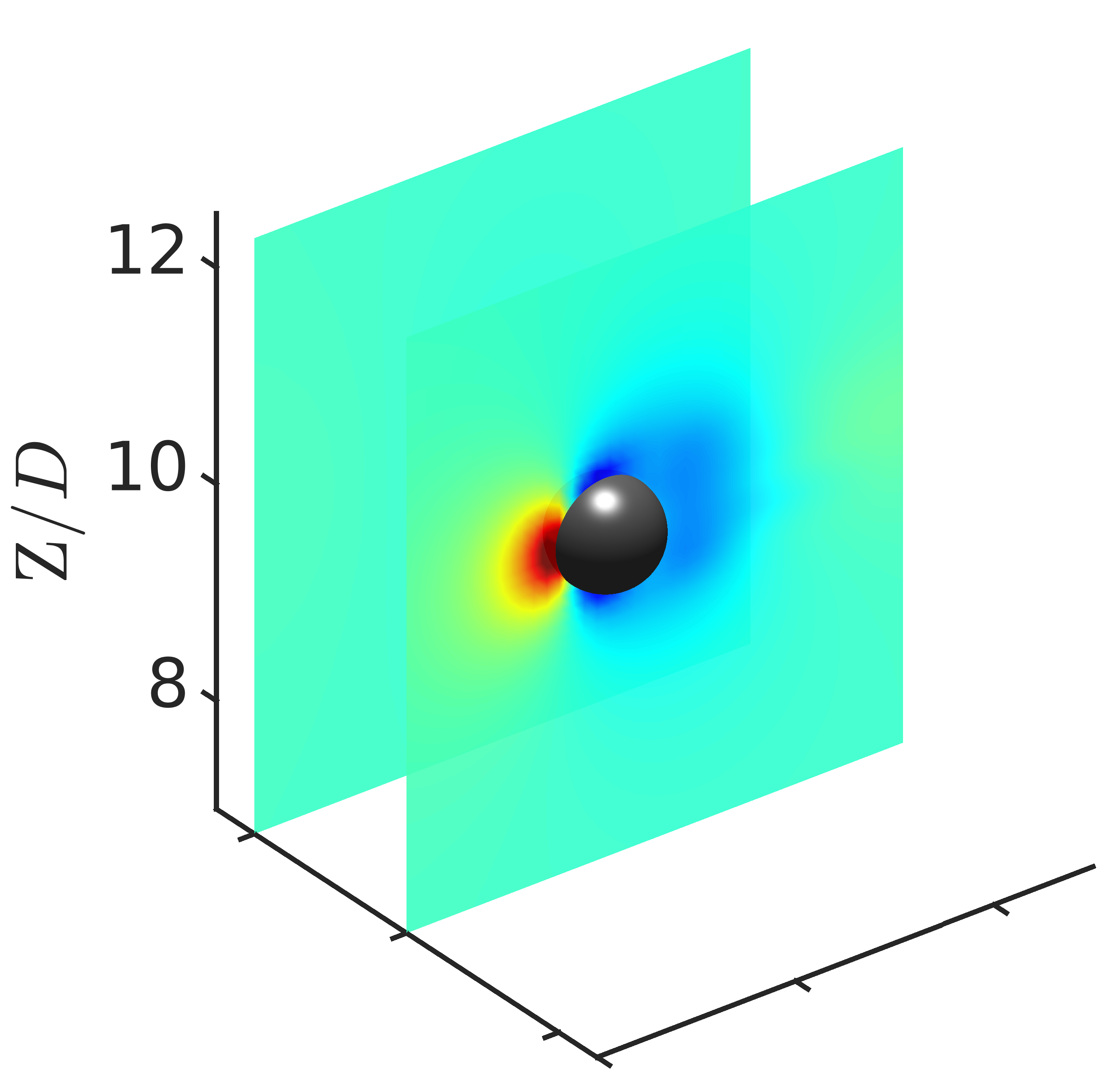}
\hspace{0.005\textwidth}
\includegraphics[width = 0.3\textwidth]{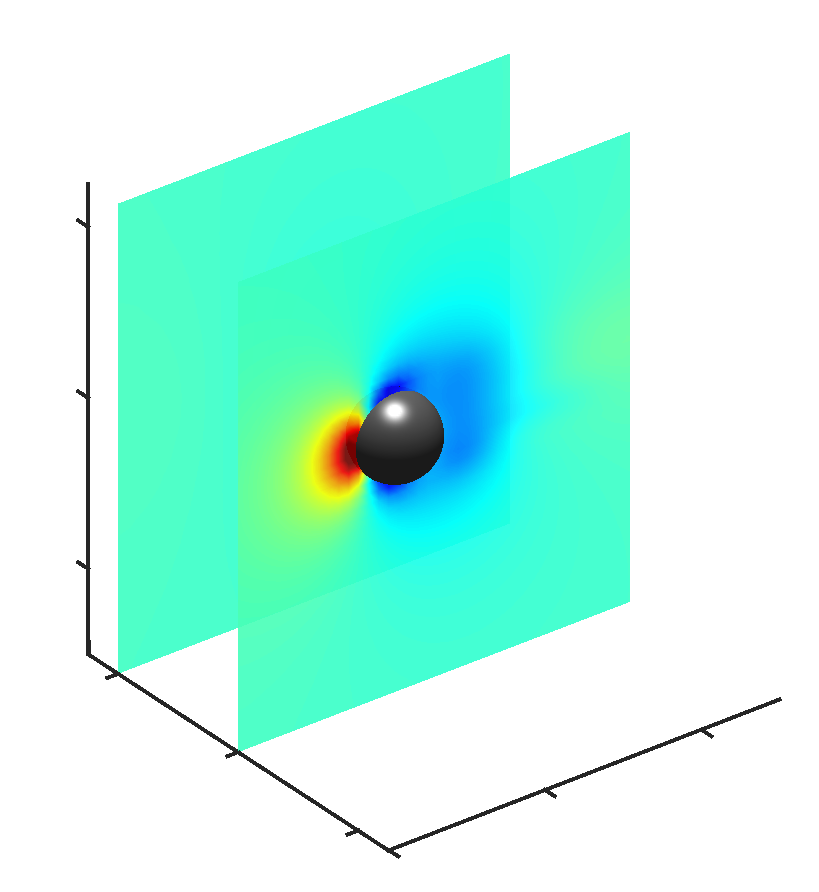}
\hspace{0.005\textwidth}
\includegraphics[width = 0.3\textwidth]{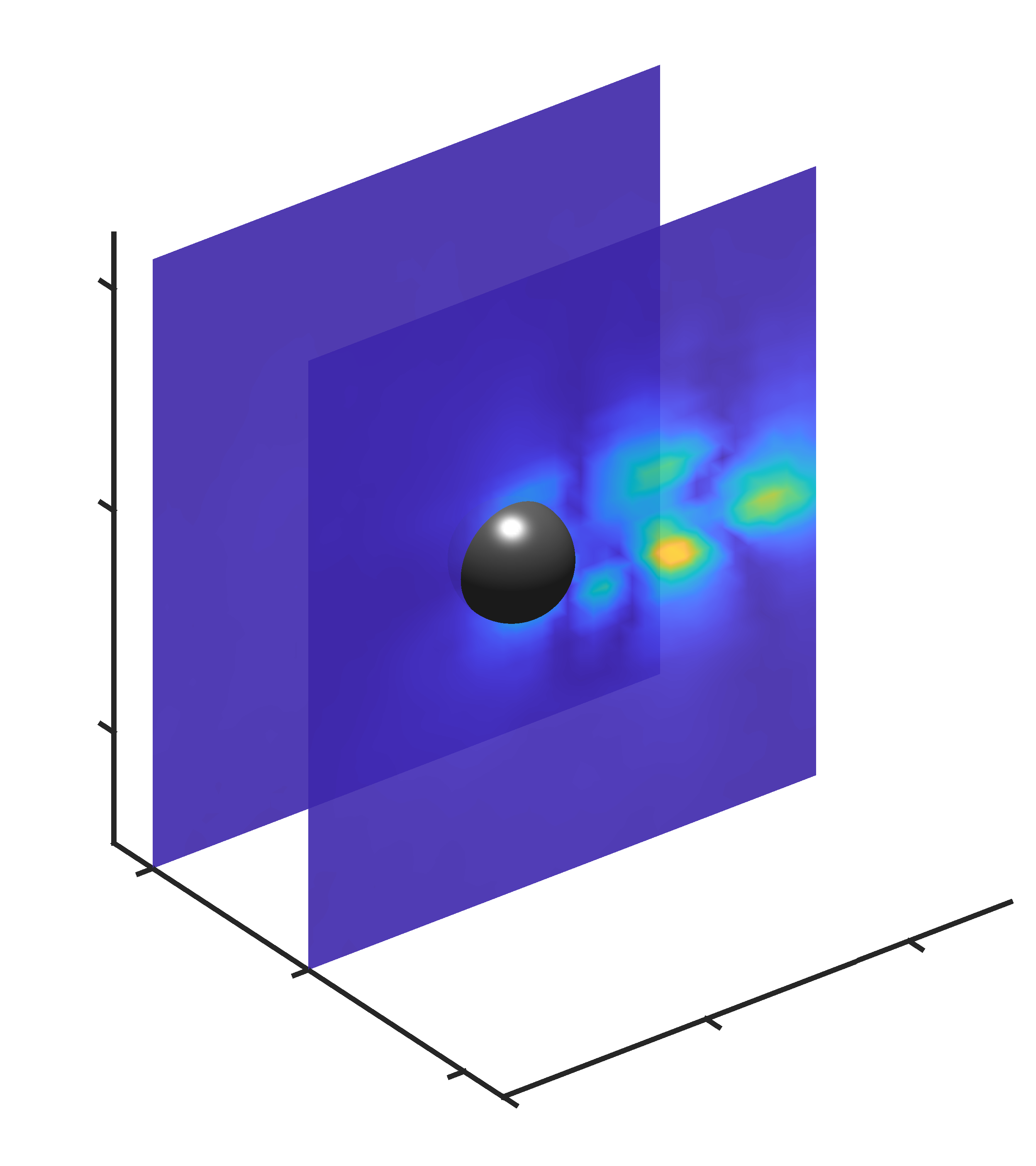}}
\\
{\includegraphics[width = 0.332\textwidth]{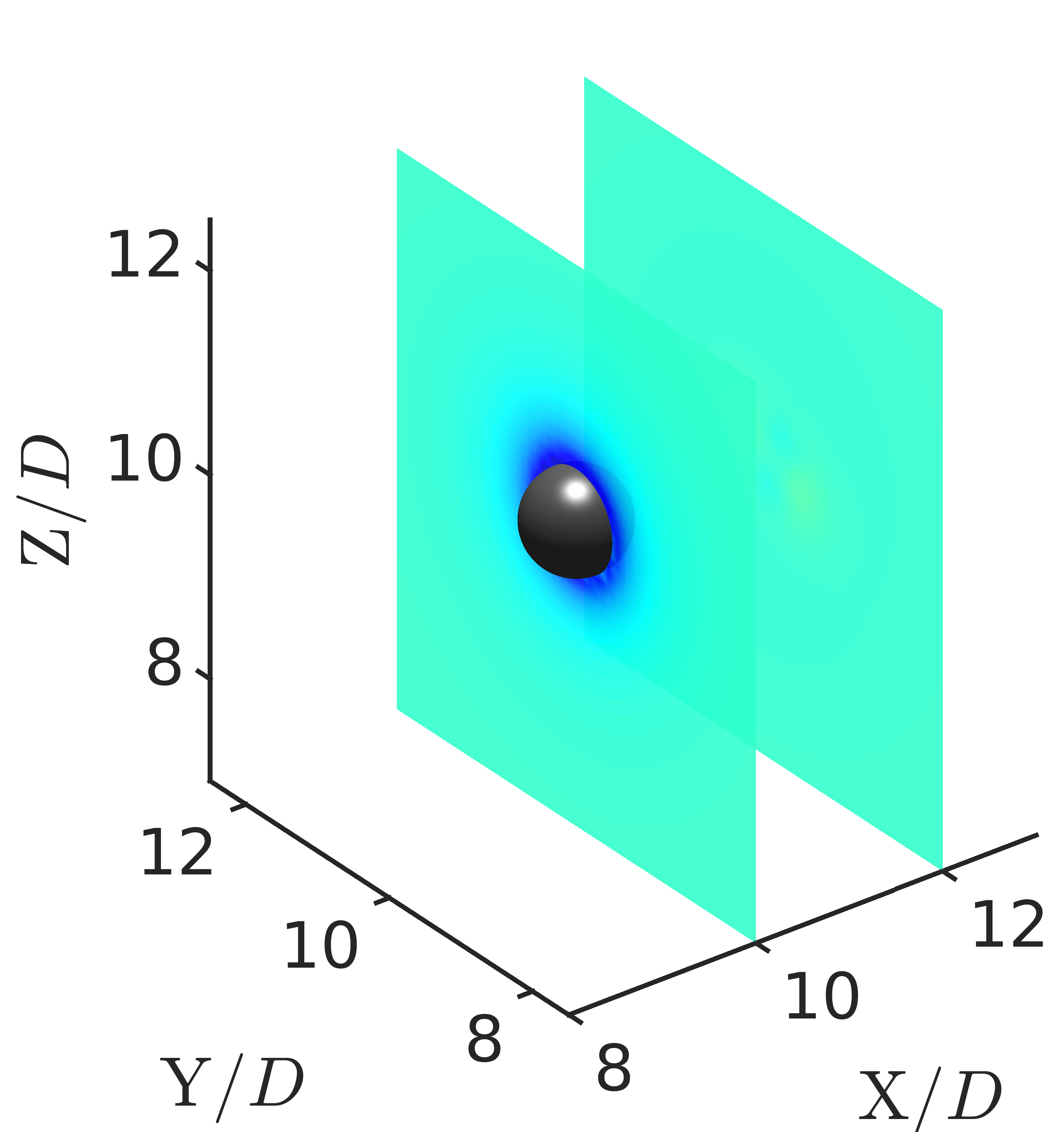}
\hspace{0.005\textwidth}
\includegraphics[width = 0.3\textwidth]{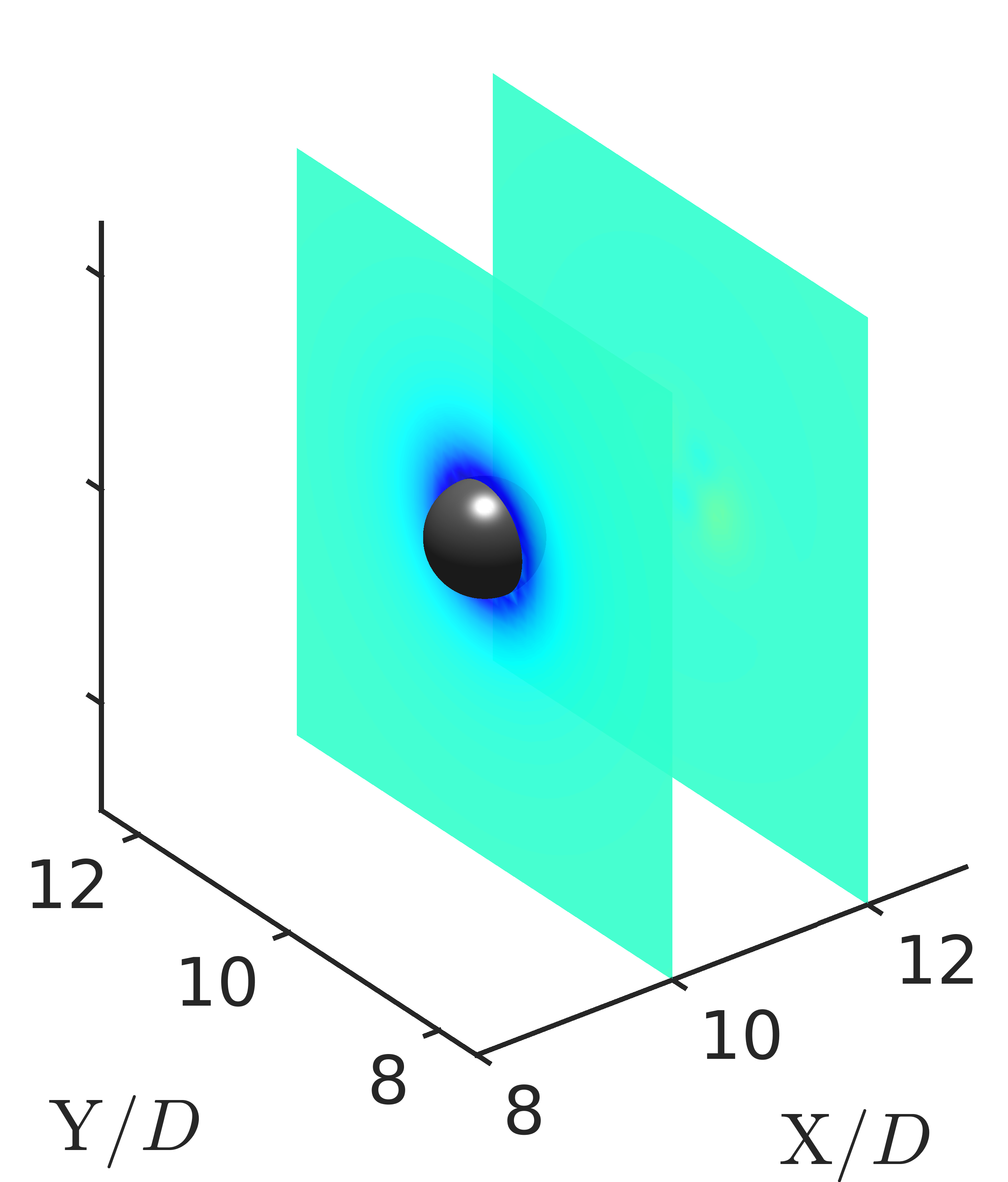}
\hspace{0.005\textwidth}
\includegraphics[width = 0.3\textwidth]{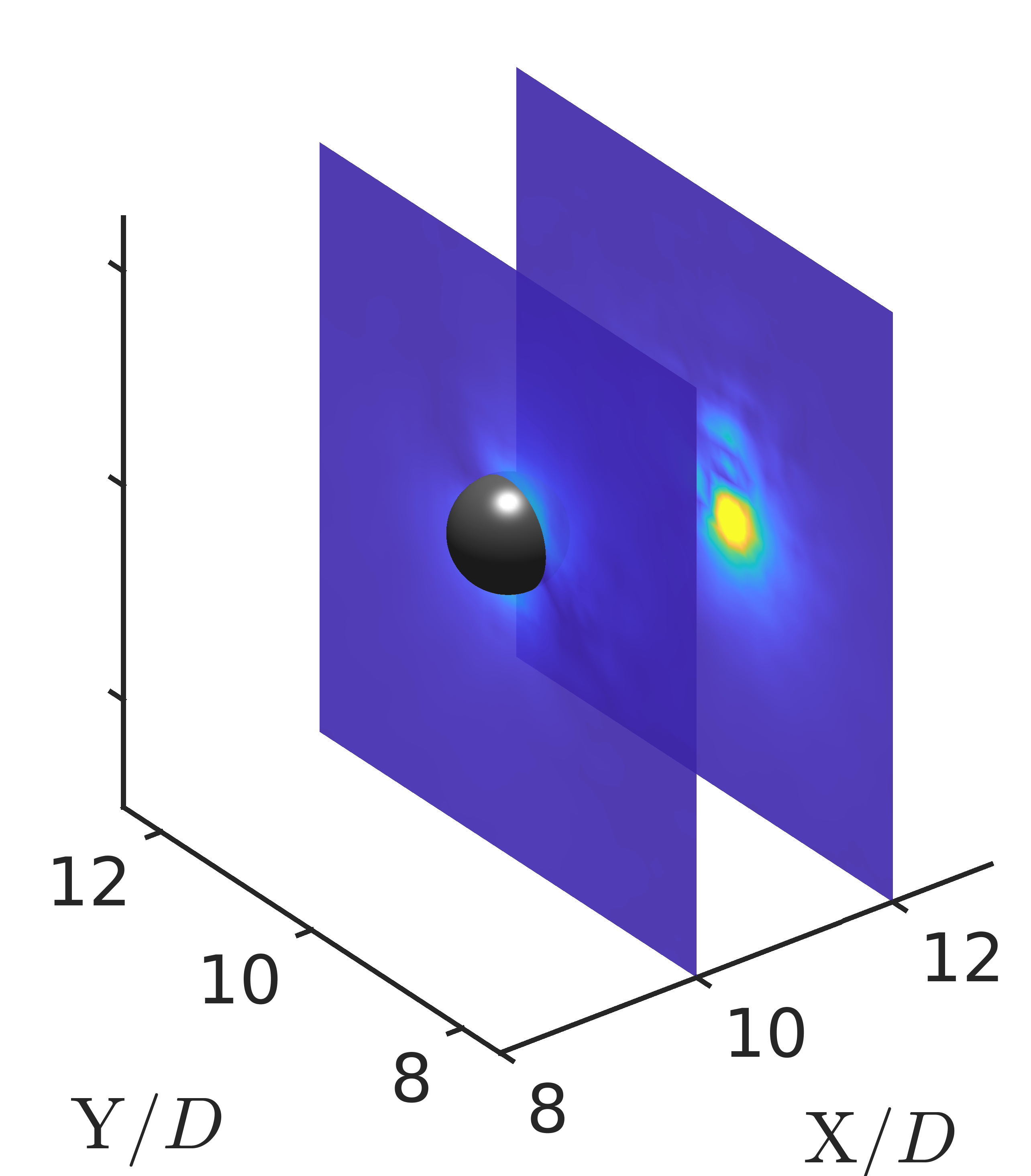}}
\caption{The flow past a sphere: Predicted and true pressure field comparison along with normalized reconstruction error $E^{i}$ at  $tU_{\infty}/D = 365$ sliced in $\mathrm{Z}/D=(8,10)$ (Row 1), $\mathrm{Y}/D=(10,12)$ (Row 2), $\mathrm{X}/D=(10,12)$ (Row 3). Left, middle and right contour plots depict the prediction, true and errors, respectively.}
\label{pres_sph_pred}
\end{figure*}

\begin{figure*}
\centering
{\includegraphics[width = 0.335\textwidth]{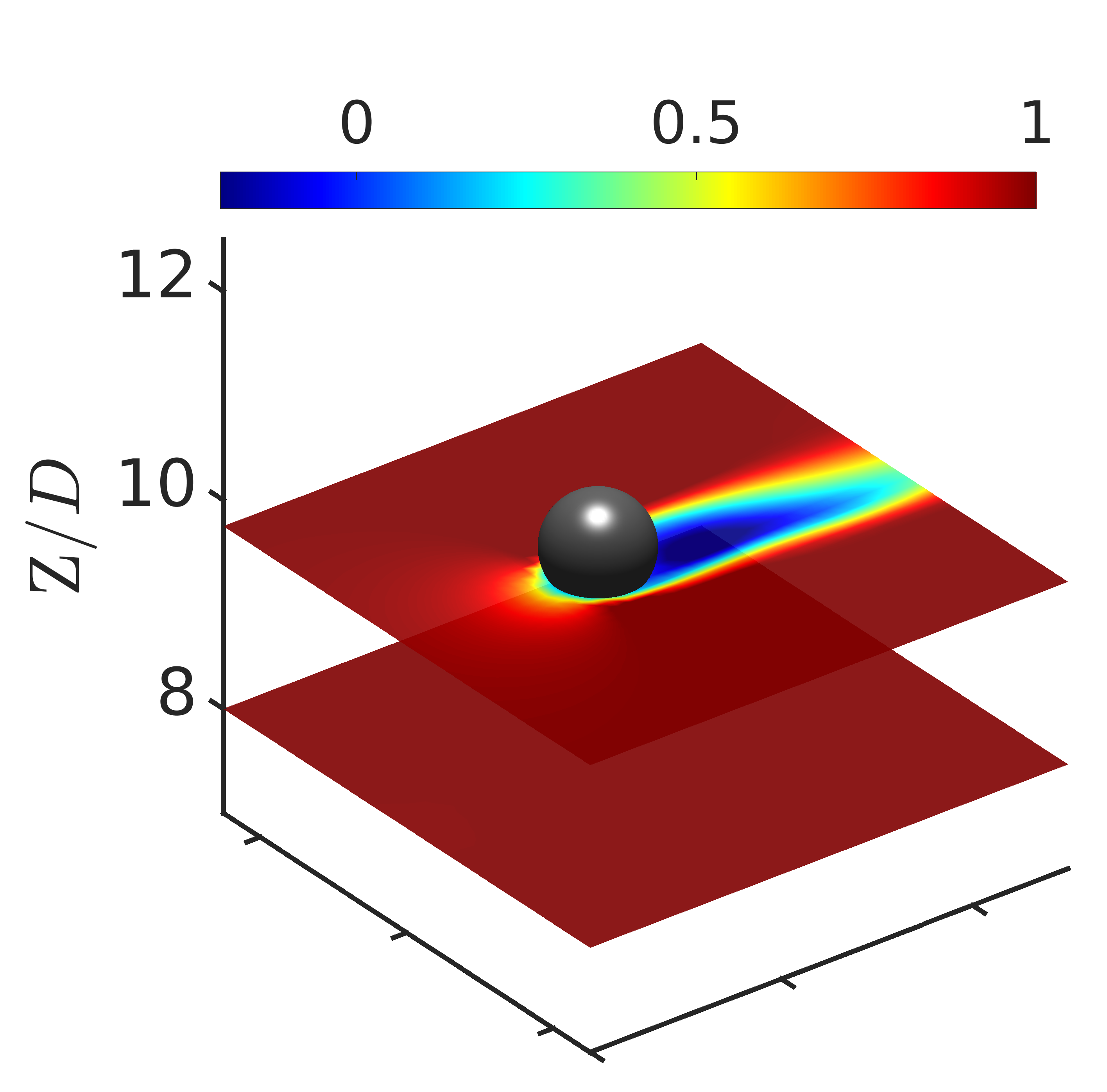}
\hspace{0.005\textwidth}
\includegraphics[width = 0.295\textwidth]{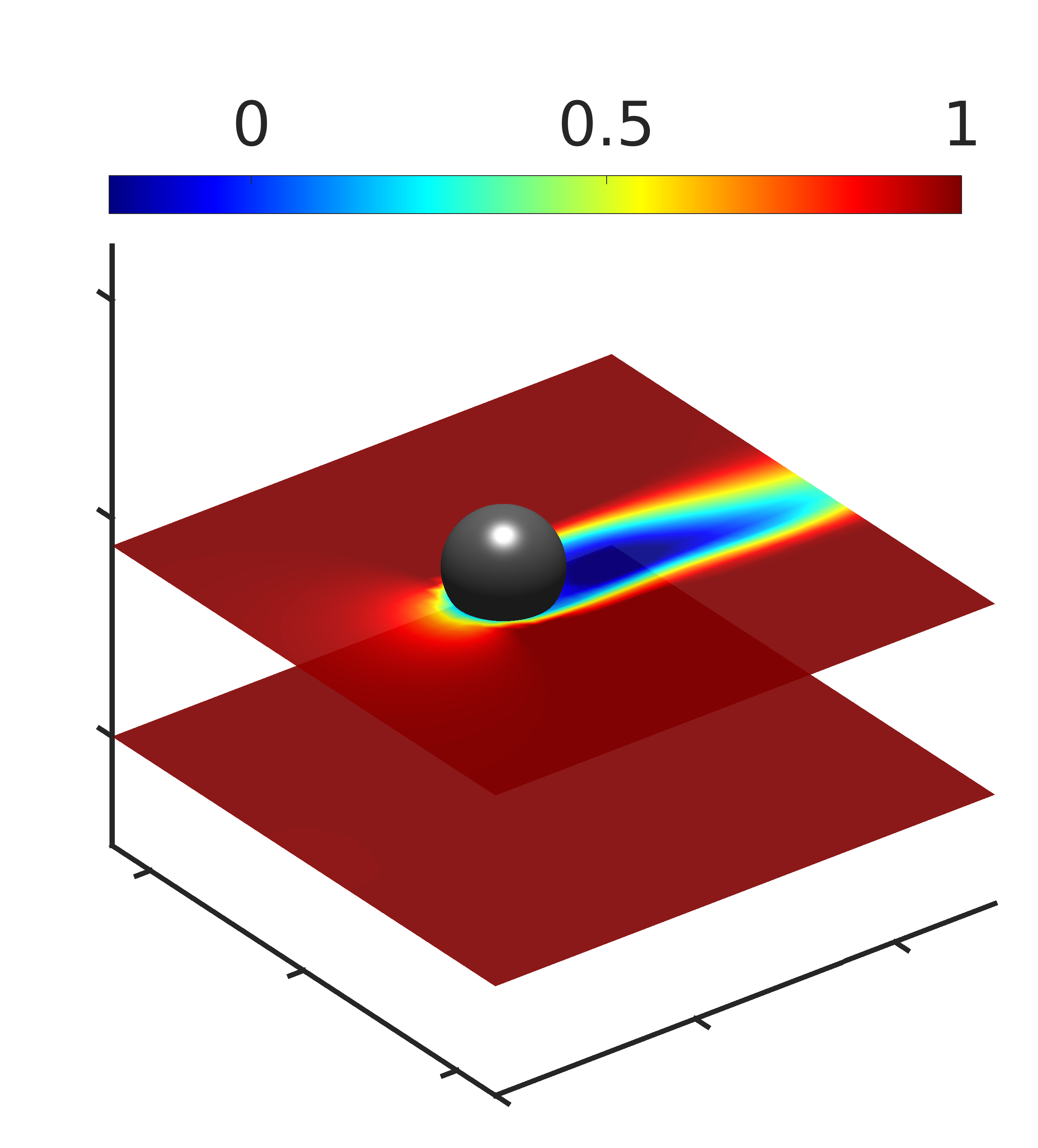}
\hspace{0.005\textwidth}
\includegraphics[width = 0.29\textwidth]{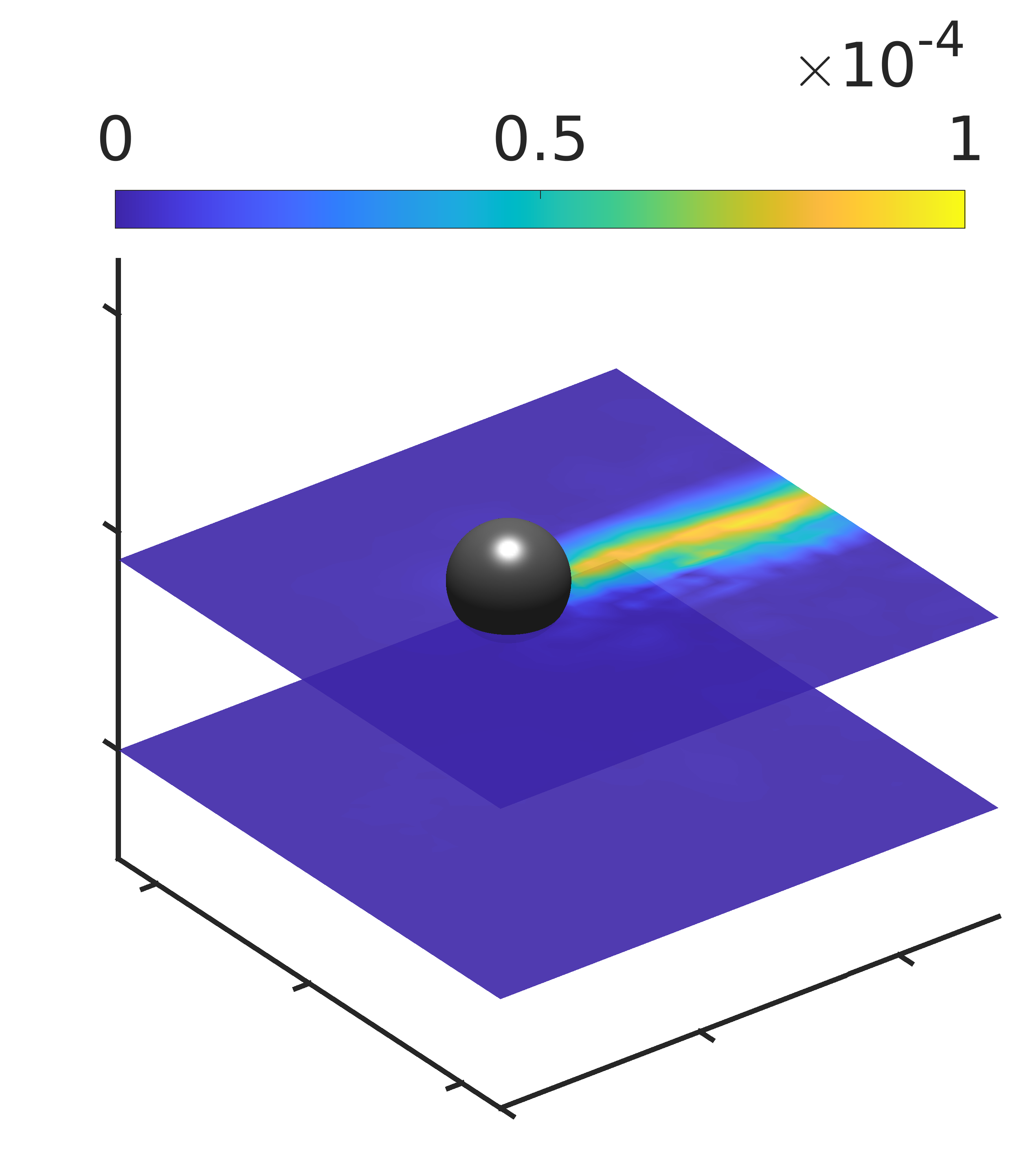}}
\\
{\includegraphics[width = 0.335\textwidth]{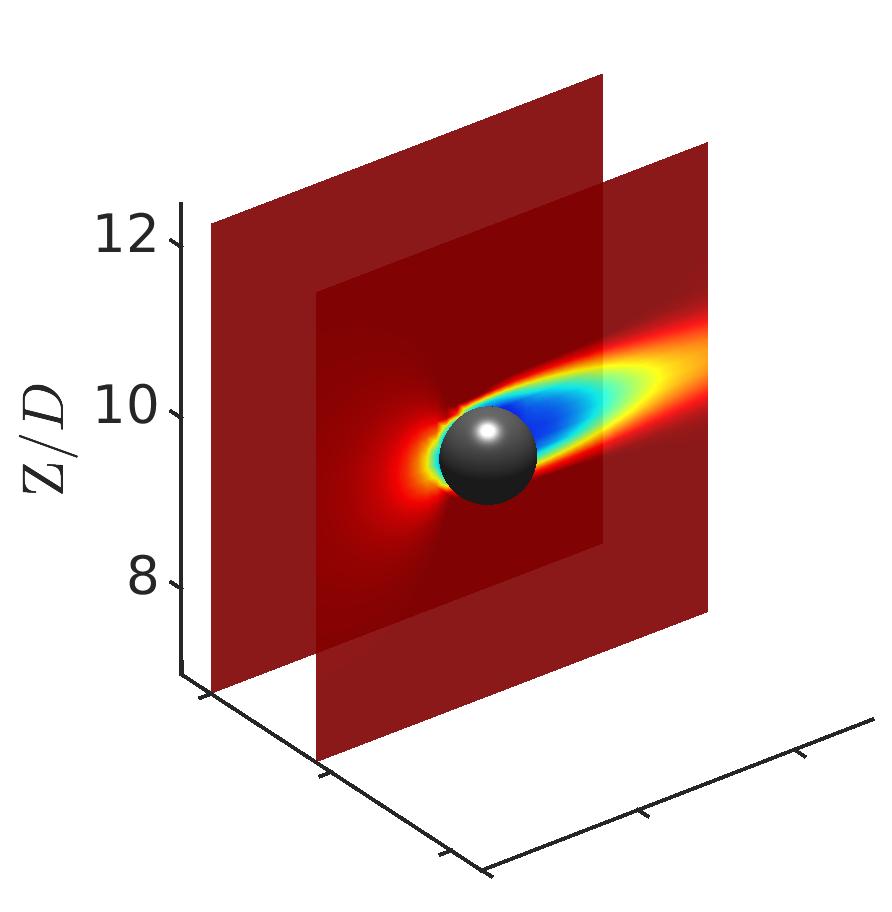}
\hspace{0.01\textwidth}
\includegraphics[width = 0.295\textwidth]{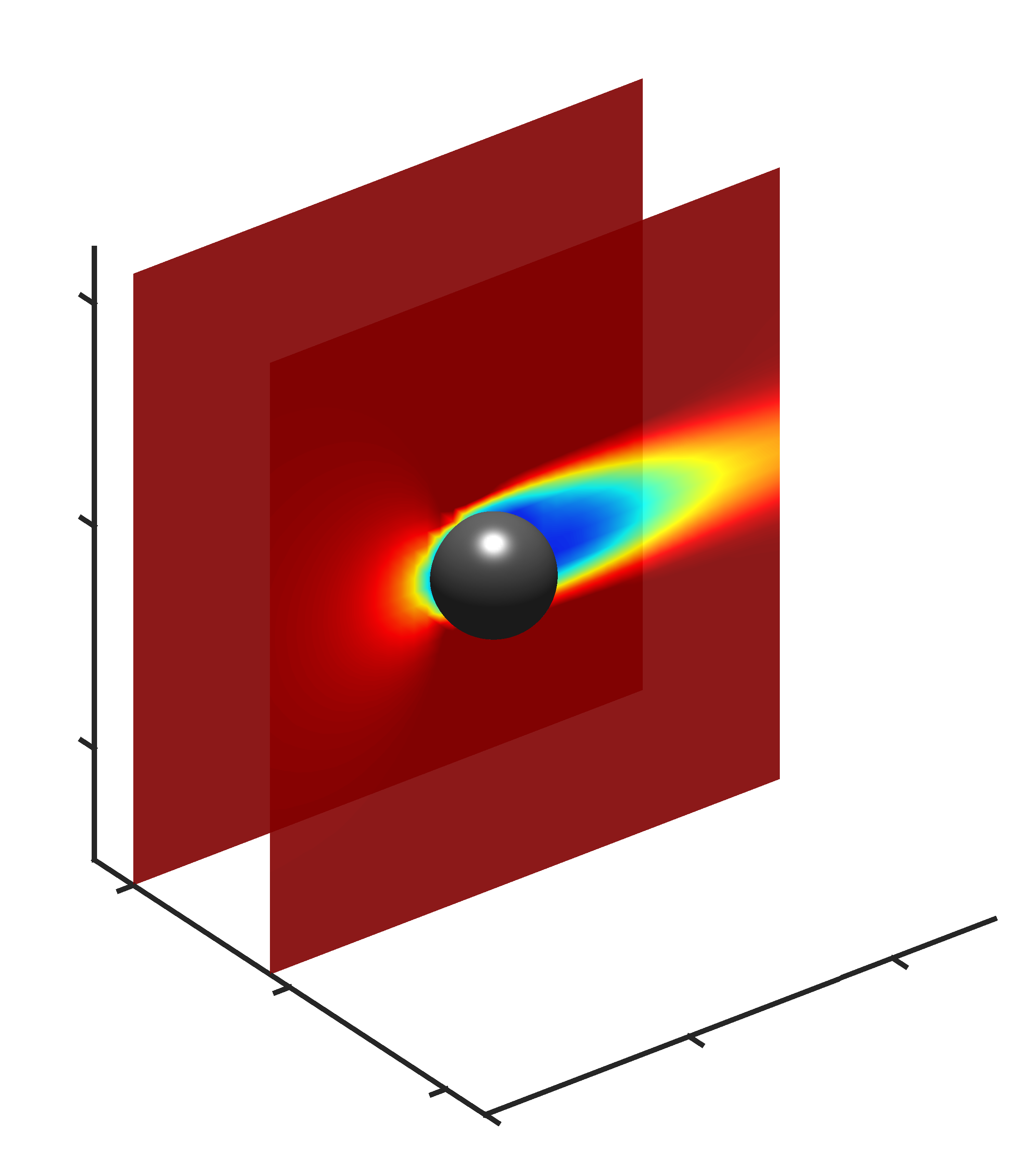}
\hspace{0.01\textwidth}
\includegraphics[width = 0.295\textwidth]{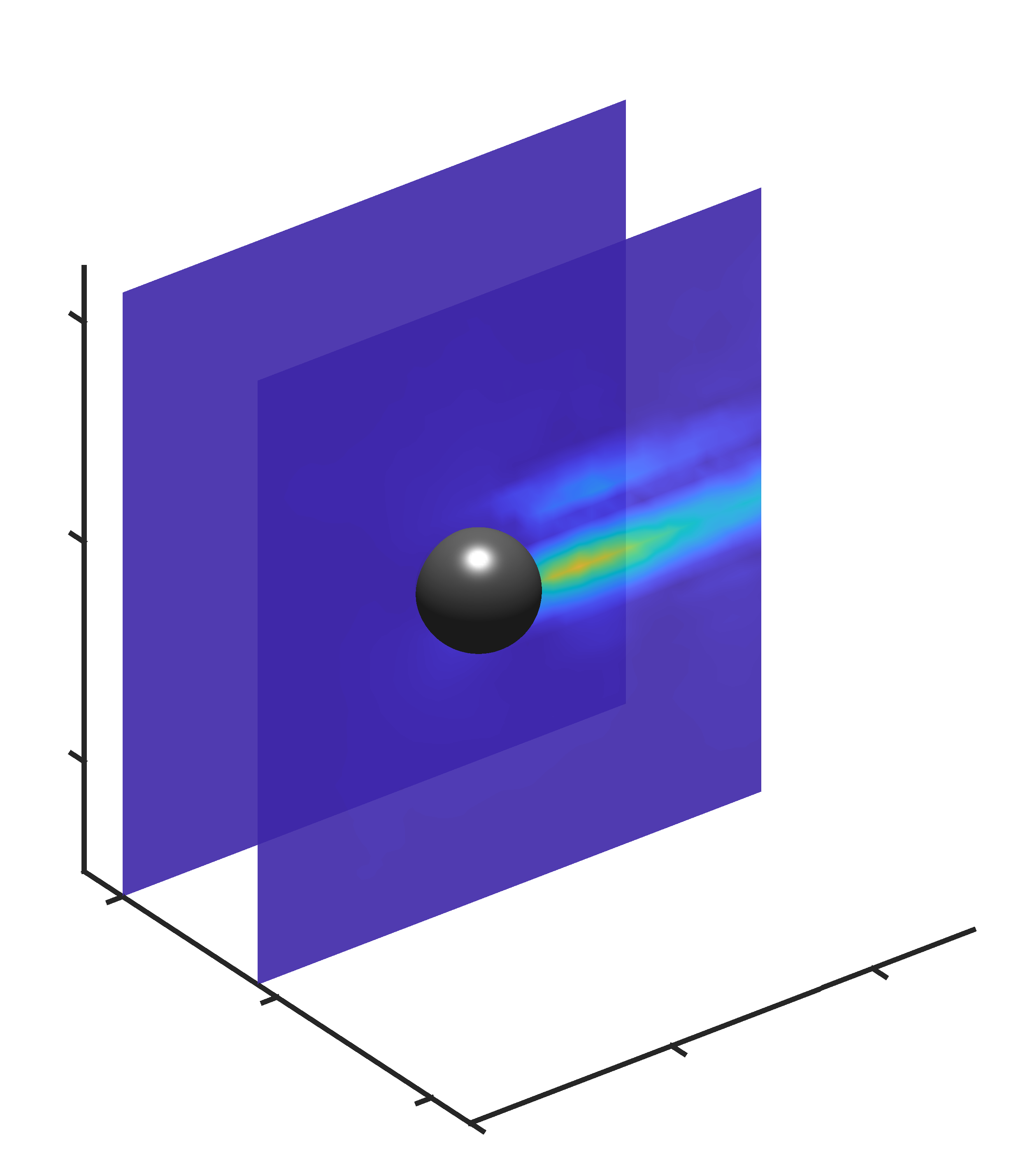}}
\\
{\includegraphics[width = 0.335\textwidth]{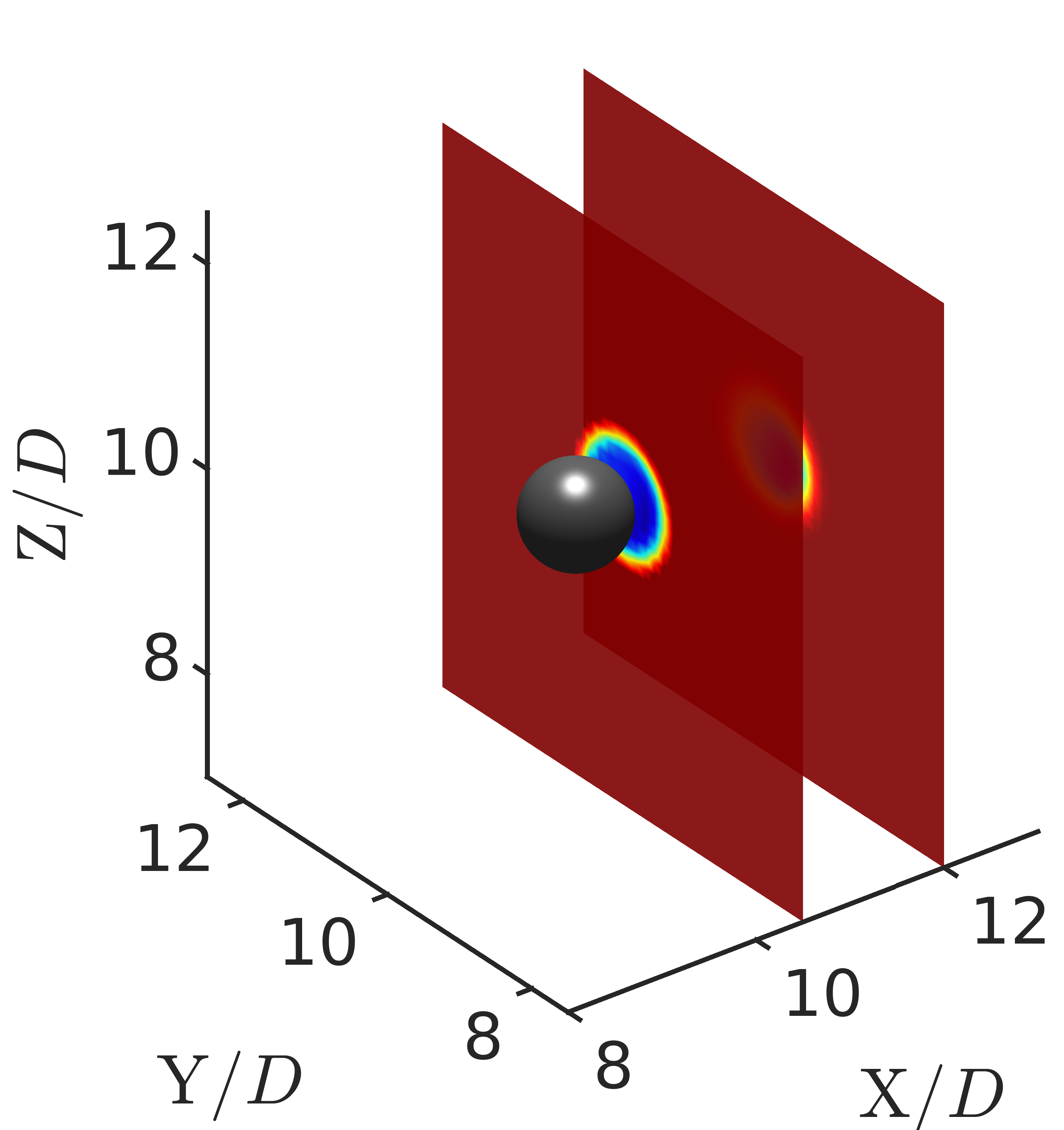}
\hspace{0.01\textwidth}
\includegraphics[width = 0.3\textwidth]{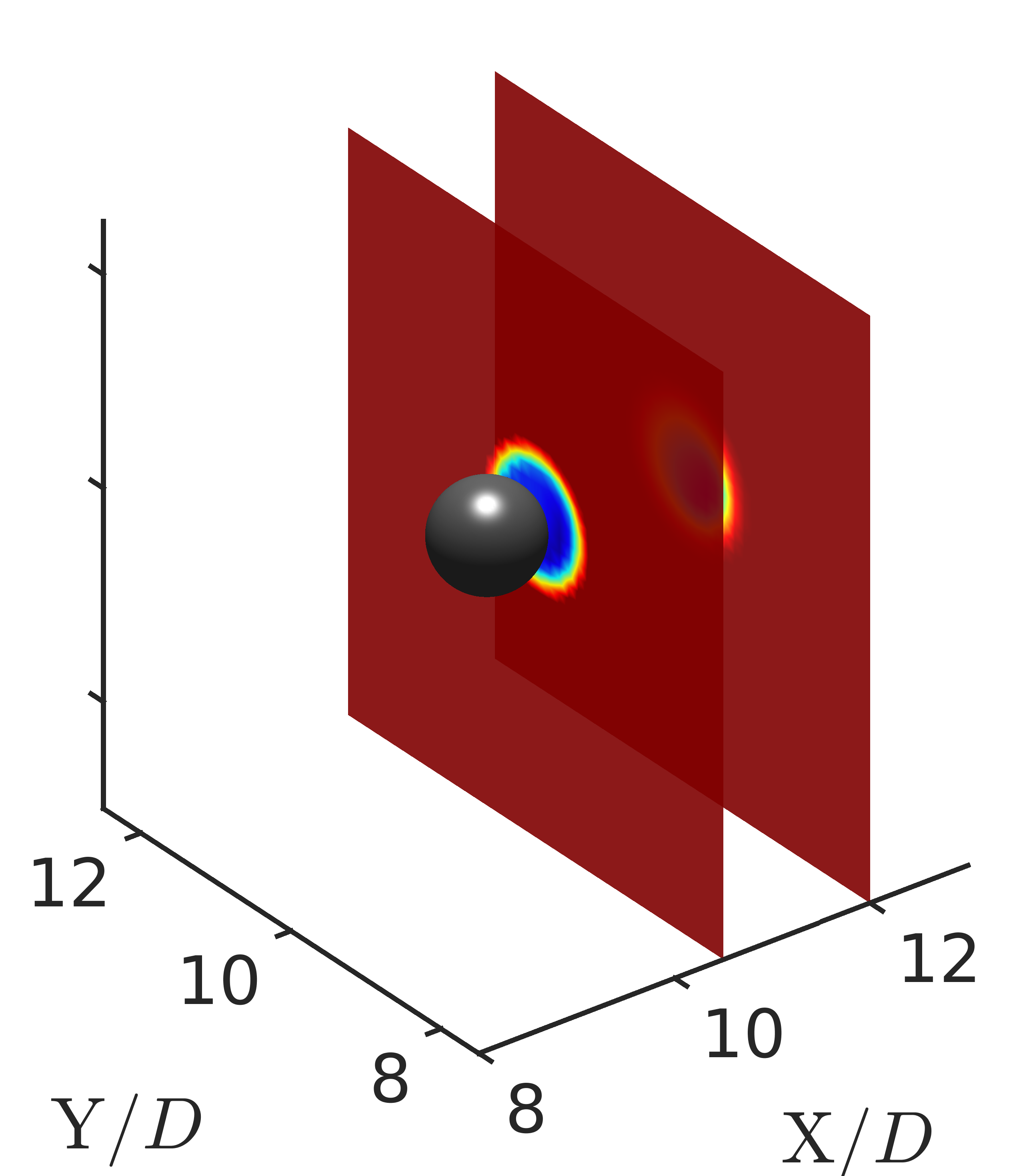}
\hspace{0.01\textwidth}
\includegraphics[width = 0.3\textwidth]{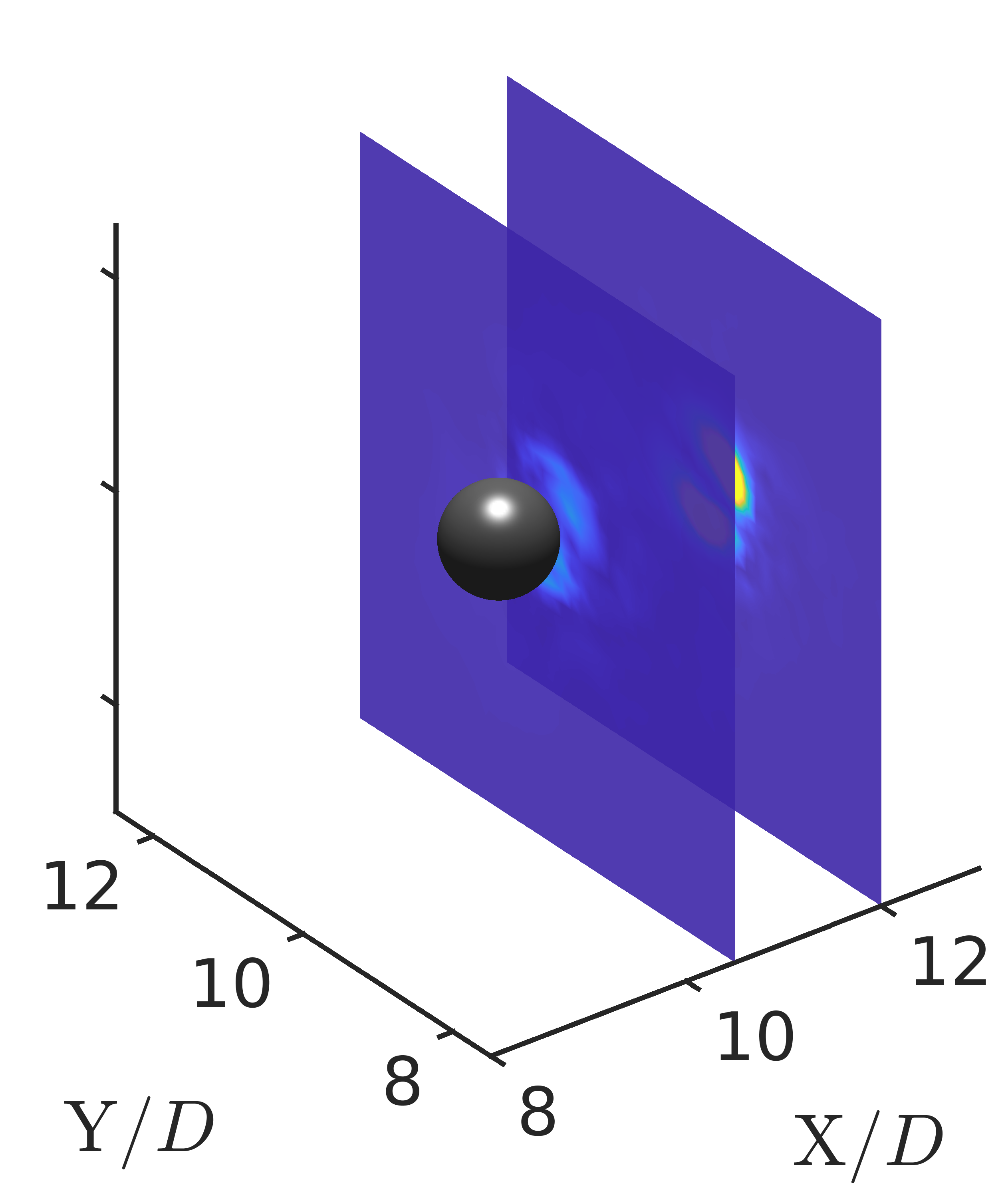}}
\caption{The flow past a sphere: Predicted and true x-velocity field comparison along with normalized reconstruction error $E^{i}$ at  $tU_{\infty}/D = 365$ sliced in $\mathrm{Z}/D=(8,9.75)$ (Row 1), $\mathrm{Y}/D=(10.25,12)$ (Row 2), $\mathrm{X}/D=(10.5,12)$ (Row 3). Left, middle and right contour plots depict the prediction, true and errors, respectively.}
\label{velx_sph_pred}
\end{figure*}

To train the 3D CRAN, we experiment with different sizes of the evolver cells $N_{h}= 64, 128, 256$ as primary tuning hyperparameters. In the 3D CRAN, the low-dimensional evolution needs to be tuned for appropriate time series learning and iterative optimization. We start by training each 3D CRAN model on a single v100 graphics processing unit (GPU) on pressure fields by instantiating with random parameter values. These parameters are updated in every training iteration. Every training iteration consists of a mini-batch of size $n_{s}=2$ randomly shuffled from the scaled featured flow input $\mathcal{S}$ and updating the neural parameters in $\approx 0.3s$. This helps speed-up the training procedure of the deep 3D CRAN architecture and lower memory usage. The objective function consists of the hybrid loss obtained from the unsupervised-supervised training as detailed in section \ref{train_predict_CRAN}. The evolution of the objective function $E_{h}$ with the training iterations $N_{train}$ is showcased in Fig.~\ref{Re300_tr_loss}.

\begin{figure*}
\centering
\subfloat[]{\includegraphics[width = 0.35\textwidth]{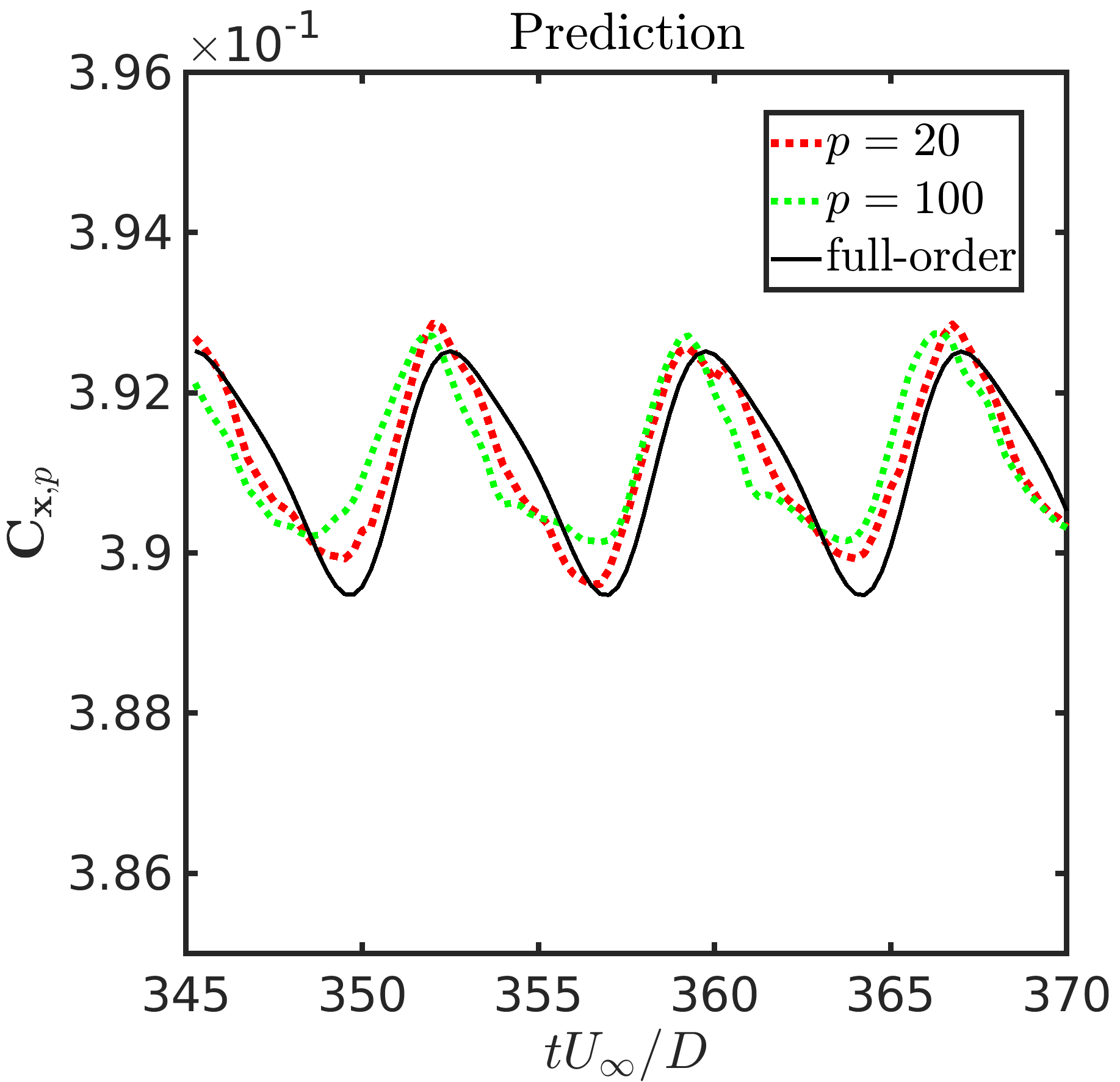}}
\hspace{0.05\textwidth}
\subfloat[]{\includegraphics[width = 0.35\textwidth]{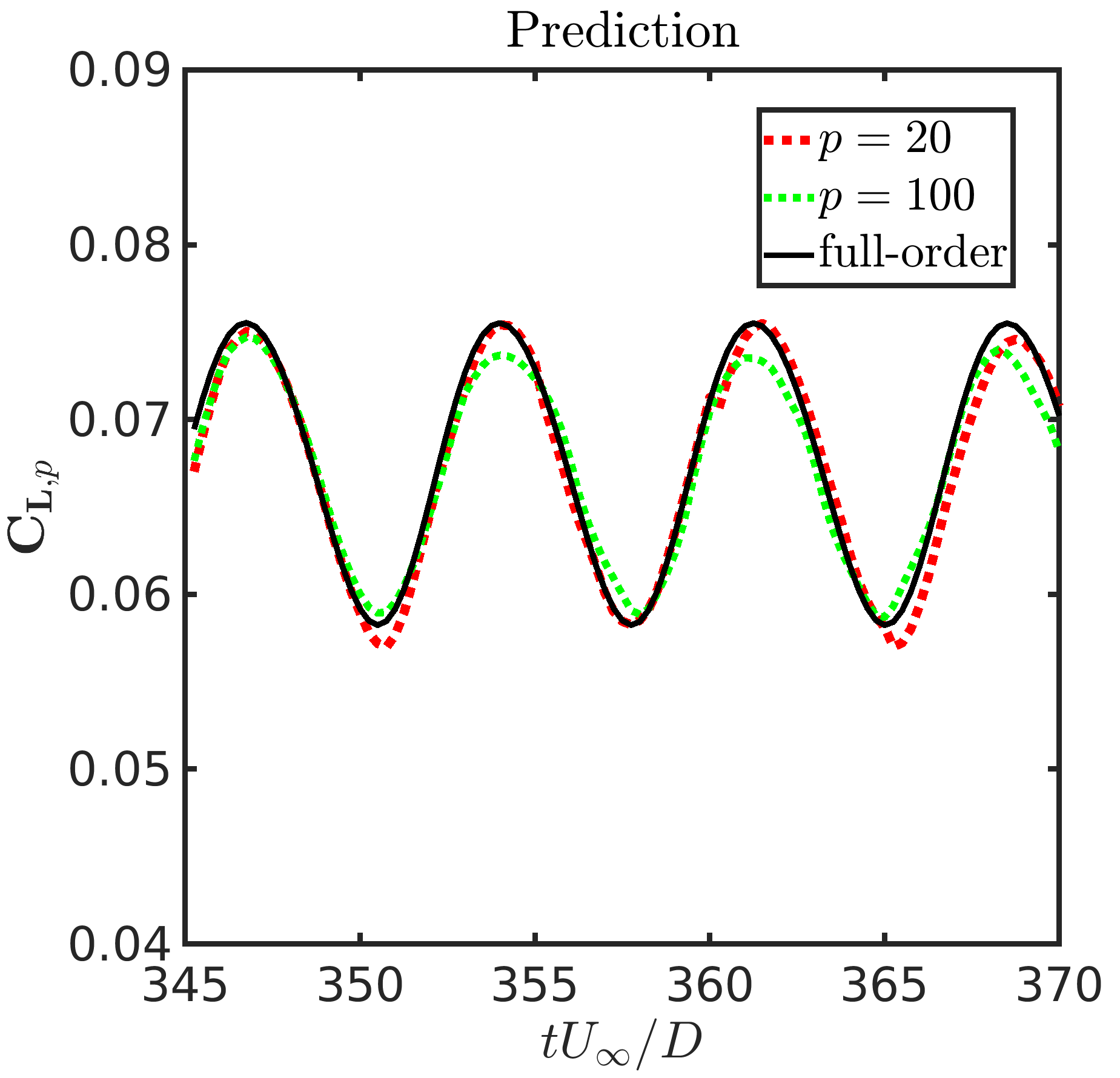}} \\
\caption{The flow past a sphere: Predicted and actual (3D CRAN model) (a) drag and (b) lift force coefficients integrated from the predicted pressure field  on the sphere for all 100 test time steps with multi-step predictive sequence $p=20$ and $p=100$.}
\label{pres_force_pred}
\end{figure*}

It can be observed that $N_{h}=256$ CRAN model tunes for the pressure dataset in $6 \times 10^{5}$ training iterations at a loss of $ 1.74 \times 10^{-6}$, which took nearly 64 hours of GPU training. At these training costs and iterations, 3D CRAN models $N_{h}=64,128$, however, do not optimise on the pressure dataset. We save the optimised 3D CRAN model ($N_{h}=256$) at $N_{train} = 600800$ on the disc memory as trained parameters. This is shown as blue dot in Fig.~\ref{Re300_tr_loss}. To avoid the expensive hyperparameter search for the velocity field training, we instead load the saved pressure 3D CRAN parameters on the velocity dataset (shown as blue cross) and optimise it further. This transfer of learning depicts that the 3D CRAN model fine-tunes on the coupled velocity dataset and mimics the dynamical model of flow past a sphere.  This initialisation of trained weights to velocity field reduces the training time to nearly 2 hours. We save the x-velocity 3D CRAN model parameters at $N_{train}= 700000$ for velocity testing. This model is depicted by red dot in Fig.~\ref{Re300_tr_loss}.  

Herein, $N_{h}=256$ trained saved instances of the 3D CRAN models are employed to analyze the field predictions for the pressure and x-velocity on the test dataset (100 time steps from $345-370\;tU_{\infty}/D$). We keep the multi-step predictive cycle length of $p=20,100$, implying that one input step infers $p$ sequence of future field predictions until a new input is fed.
Figs.~\ref{pres_sph_pred} and \ref{velx_sph_pred} depict a comparison 
of the predicted and true values of pressure and x-velocity fields, respectively, at $tU_{\infty}/D = 365$ sliced in various orthogonal planes with $p=100$. 
The normalized reconstruction error $E^i$ is calculated by taking the absolute value of the difference between the real and predicted fields and then normalizing it with the truth's $L_2$ norm. It can be observed that the majority of these errors are located in the nonlinear flow separation region and near-wake of the sphere. These 3D reconstruction errors are in the order of $10^{-3}$ for pressure and $10^{-4}$ for $\mathrm{x}$-velocity predictions. This demonstrates the high accuracy of the 3D CRAN for reconstruction and time series extrapolations if properly trained.  

\begin{table}
\centering
\caption{Summary of the offline and online times for 3D CRAN vs. 3D FOM simulations.}
\begin{tabular}{p{2.8cm}|p{2.0cm}|p{2.0cm}}
\toprule
\toprule
 &  FOM-HPC &  3D CRAN-PC  \\
\bottomrule
\bottomrule
Processor number & 32 CPUs & 1 GPU\\
Offline time$^{*}$ & $\approx 10 \; \mathrm{h}$  &  $\approx 64 \; \mathrm{h}$ \\
Online time$^{**}$  & $\approx 1 \; \mathrm{h}$  & $\approx 1.99 \; \mathrm{s}$ \\
Offline speed-up   & $1$ & 0.1563 \\
Online speed-up & $1$ &  1800 \\
\bottomrule
\end{tabular}\\ 
$^{*}$ Elapsed time 1000 training steps.\\
$^{**}$ Elapsed time 100 test steps.
\label{tab:cran_pred_single_Re}
\end{table}

The predicted coarse-grain flow fields are directly integrated and corrected using the snapshot-FTLR to get the pressure loads over the sphere. Evolution of the drag $\mathrm{C}_{\mathrm{x},{p}}$ and lift $\mathrm{C}_{\mathrm{L},{p}}$ loads are depicted in Fig.~\ref{pres_force_pred}. The red line in the figure depicts force calculation from pressure fields with sequence prediction length of $p=20$ steps in a closed-loop recurrence. This helps reduce the compounding effect of the errors with slight improvements compared to $p=100$ steps predicted from one time instant. 
In Table~\ref{tab:cran_pred_single_Re} we provide an estimate of the computational costs for the 3D CRAN together with the full-order simulations for flow past a sphere. We recall that the DL-ROM solution can offer remarkable speed-ups in online predictions compared with the FOM by nearly 1800 times. However, the offline training time of 3D CRAN is expensive compared with a similar FOM. 
The following sub-section extends our DL-ROM methodology for predicting unsteady flow fields with variable $Re$ while focusing on speeding up both offline and online computational times.  

\subsection{Variable flow past a fixed sphere}

\begin{figure*}
\centering
{\includegraphics[width = 0.329\textwidth]{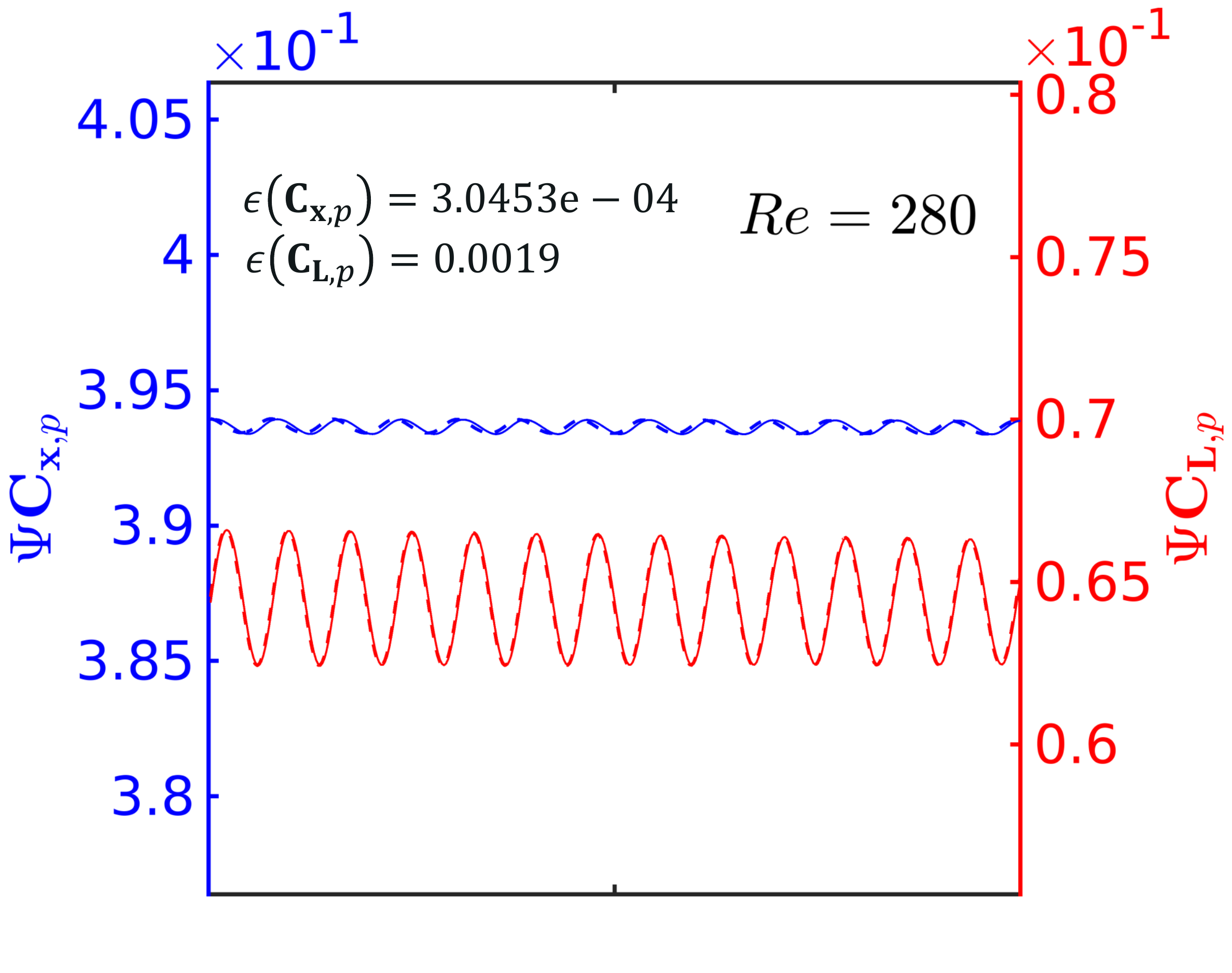}} 
\hspace{0.005\textwidth}
{\includegraphics[width = 0.315\textwidth]{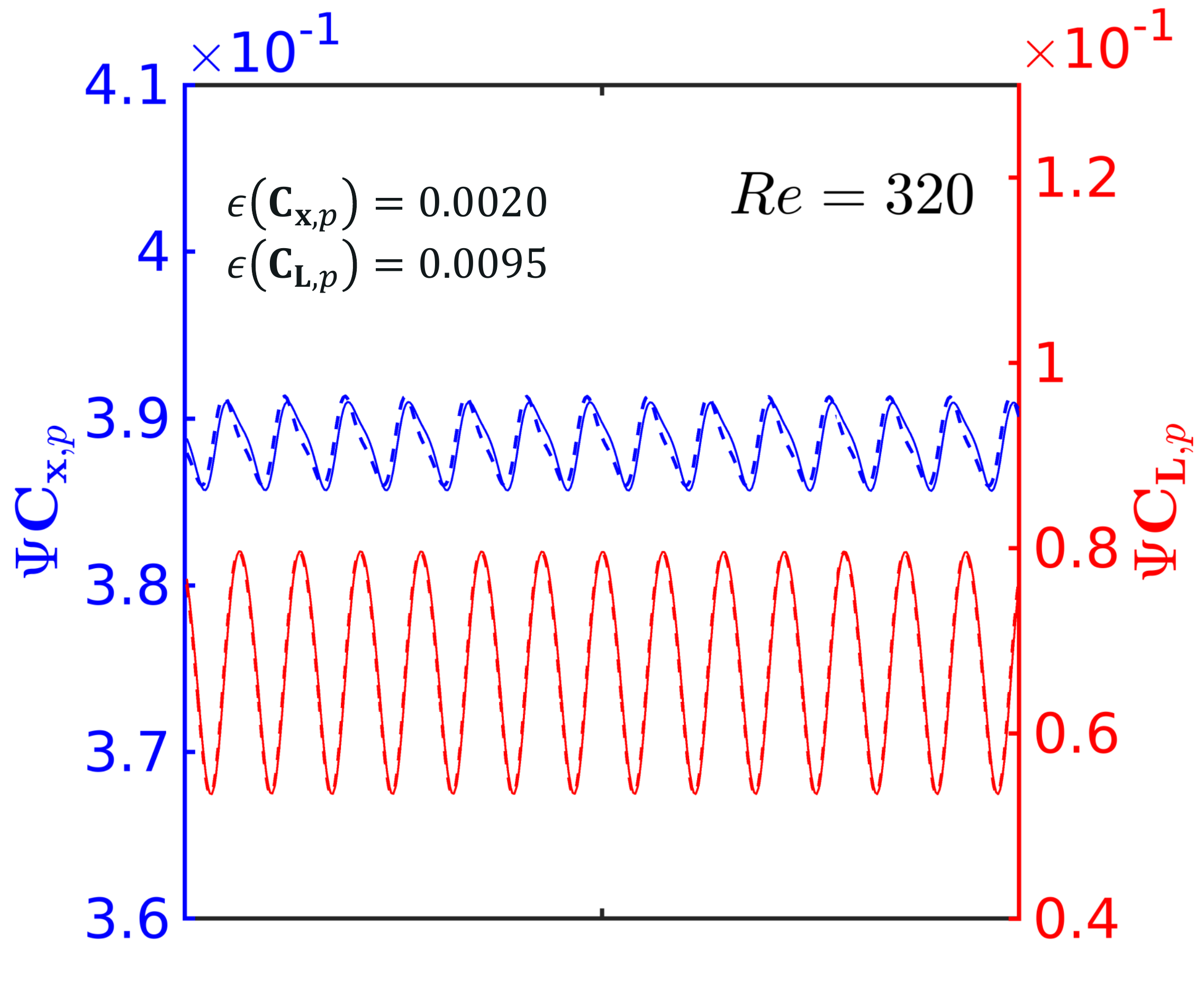}} 
\hspace{0.005\textwidth}
{\includegraphics[width =  0.316\textwidth]{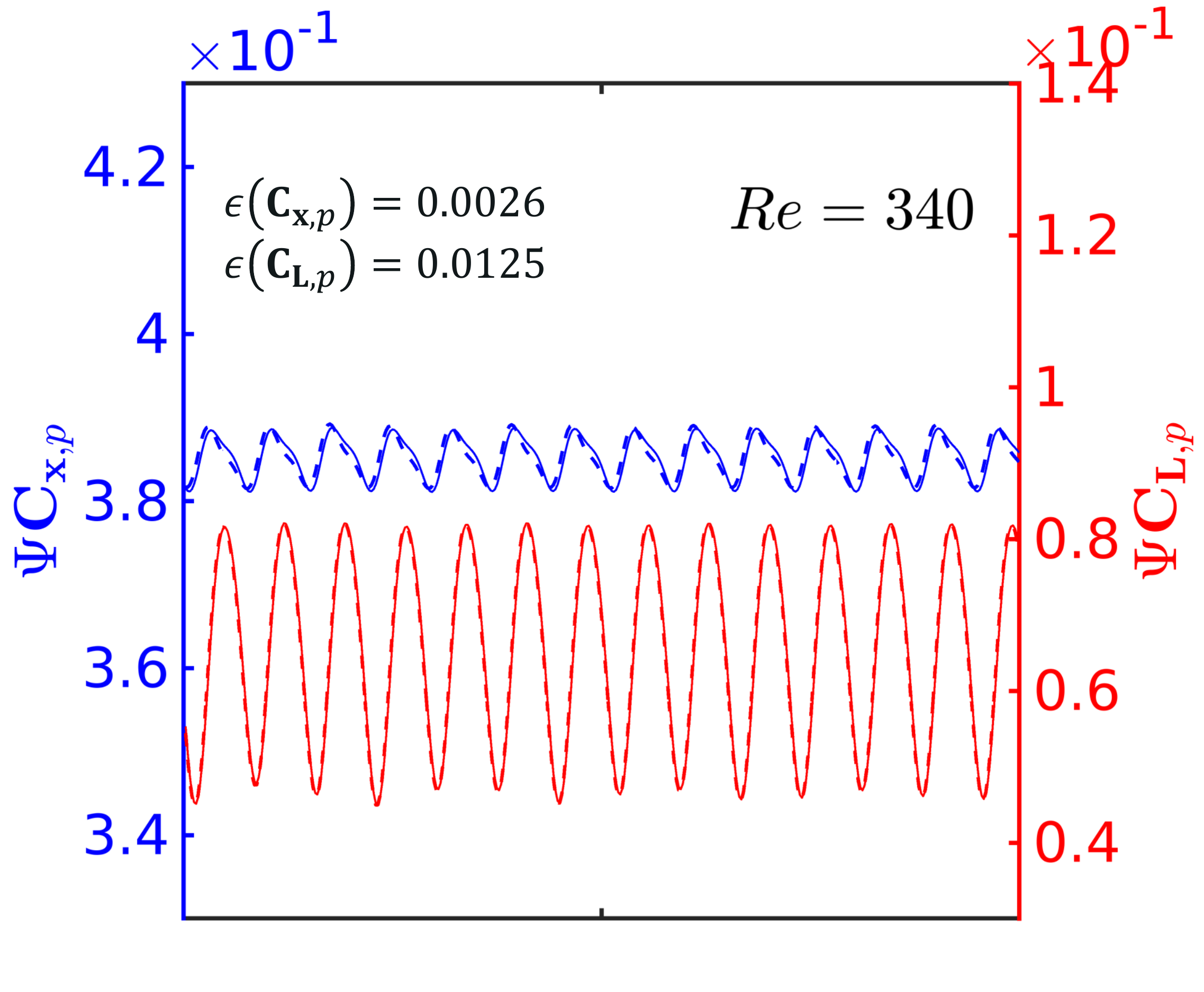}}
\\ 
{\includegraphics[width = 0.32\textwidth]{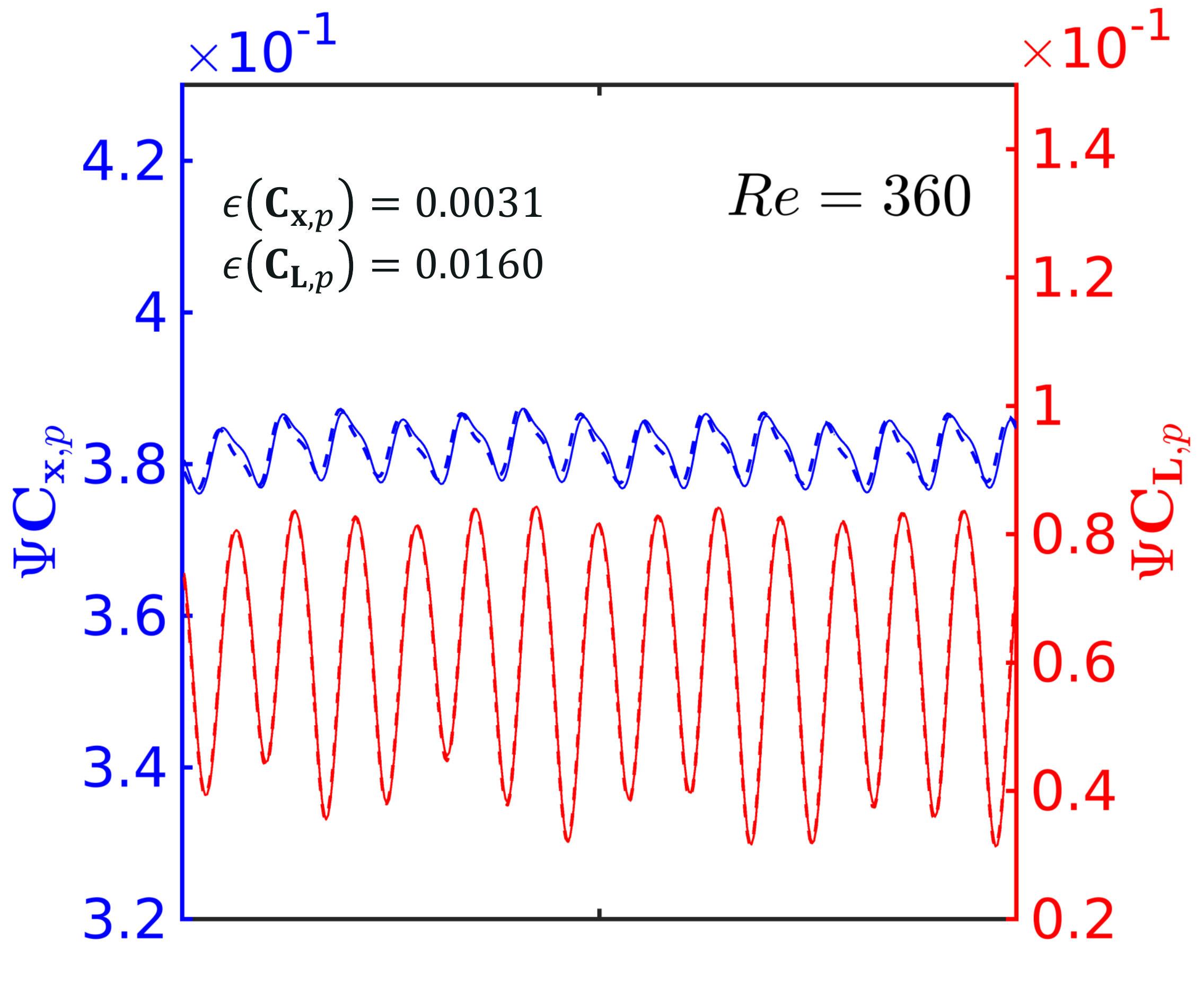}} 
\hspace{0.005\textwidth}
{\includegraphics[width = 0.319\textwidth]{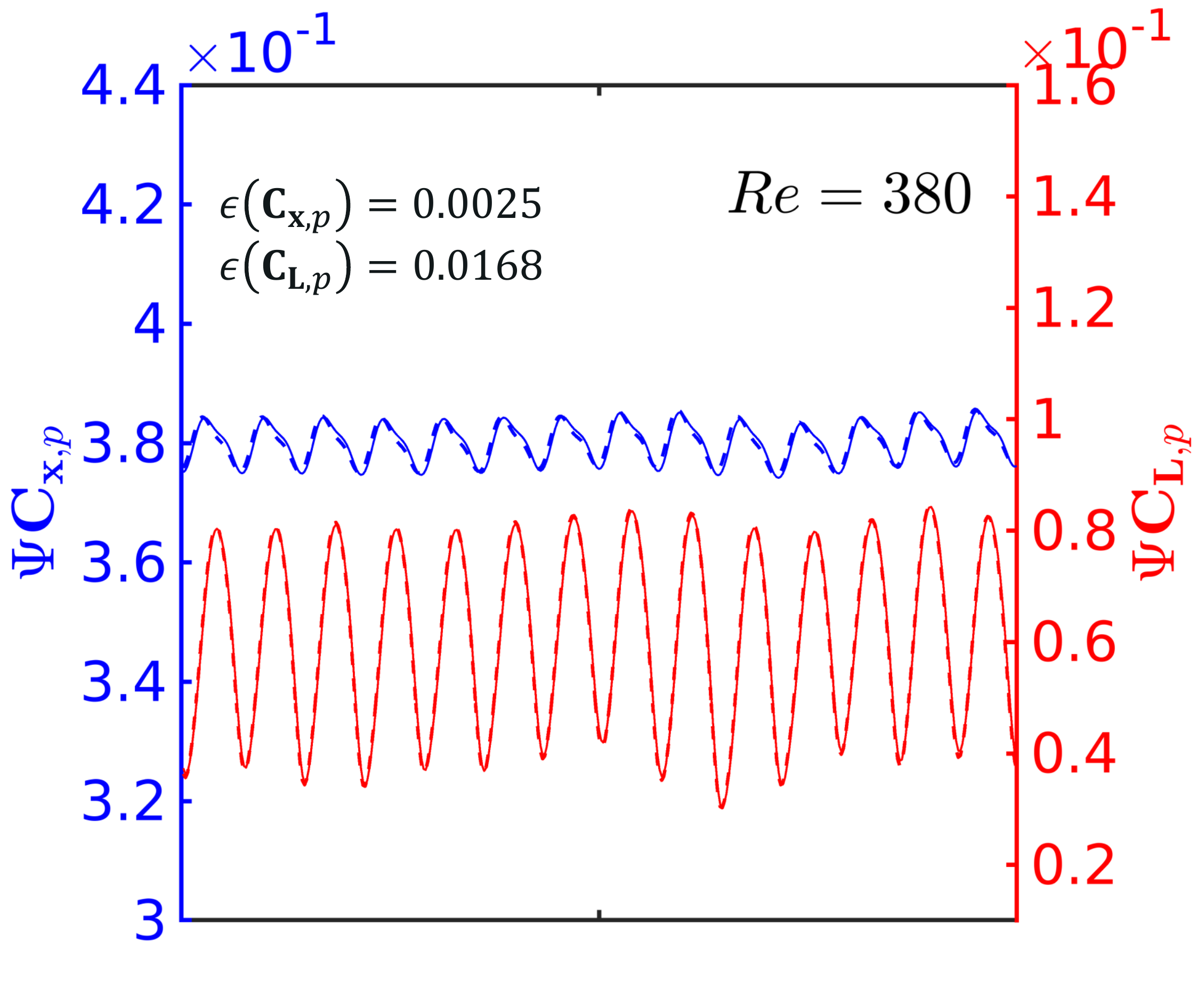}} 
\hspace{0.005\textwidth}
{\includegraphics[width =0.314\textwidth]{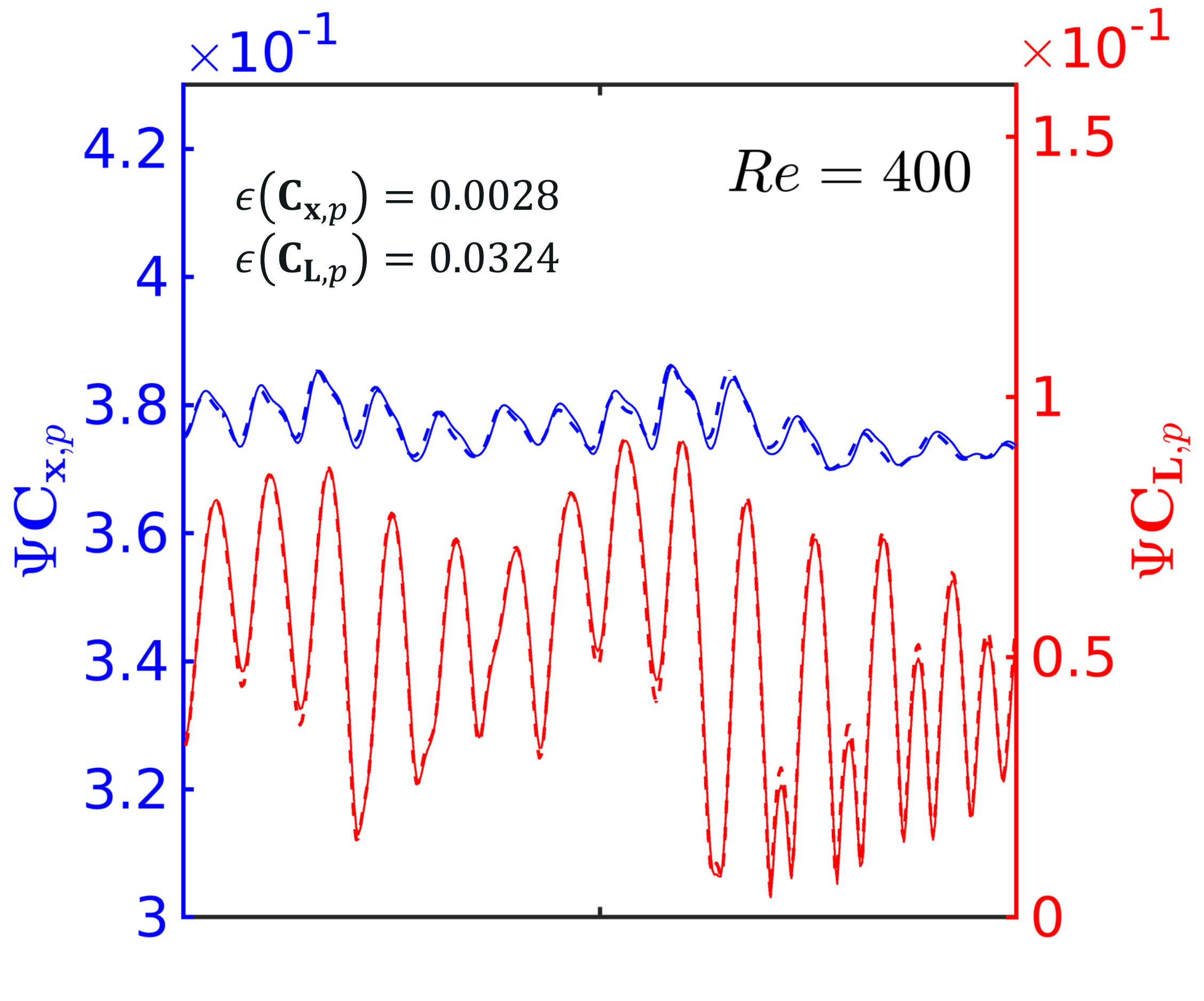}} 
\\ 
{\includegraphics[width = 0.32\textwidth]{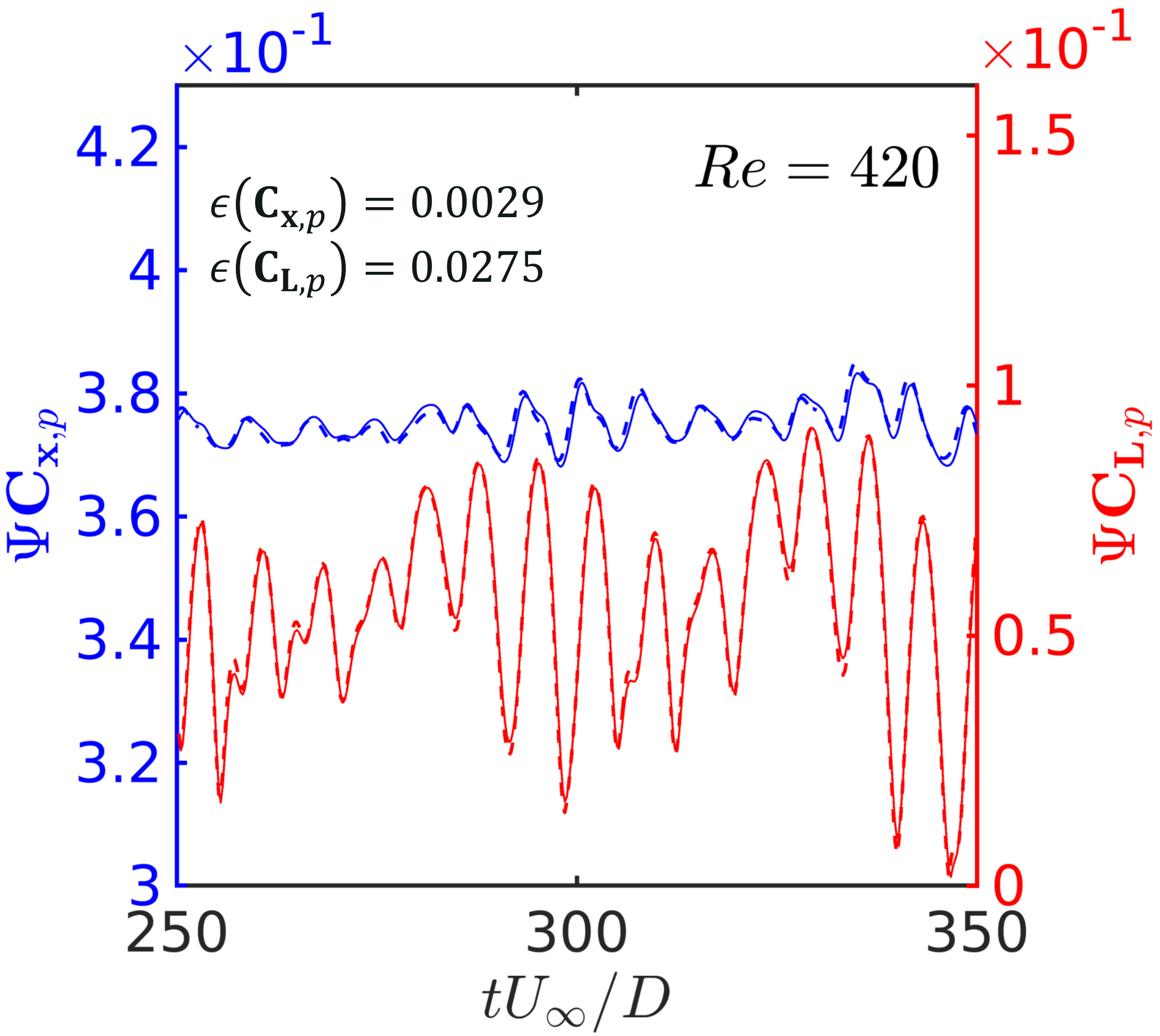}} 
\hspace{0.005\textwidth}
{\includegraphics[width = 0.318\textwidth]{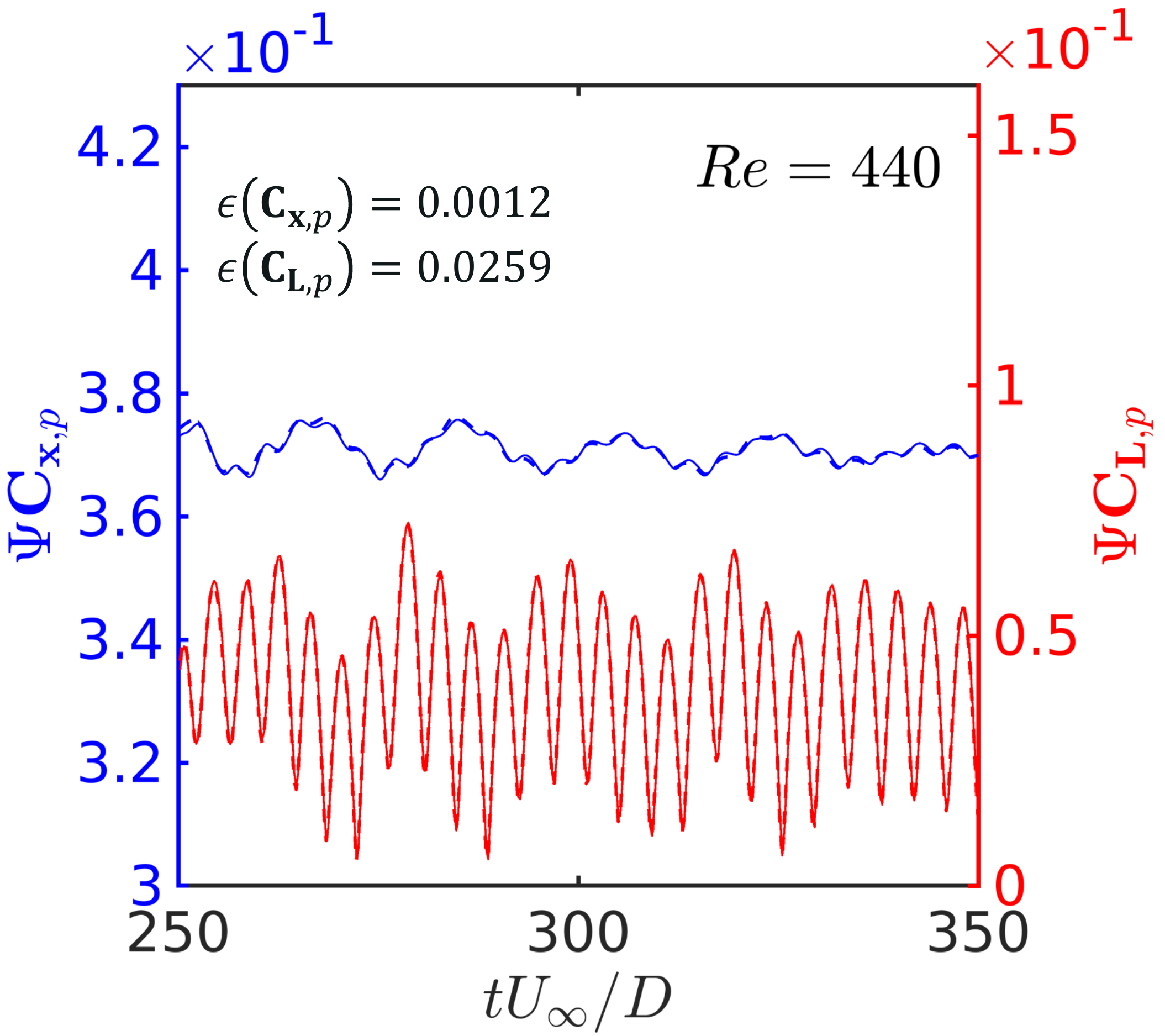}} 
\hspace{0.005\textwidth}
{\includegraphics[width =0.318\textwidth]{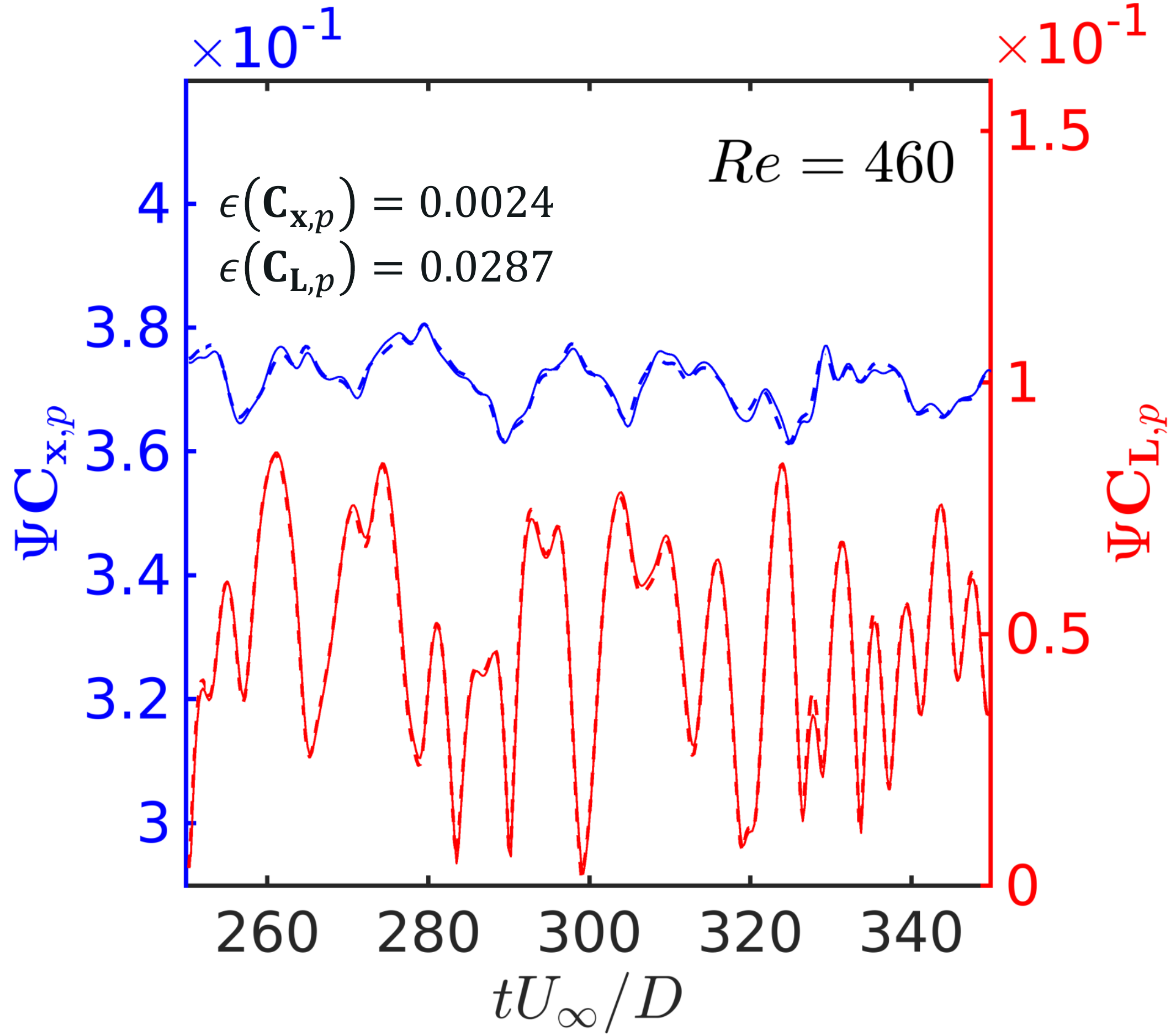}} \\ 
\caption{The variable flow past a sphere: Recovered voxelated load propagation (pressure drag and lift) on DL grid $64 \times 64 \times 64$ vs full-order CFD grid for variable $Re$ flows. Dashed and solid lines indicate the recovered DL grid loads and full-order loads, respectively.}
\label{forces_behavior_multiRe_voxel}
\end{figure*}

As seen previously, a trained 3D CRAN model and snapshot-FTLR offer fast and accurate data-driven field predictions and physical force integration.
By fast and accurate, we mean that one can avoid running the full-order model at a specific $Re$ and replace it with an optimized 3D CRAN framework. 
However, the major bottleneck of this deep learning architecture is the task of hyperparameter tuning and training, even for learning a periodical vortex shedding phenomenon at constant $Re$. 
The expensive offline training for another $Re$-dependent flow is, hence, less appealing for data-driven prediction from a practical standpoint. 
Moreover, the training can become challenging in flow scenarios that involve multi-$Re$ information. 
Subsequently, in this sub-section, we explore the training and predictive abilities of the 3D CRAN framework with multiple $Re$-based flow patterns. 
Of particular interest is to optimize a 3D CRAN framework on a variable $Re$ flow dataset within acceptable training cost and accurate predictive abilities.

We start by generating the full-order unsteady point cloud dataset for a variable $Re$-based 3D flow regime. 
We utilize the CFD domain in Fig.~\ref{setup_single} to generate flow snapshots for $Re_m=[280 \; 300 \; 320 ... 460]$ with a time step $0.25\;tU_{\infty} /D$.
For every $Re \subset Re_m$, we select a reduced time series training dataset $n_{\mathrm{v}}=400$ (from 250 till 350 $tU_{\infty} /D$). 
However, we maintain the same number of testing dataset $n_{ts}=100$ (from 350 till 375 $tU_{\infty} /D$) as compared to a single $Re$ scenario. 
%
While the hairpin shaped vortices are periodically shed  for the unsteady planar-symmetric flow regime $280 \leq Re \leq 370$, as the $Re$ is increased, the shedding orientation of the unsteady hairpin vortices becomes asymmetric in $370 \leq Re \leq 460$. 
The particular flow regime makes the problem challenging and is a good test case to replicate complexities in flow phenomenon where $Re$ can change. 
In the present case, we are interested in learning the strength and shedding of 3D flow patterns from unsteady planar-symmetric to asymmetric flows for $280 \leq Re \leq 460$.

\begin{figure*}
\centering
\includegraphics[width = 0.75\textwidth]{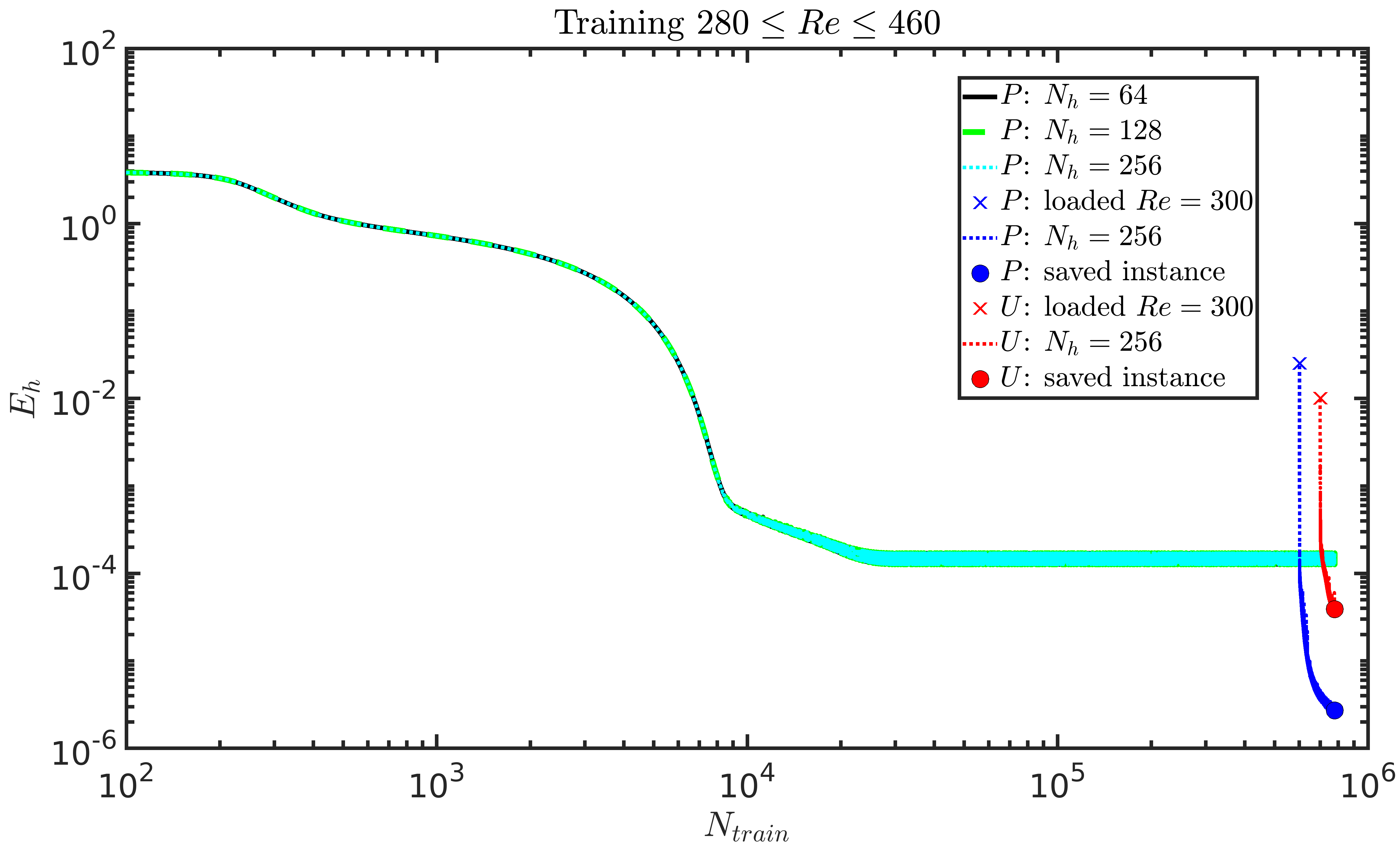}
\caption{The variable flow past a sphere: Evolution of the loss function $E_{h}$ with training iterations for different evolver cell sizes $N_h$. $P$ and $U$ denote the 3D CRAN models trained with variable $Re$-based pressure and x-velocity datasets, respectively. The blue cross and red cross represent the initialization of pressure and x-velocity training from optimized single $Re$ 3D CRAN model from Fig.~\ref{Re300_tr_loss}. Blue and red dots depict the new saved instances of the 3D CRAN parameters.}
\label{Revariable_tr_loss}
\end{figure*}

The point cloud field dataset is processed by interpolating and projecting in the same uniform DL space $\mathrm{\Omega}^{DL}$ of size $8D \times 8D \times 8D$ using the snapshot-FTLR. Like the single-$Re$ flow scenario, the coarse-grain interpolation and projection of the unstructured dataset are achieved using the linear method on a voxel grid $(N_x \times N_y \times N_z) = (64 \times 64 \times 64)$. The voxel forces from the 3D DL space are corrected by observing functional corrective mapping $\Psi$ for every $Re$-based flow information. 
Fig.~\ref{forces_behavior_multiRe_voxel} depicts the voxel force corrections on the training forces, with respective reconstruction error $\epsilon(.)$ for the drag and lift signals over the sphere. It can be interpreted that the mean and derivative error corrections (via Eq.~(\ref{eq15})) over various $Re$ numbers on the same 3D DL grid account for the generality of the snapshot-FTLR data recovery process. Irrespective of the flow patterns, the FTLR method recovers the bulk forces within $\epsilon(\mathrm{C}_{\mathrm{x},p}) = 0.003$ for drag and  $\epsilon(\mathrm{C}_{\mathrm{L},p}) = 0.03$ for lift, without requiring a change in the DL grid or grid resolution. Analogous to a single mesh generation process in CFD applications, the snapshot-FTLR method potentially generates a uniform DL grid for the domain-specific problem. Moreover, the inherent unstructured mesh complexity can be bypassed by focussing on a uniform Eulerian grid and 3D CNN operations.

The full-order training dataset matrix $\textbf{S}(Re_{m}) \in \mathbb{R}^{N_x \times N_y \times N_z \times (\mathrm{v} n_{\mathrm{v}})}$ consists of $\mathrm{v}=10$ $Re$ numbers and each $Re$ consisting $n_{\mathrm{v}}=400$ flow snapshots. The scaled flow trainable input 
$\mathcal{S}(Re_m)=\left\{\mathcal{S}_{s}^{\prime 1}\; \mathcal{S}_{s}^{\prime 2}\; \ldots\; \mathcal{S}_{s}^{\prime N_{s} } \right\} \in[0,1]^{N_{x} \times N_{y} \times N_{z} \times N_{t} \times N_{s}}$
is generated using the basic principles of normalisation and batch-wise arrangement with $N_s = 160,\;N_t=25$. Note that $\mathcal{S}_{s}^{\prime j}=\left[\mathbf{S}_{s, j}^{\prime 1} \; \mathbf{S}_{s, j}^{\prime 2} \ldots \mathbf{S}_{s, j}^{N_{t}}\right]$ is a time series data at a particular $Re$ value. The complete spatio-temporal training dataset for the present case is in the order of $1.04 \times 10^{9}$. The dataset preparation is detailed in section~\ref{Tansfer_learning_VAR_RE}. 
To train such a big spatio-temporal dataset on a deep 3D CRAN architecture, we initialize the network training from saved $Re=300$ parameters as source domain. 
This is done to gain the advantage of fine-tuning the 3D CRAN for variable $Re$ flow data from single $Re$ flow parameters and bypassing the expensive hyperparameter search.
This is further elaborated in  Fig.~\ref{Revariable_tr_loss} where the evolution of the hybrid loss function $E_h$ is showcased with the training iterations. Every training iteration consists of a mini-batch of size $n_{s}=1$ randomly shuffled from the scaled flow input $\mathcal{S}(Re_m)$ and updating the neural parameters in $\approx 0.15s$. 

As shown in Fig.~\ref{Revariable_tr_loss}, the 3D CRAN architecture with $N_{h} = 64,128,256$ does not optimise with a random parameter search for the pressure dataset even after training for $\approx 64$ hours on a single GPU. However, the transfer of learning improves traditional learning by transferring knowledge learned in a single $Re$ flow scenario and improving learning in variable $Re$ flow scenarios. With transfer learning, the 3D CRAN starts to optimize from $N_{train}=10000$ iterations. We save the new 3D CRAN model parameters after $N_{train} = 80000$ for the pressure and $\mathrm{x}$-velocity testing with $E_{h} = 2.73 \times 10^{-6}$ and $3.90 \times 10^{-5}$, respectively. The total offline training time for learning variable $Re$ flow regime took around 3 hours of single GPU training by leveraging single $Re$ flow domain knowledge. 

\begin{figure*}
\centering
\includegraphics[width = 0.325\textwidth]{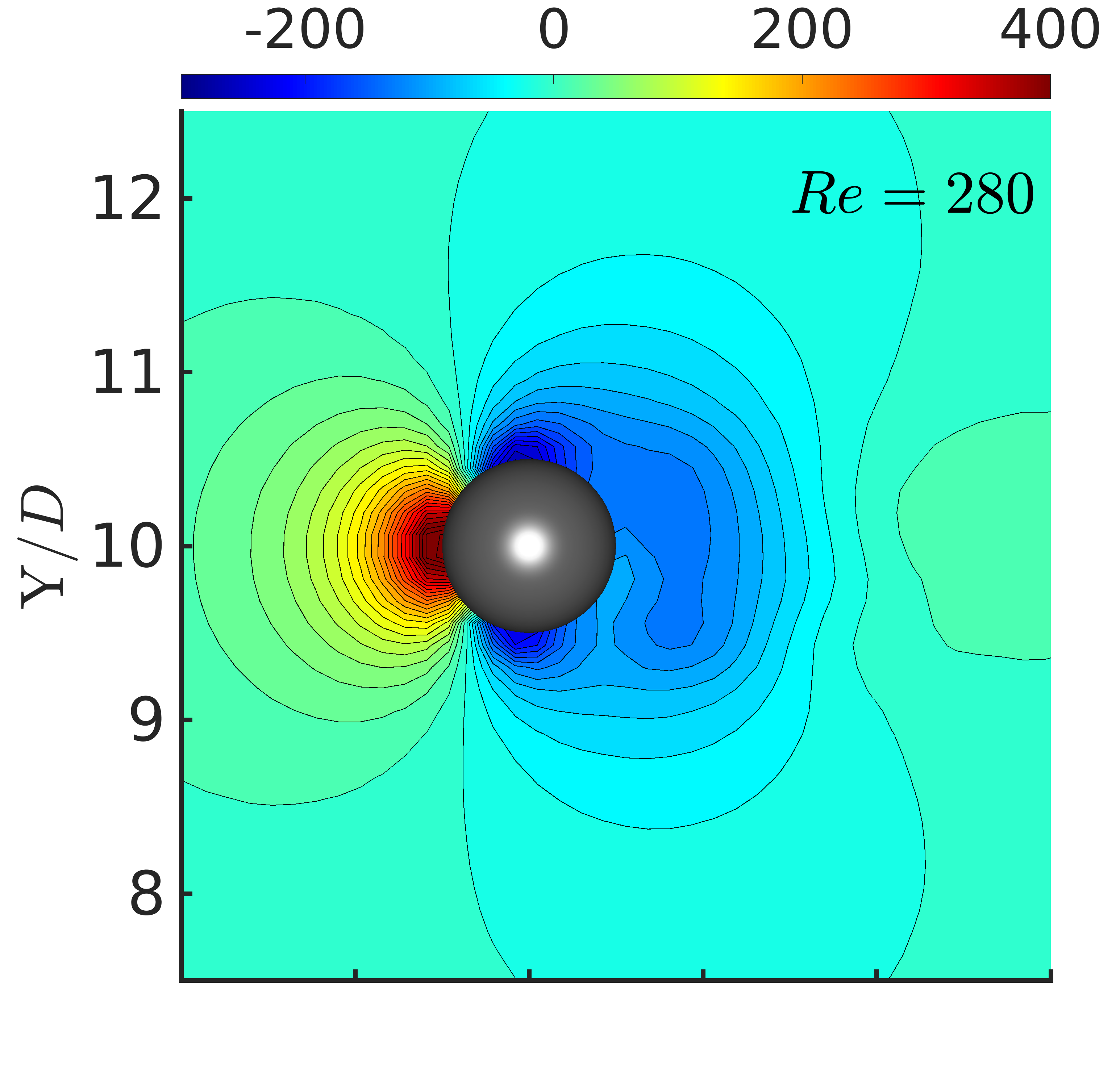}
\hspace{0.005\textwidth}
\includegraphics[width = 0.3\textwidth]{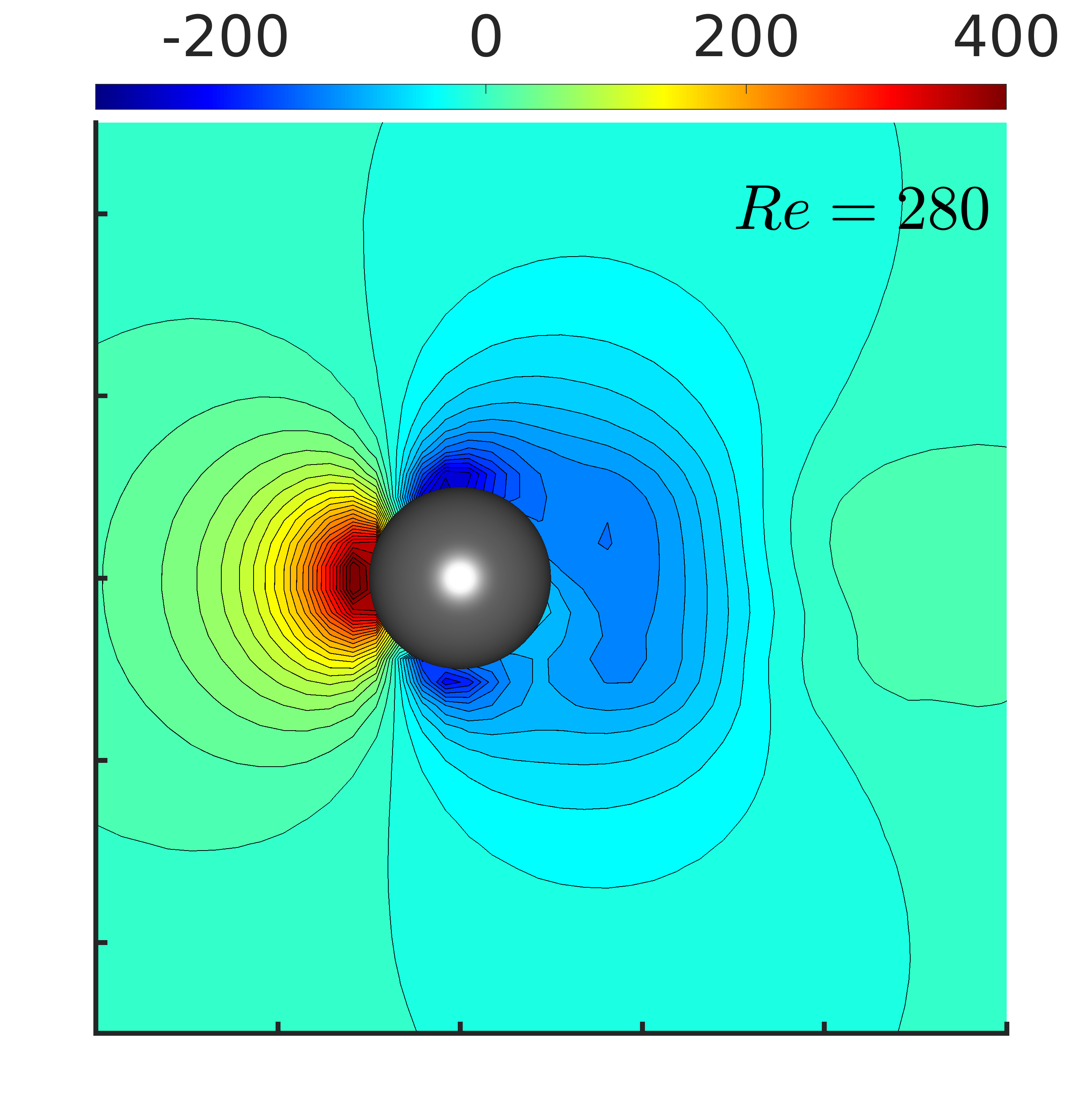}
\hspace{0.005\textwidth}
\includegraphics[width = 0.315\textwidth]{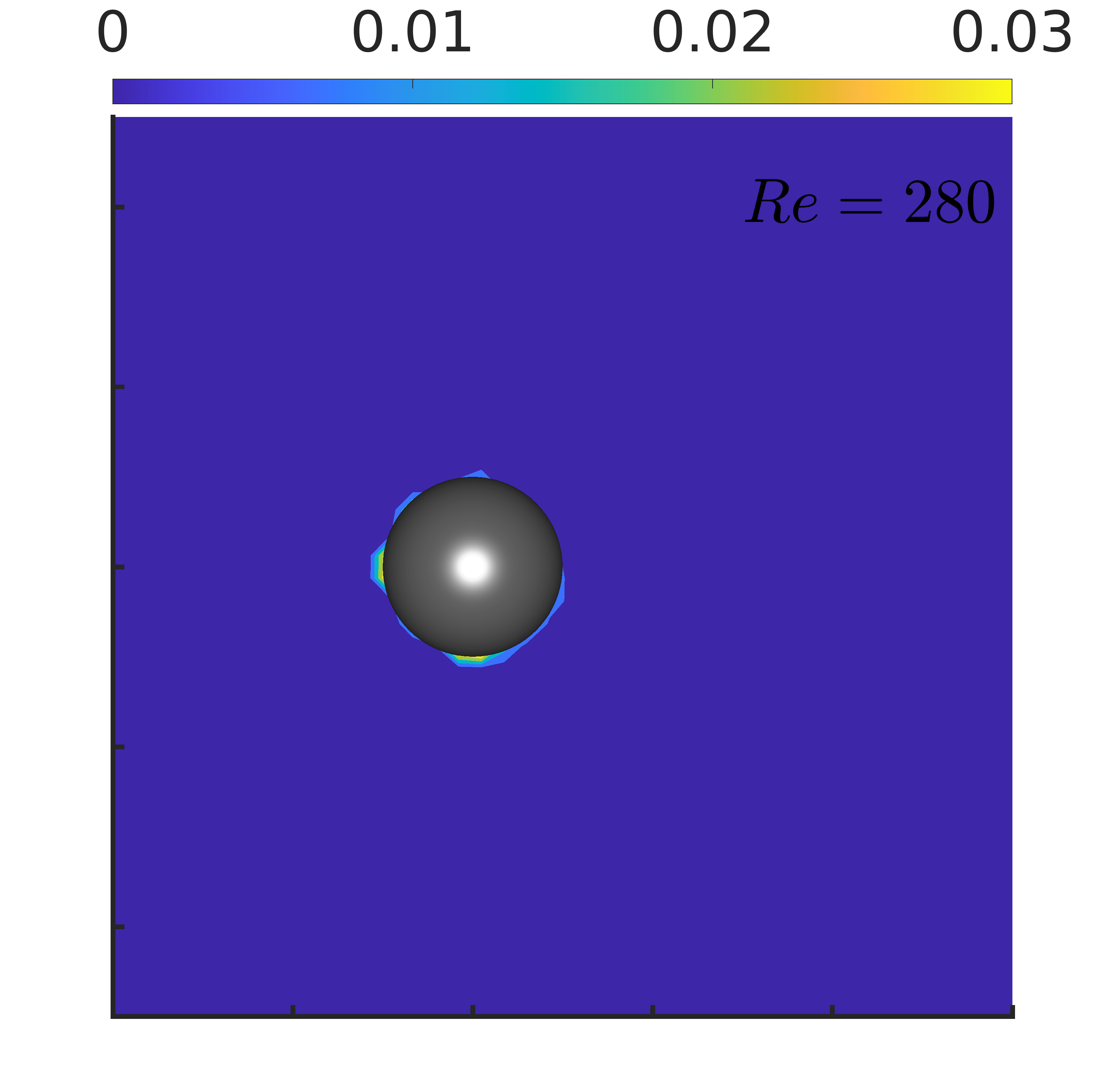}
\\
{\includegraphics[width = 0.325\textwidth]{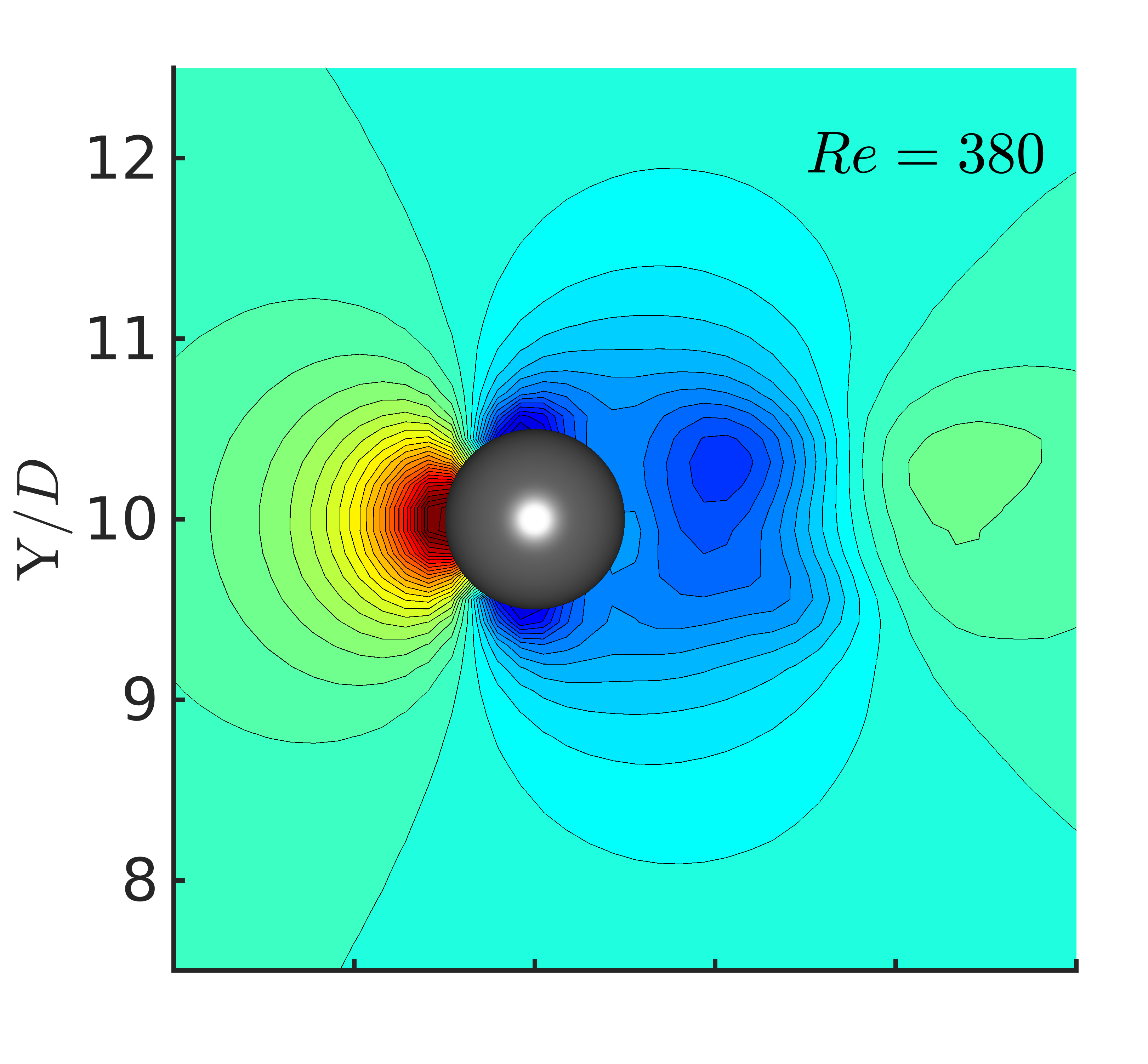}
\hspace{0.005\textwidth}
\includegraphics[width = 0.3\textwidth]{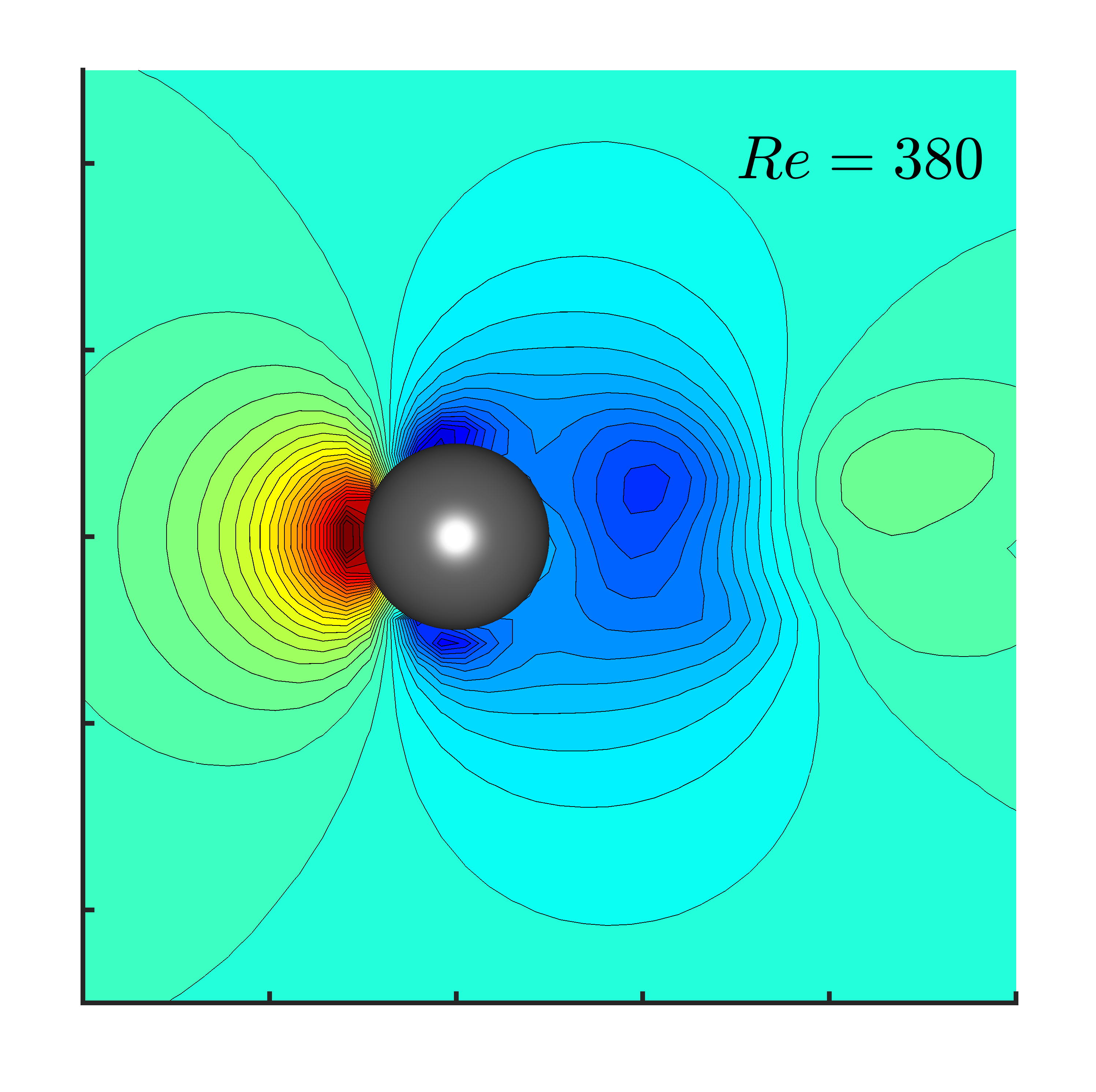}
\hspace{0.005\textwidth}
\includegraphics[width = 0.305\textwidth]{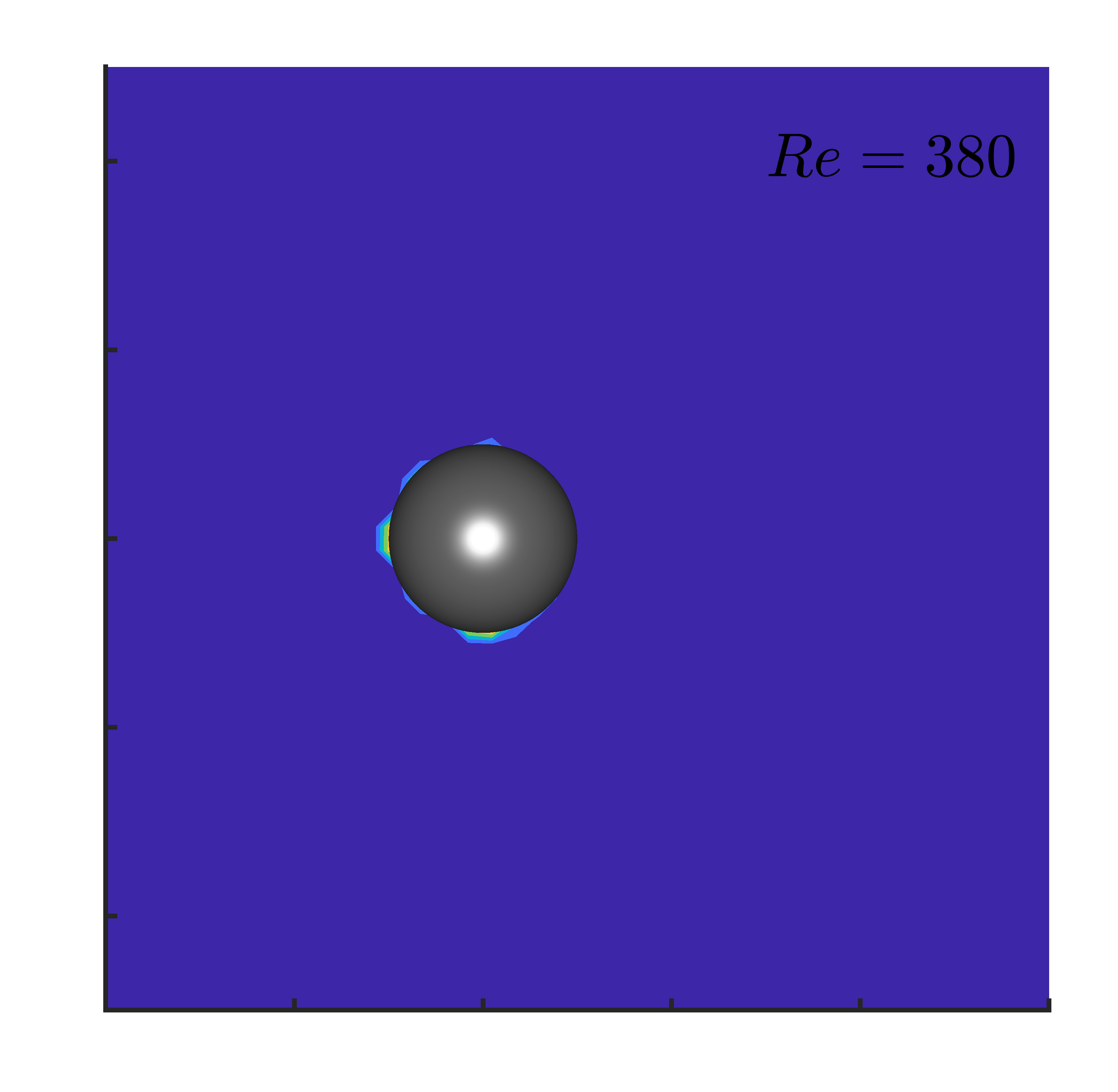}}
\\
{\includegraphics[width = 0.33\textwidth]{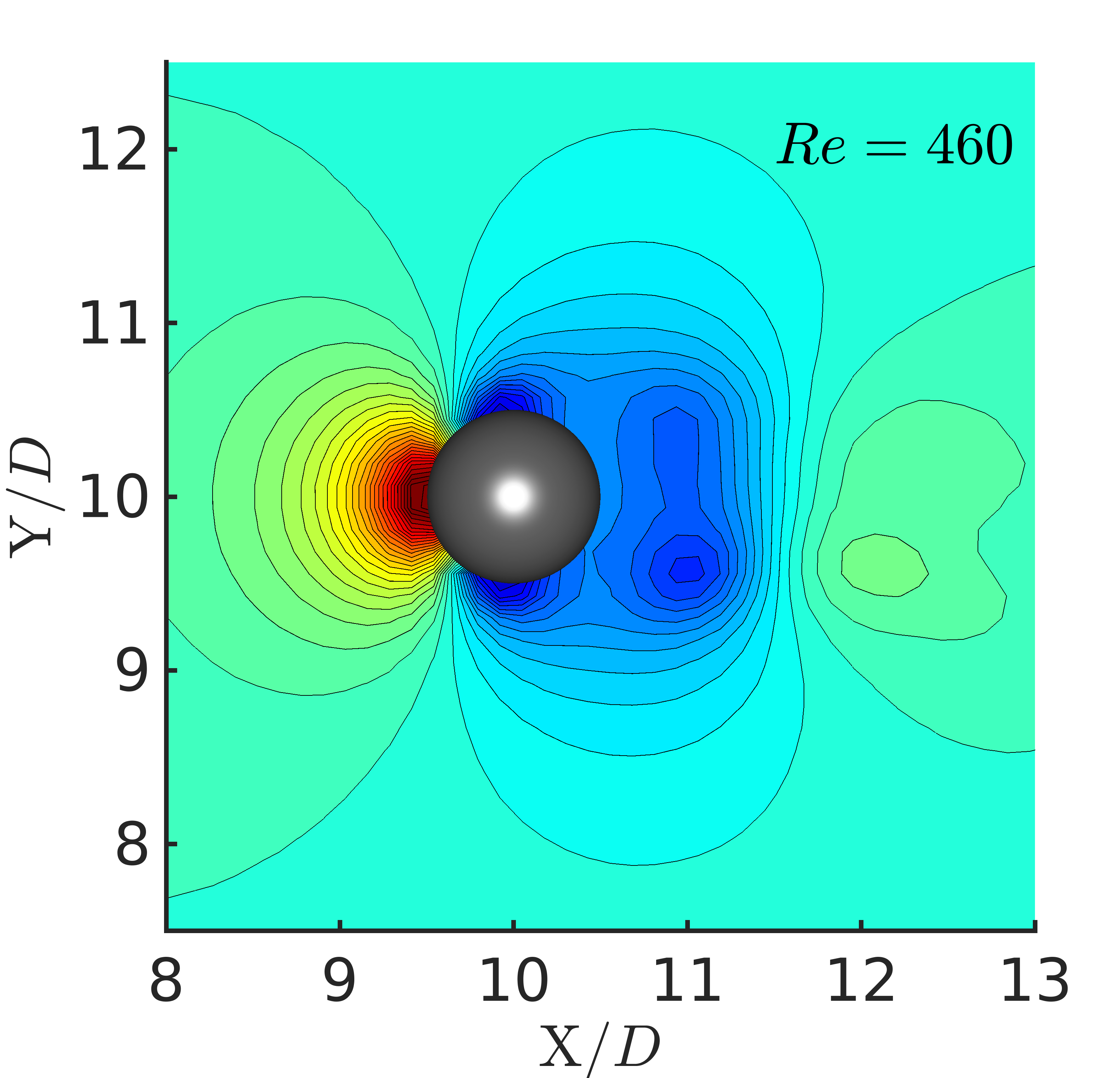}
\hspace{0.005\textwidth}
\includegraphics[width = 0.307\textwidth]{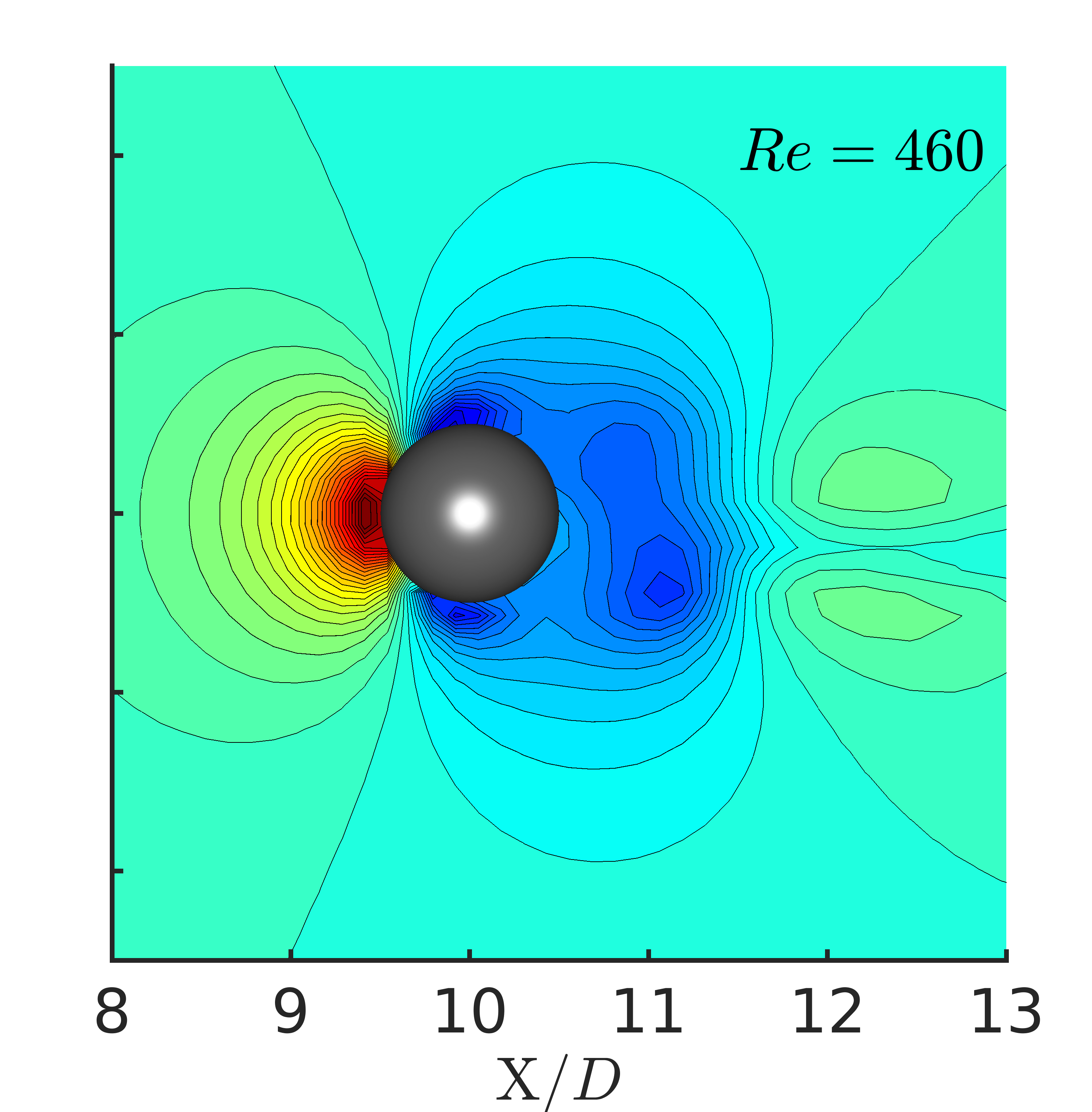}
\hspace{0.005\textwidth}
\includegraphics[width = 0.302\textwidth]{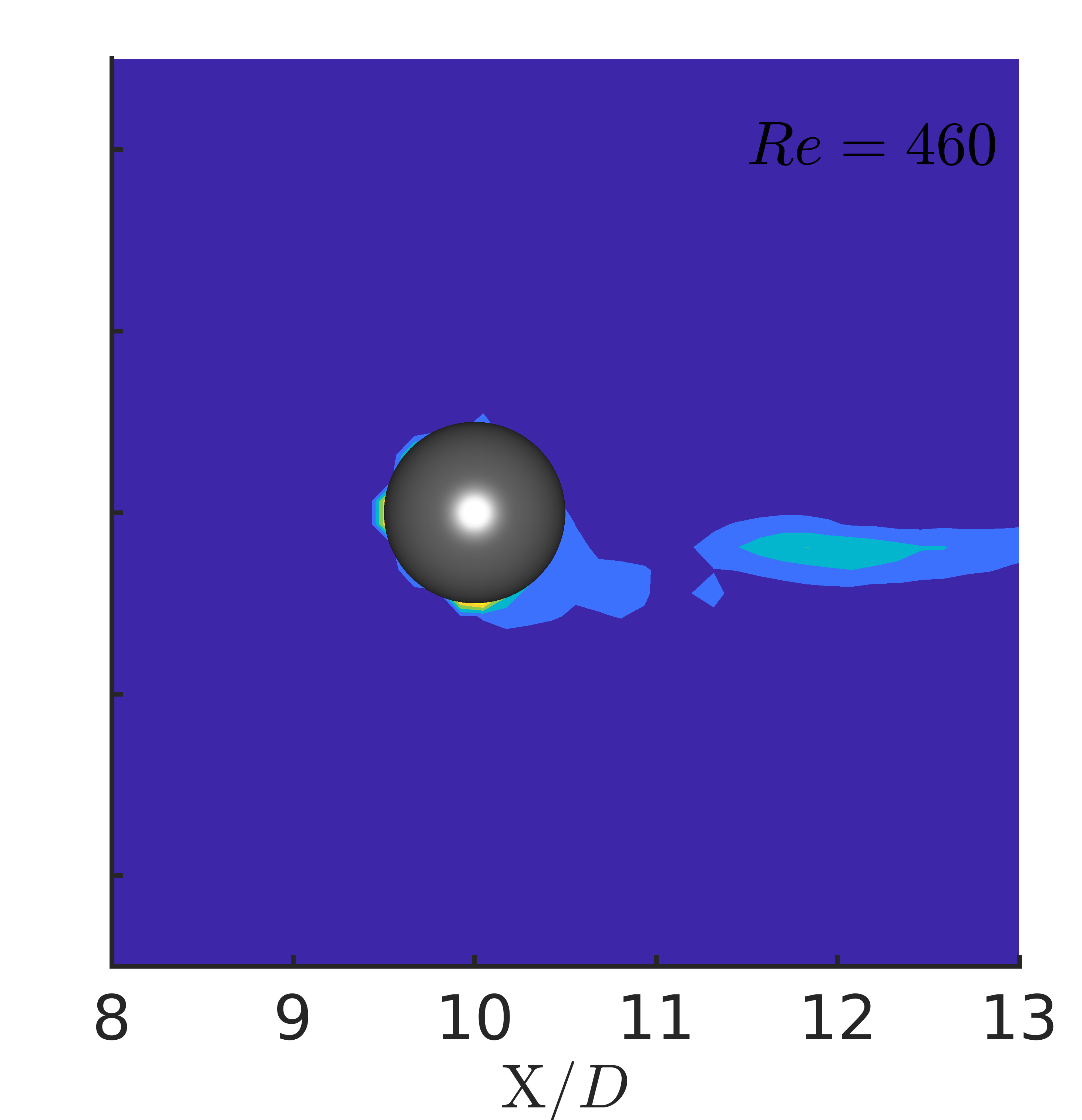}} 
\caption{The variable flow past a sphere: Predicted and true pressure field comparison with normalized reconstruction error $E^{i}$ at $tU_{\infty}/D = 372.5$ sliced in $\mathrm{Z}/D=10$ plane. Left, middle and right contour plots depict the prediction, true and errors, respectively.}
\label{Re_variable_pres_sph_pred}
\end{figure*}

For the data-driven predictions of the pressure and x-velocity, the trained 3D CRAN models with $N_h=256$ are utilized. We keep the multi-step predictive cycle length of $p=25$. 
The predicted and true values of the pressure field at the test time $372.5\;tU_{\infty}/D$ (80th step) are compared in Fig.~\ref{Re_variable_pres_sph_pred} where the contour plots for $\mathrm{Z}/D = 10$ plane are shown.
From the reconstruction and inference, it is interesting to observe that the network differentiates and infers in time a specific $Re$-based field that it is instantiated with.
The reconstruction error $E^i$ is calculated by taking the absolute value of the difference between the true and predicted fields and then normalizing it with the truth's $L_2$ norm for the 3D DL space. 
The errors are in the order of $10^{-2}$ near the interface and $10^{-4}$ elsewhere
and are found to increase slightly in the nonlinear wake as the flow becomes asymmetric from $Re \geq 360$. 
These predictions imply that the network accurately learns the low-dimensional patterns for the variable $Re$-based flow with limited and unlabelled information. 

\begin{figure*}
\centering
\includegraphics[width = 0.65\textwidth]{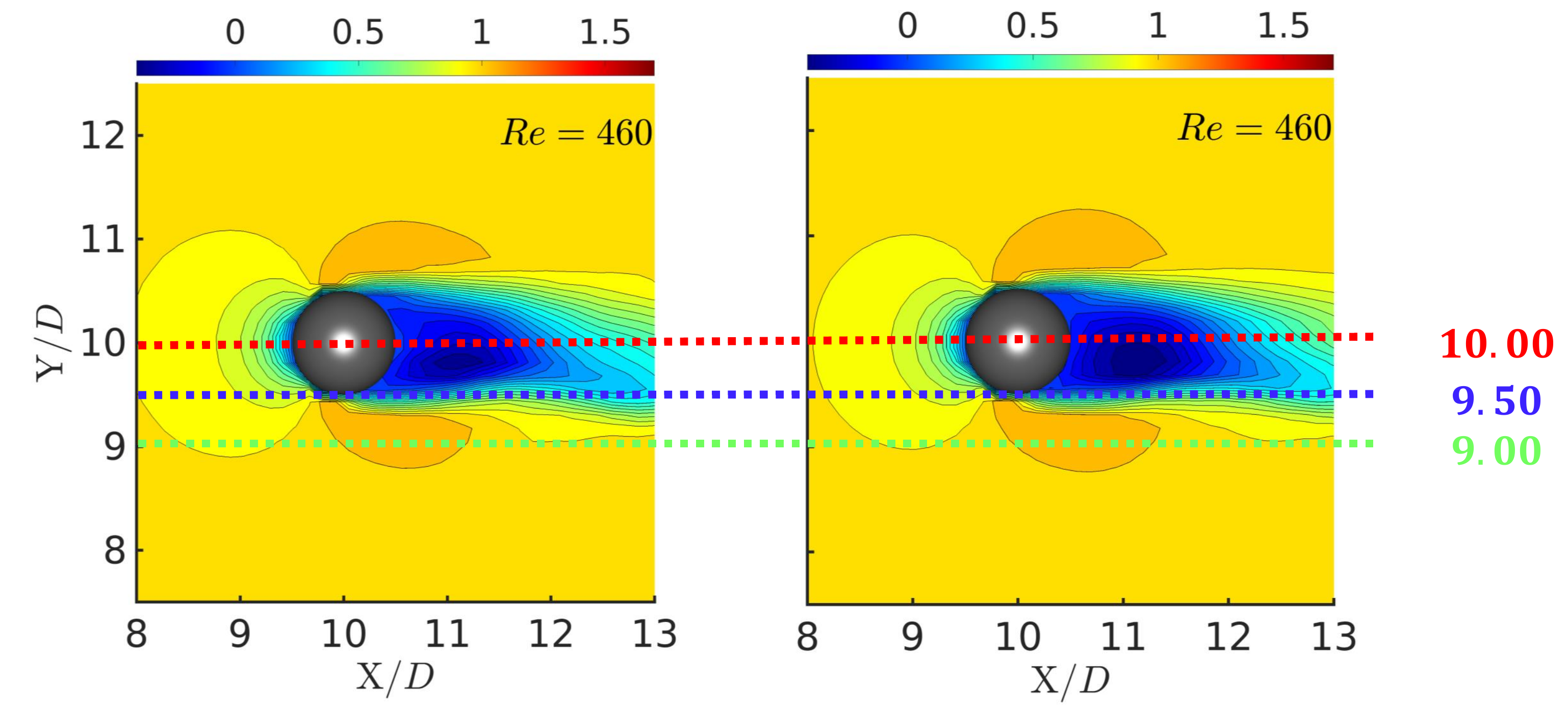}\\ 
(a)\\
{\includegraphics[width = 0.28\textwidth]{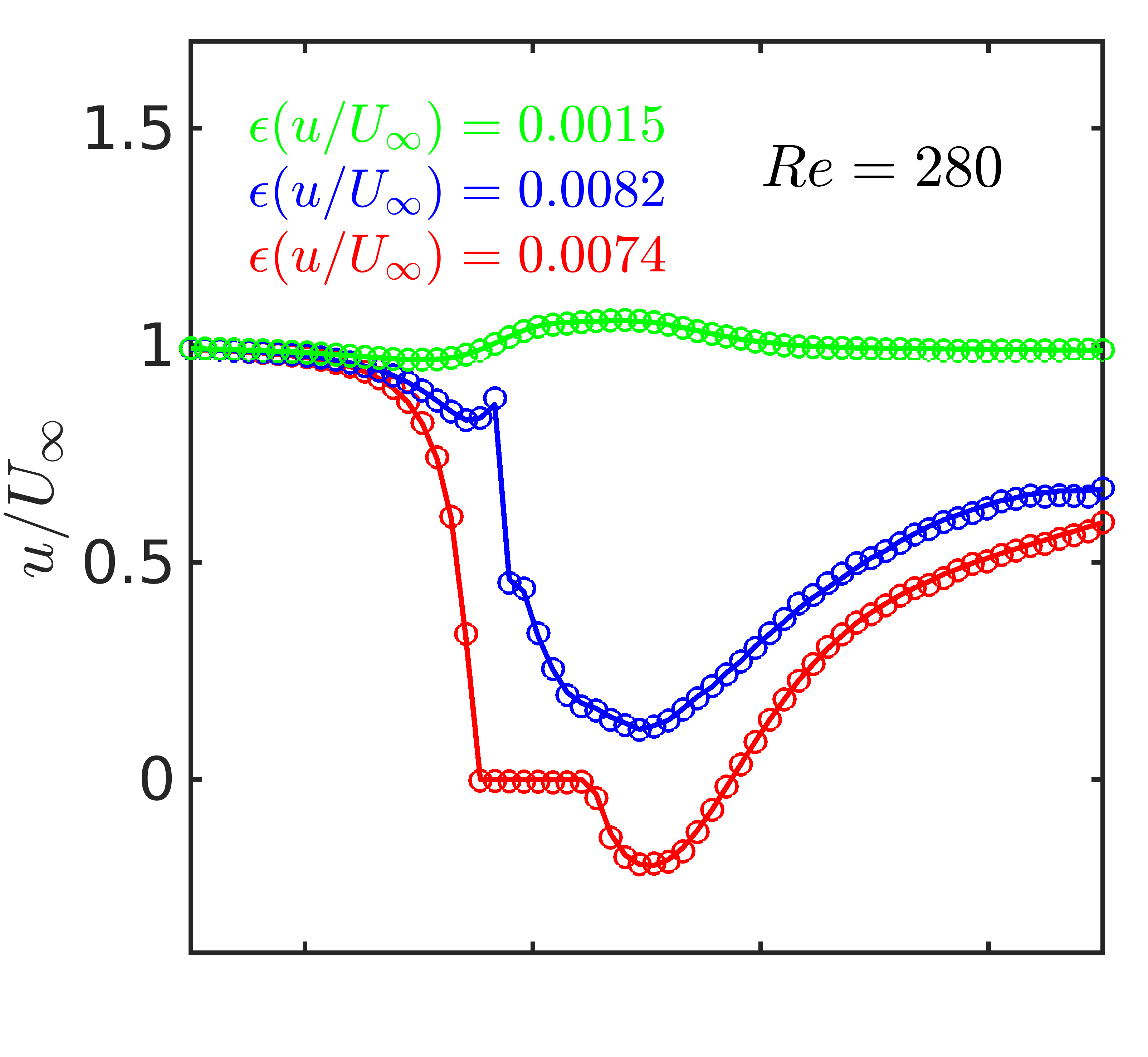}} 
\hspace{0.005\textwidth}
{\includegraphics[width = 0.285\textwidth]{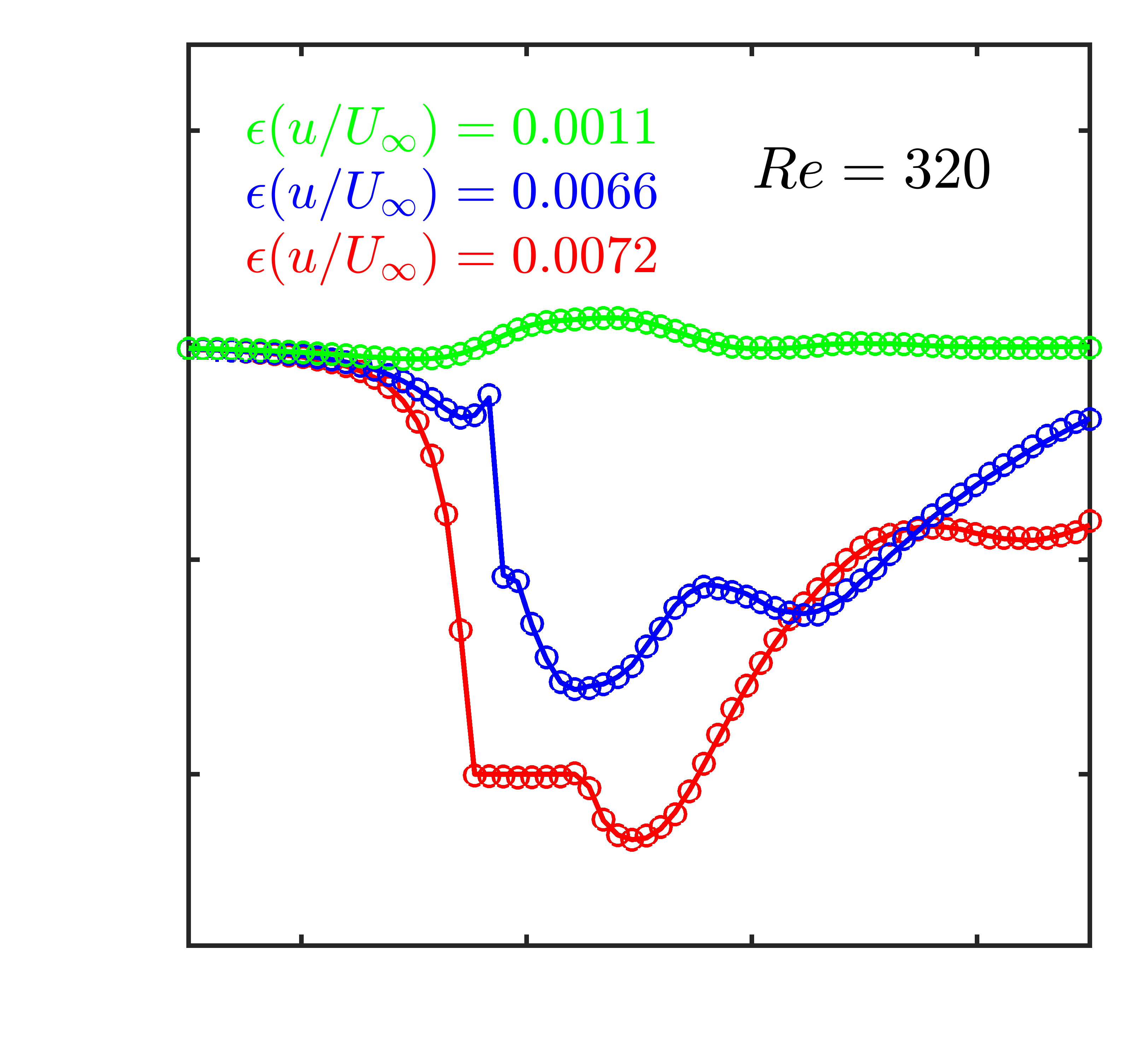}} 
\hspace{0.005\textwidth}
{\includegraphics[width =  0.28\textwidth]{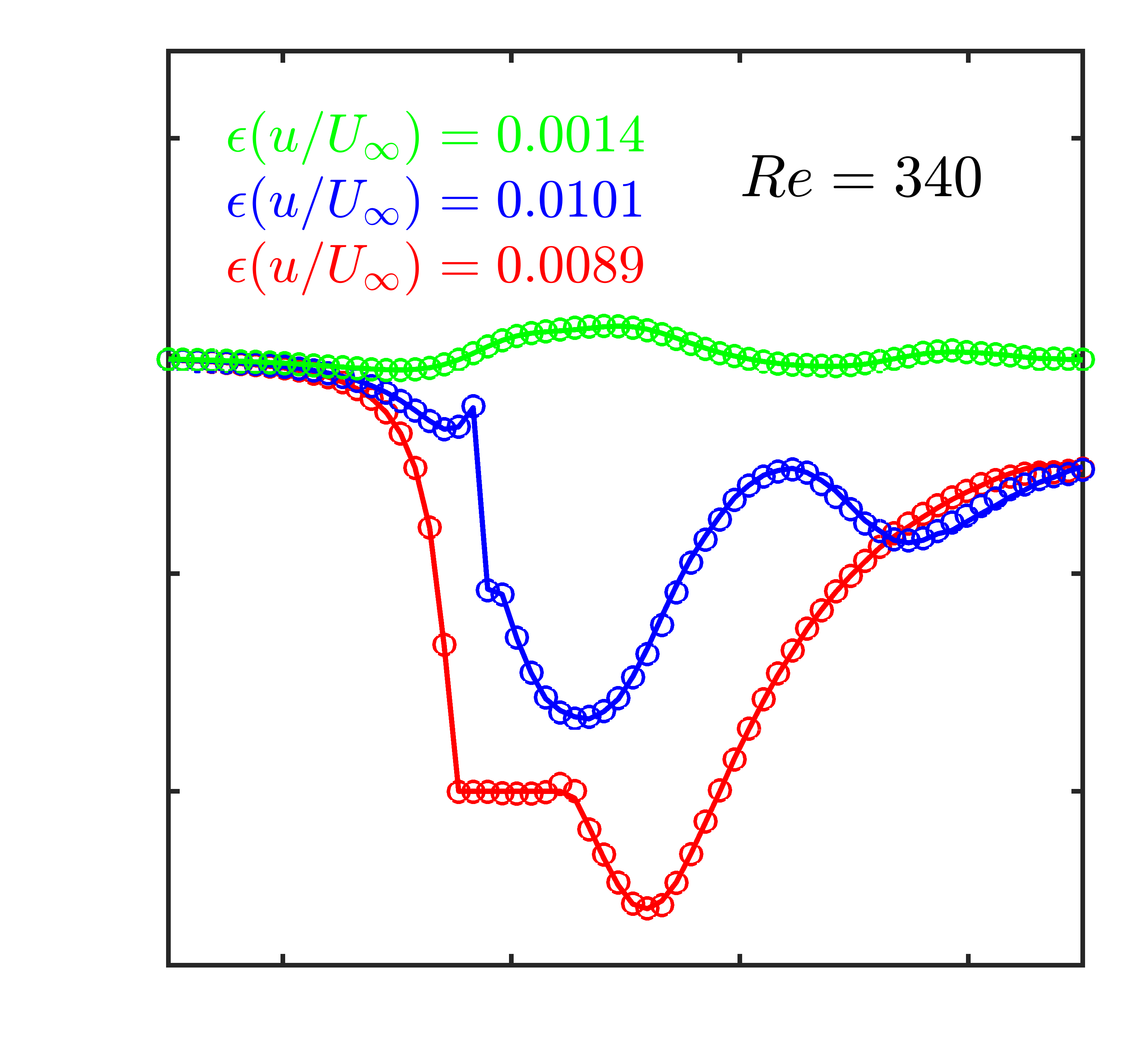}} 
\\ 
{\includegraphics[width = 0.28\textwidth]{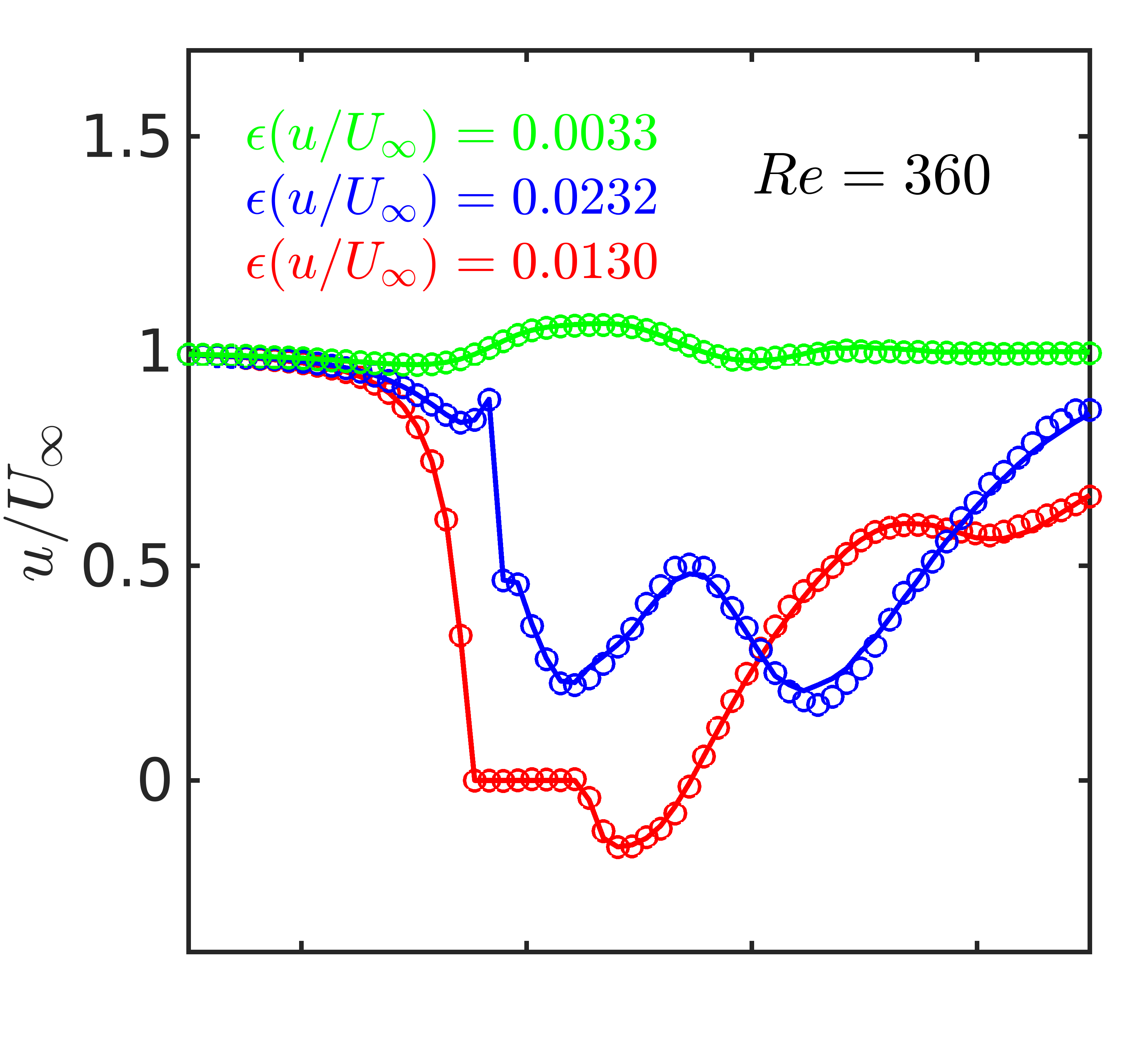}} 
\hspace{0.005\textwidth}
{\includegraphics[width = 0.278\textwidth]{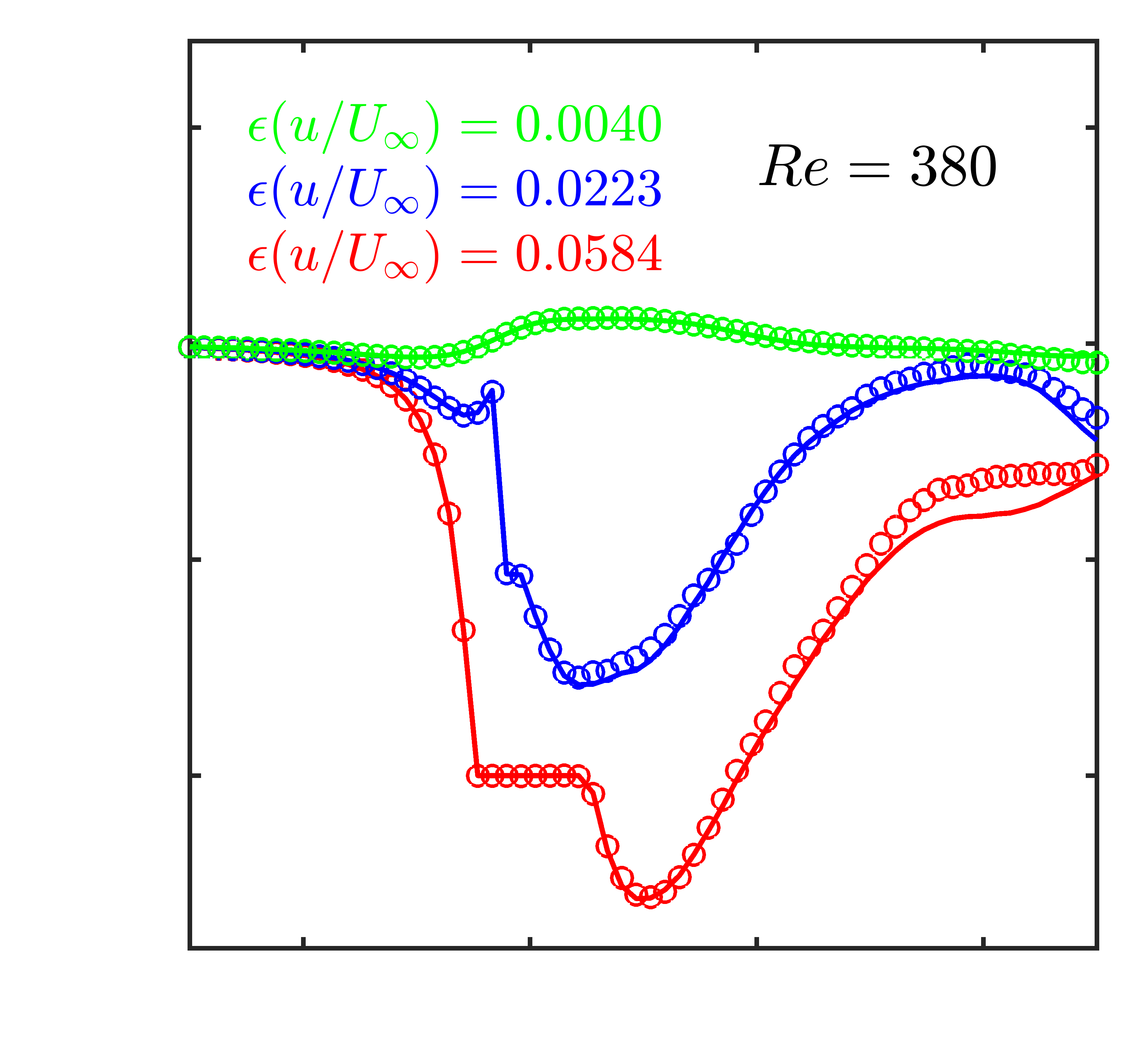}} 
\hspace{0.005\textwidth}
{\includegraphics[width =0.278\textwidth]{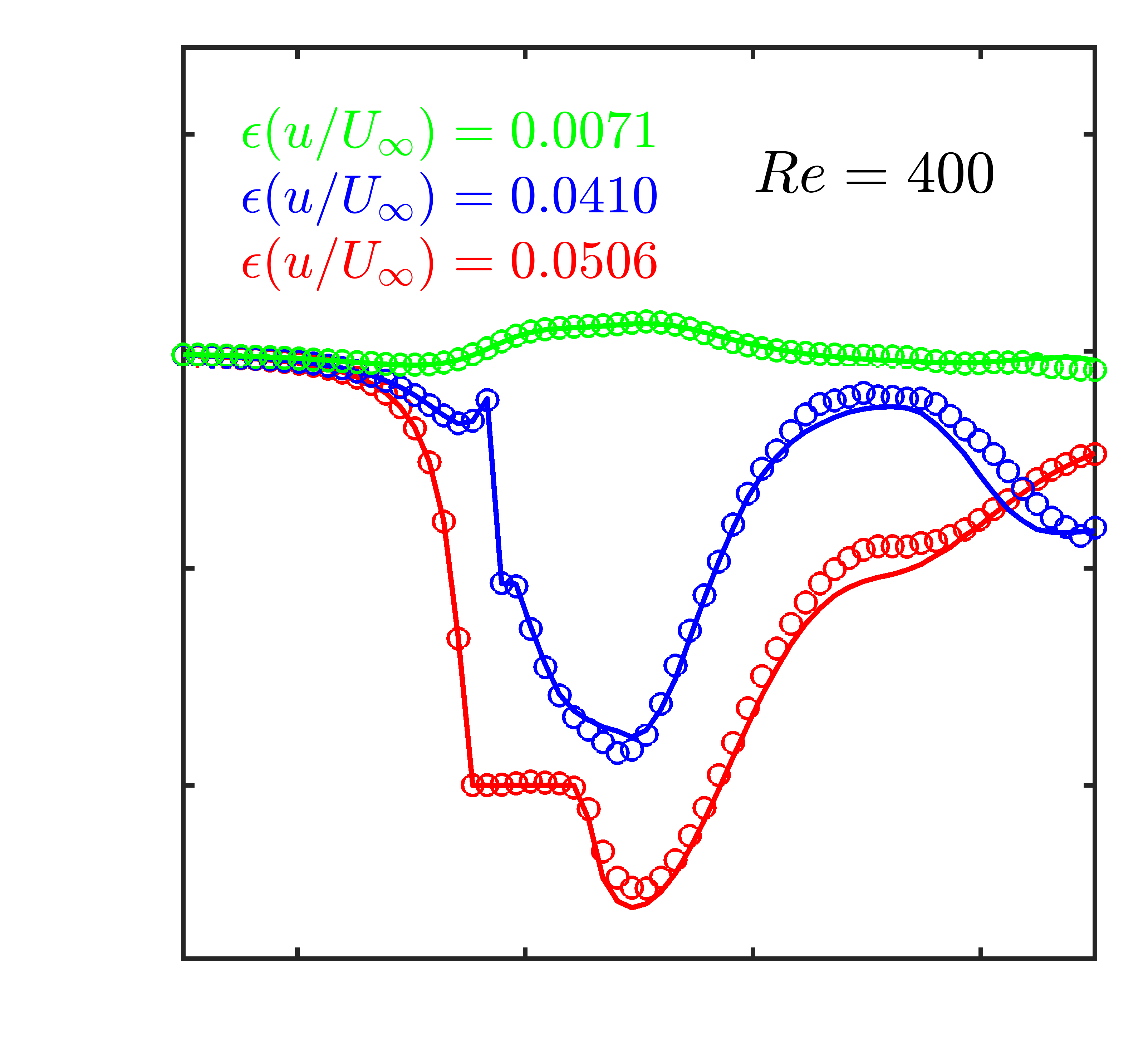}} 
\\ 
{\includegraphics[width = 0.28\textwidth]{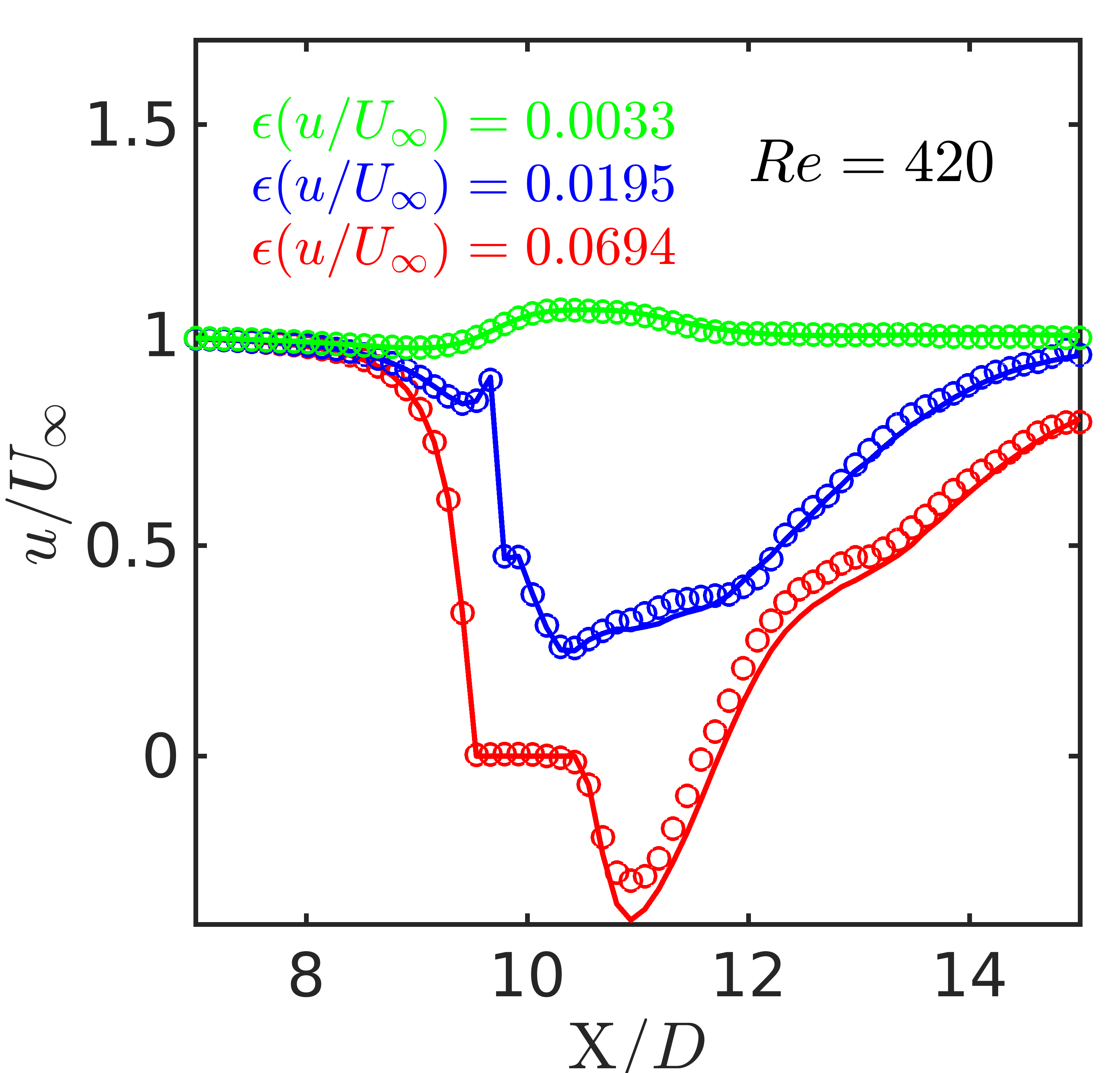}} 
\hspace{0.005\textwidth}
{\includegraphics[width = 0.278\textwidth]{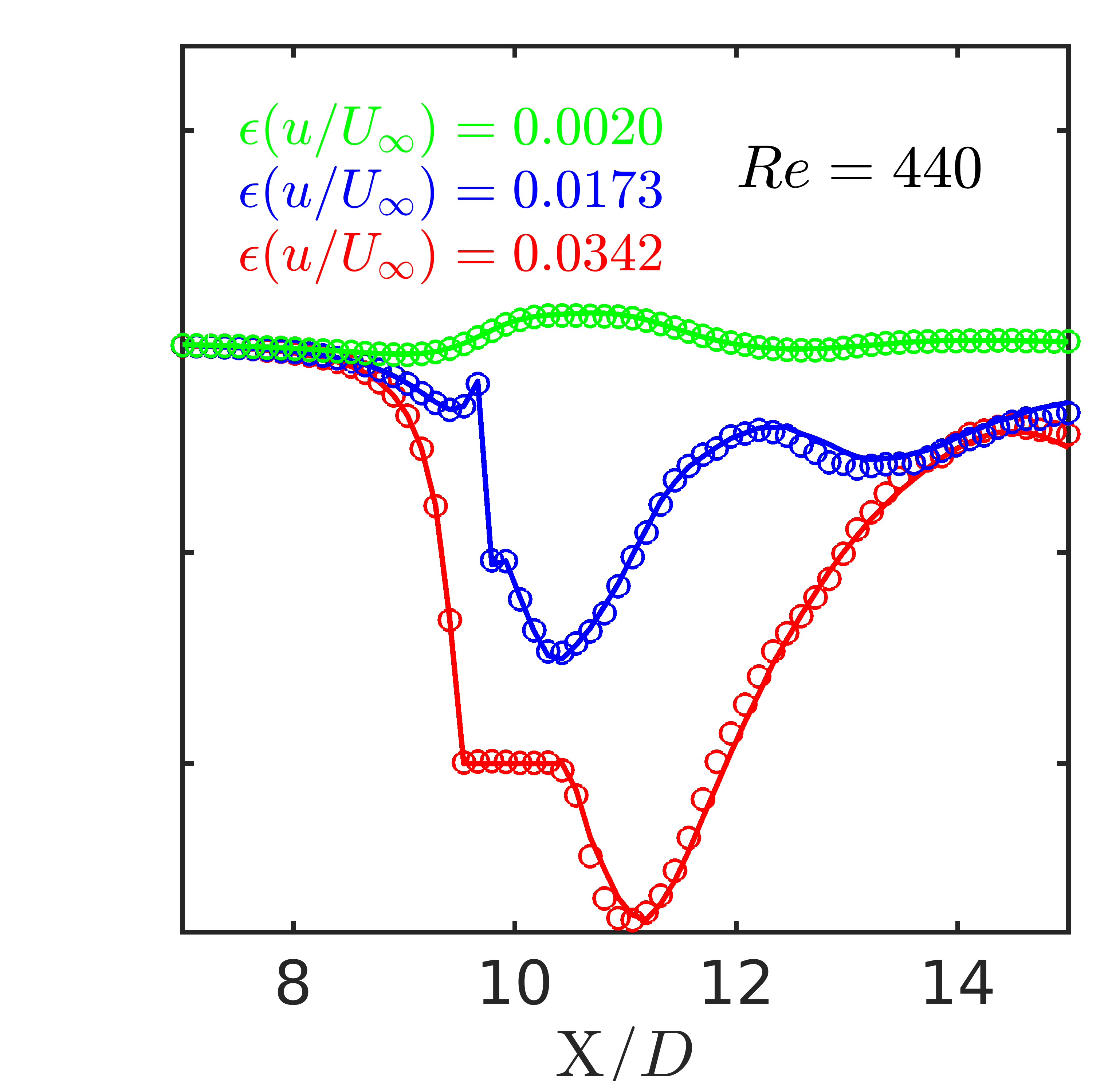}} 
\hspace{0.005\textwidth}
{\includegraphics[width =0.277\textwidth]{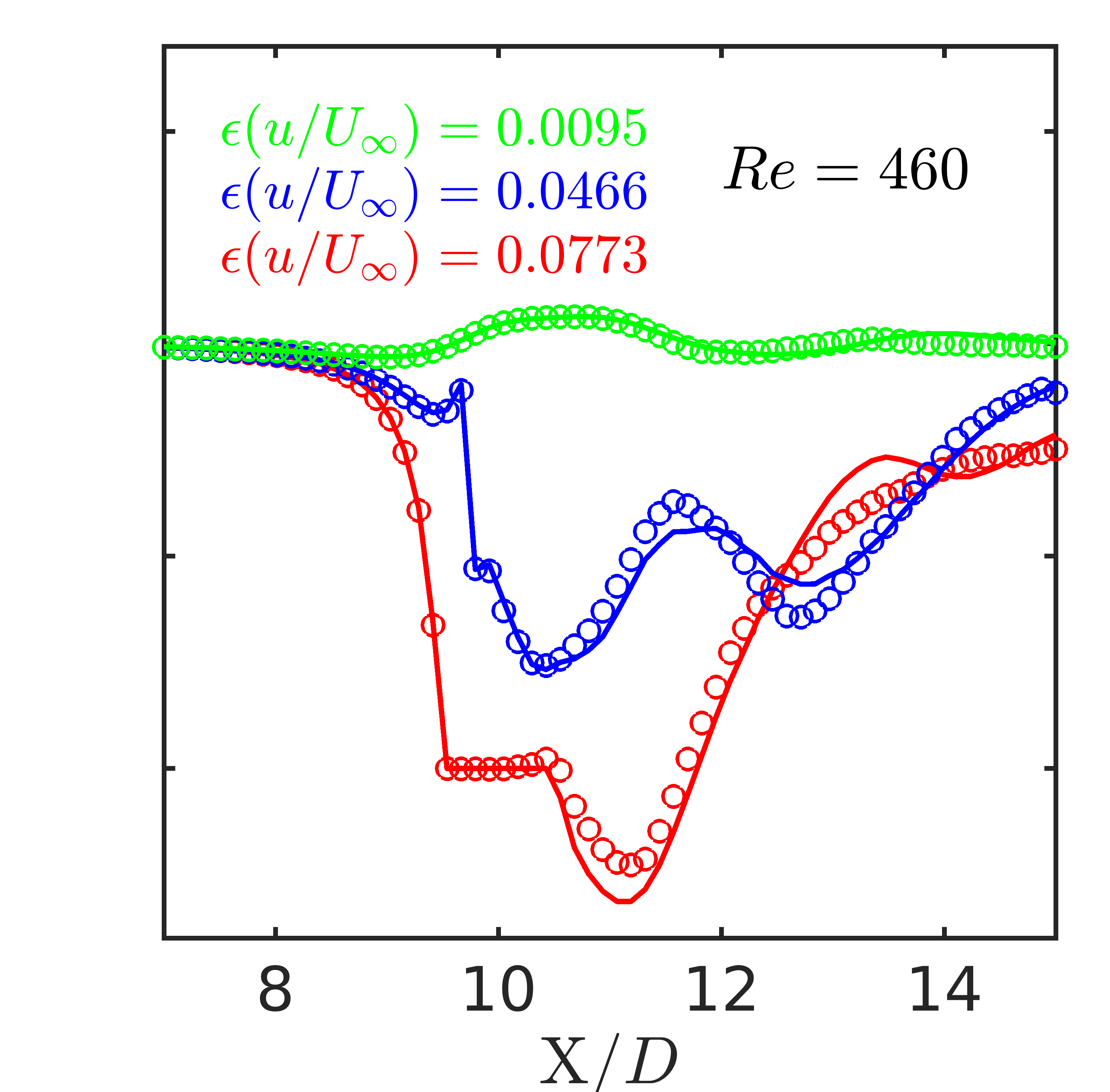}}\\ 
(b)
\caption{The variable flow past a sphere: (a) Predicted (left) and true (right) x-velocity field comparison. (b) Comparison of the streamwise velocity profiles of the 3D CRAN prediction and ground truth at three locations $\mathrm{Y}/D=9.0,9.5,10$ for all Reynolds numbers. Results are plotted at test time $tU_{\infty}/D = 372.5$ sliced in $\mathrm{Z}/D=10$ plane. Circles indicate the 3D CRAN predictions, and solid lines represent the ground truth.}
\label{Re_streamwise_velo_variable}
\end{figure*}

Similarly, the profiles of the streamwise velocity from the predicted fields and ground truth are compared in Fig.~\ref{Re_streamwise_velo_variable} at test time step $372.5\;tU_{\infty}/D$ in $\mathrm{Z}/D = 10$ plane. 
Closed-loop predictions at all $Re$ are in good agreement with the ground truth velocity in terms of the peak, width, and shape of the streamwise velocity profiles.
Velocity profiles at $\mathrm{Y}/D= 9.0, 9.5$ show no identifiable differences
between the ground truth and 3D CRAN predictions at all Reynolds
numbers. This is because flow at $\mathrm{Y}/D= 9.0, 9.5$ is almost laminar layer flow, the characteristics of which are relatively easily 
trained by the network. Minor differences in the velocity deficit are observed for $\mathrm{Y}/D= 10.0$ in the nonlinear wake region for $Re \geq 380$ where
the 3D CRAN does not accurately capture small-scale oscillatory motions. 

Based on the predicted pressure flow fields and the snapshot-FTLR force integration, we discuss the comparison of the mean drag and lift forces from the 3D CRAN prediction and ground truth.
Fig.~\ref{forces_behavior_Re} shows the performance of the CRAN-based force predictions when fed with different 3D flow snapshots for Reynolds number $Re \subset Re_{m}$. 
To compare the accuracy of the force predictions, we report the $\mathrm{R}^2$ error between the true $\mathrm{C}_i$ and predicted $\hat{\mathrm{C}}_i$ mean force coefficients calculated using 
\begin{equation}
\mathrm{R}^{2}= 1 - \frac{\sum\left(\hat{\mathrm{C}}_{i}-\bar{\mathrm{C}}\right)^{2}}{\sum\left(\mathrm{C}_{i}-\bar{\mathrm{C}}\right)^{2}}.
\end{equation}
Here, $\mathrm{C}_i$ can be the mean drag or lift for a particular Reynolds number. 
The $\mathrm{R}^2$ errors for the mean drag and lift fit for different Reynolds numbers are $98.58 \%$ and $76.43\%$, respectively, which demonstrates the high efficiency of the CRAN-based prediction process. 
We find that the predictions perform the best when
the field values correspond to $Re \leq 380$, which characterizes a 3D symmetric shedding of the unsteady vortex patterns. 
Furthermore, when $Re \geq 440$, all the predictions are accurate within a $5\%$ error margin of the FOM results. 
The performance of the 3D CRAN-based deep learning becomes slightly deficit over the 3D transitional flow regime consisting of $Re = 400,420$. Interestingly, the maximum prediction errors correspond to this complicated flow from symmetric to asymmetric unsteady patterns.  
The data of a similar problem may enhance the accuracy of predictions in this transitional flow regime. 
The most significant result is that the 3D CRAN has accurately captured the maximum and minimum mean drag and lift coefficients for the chosen flow regime $280 \leq Re \leq 460$ in a dearth of training data and on limited training time. 
Accurate force predictions correspond to a proper field inference from the 3D CRAN framework.

\begin{figure}
\centering
\subfloat[]{\includegraphics[width = 0.48\textwidth]{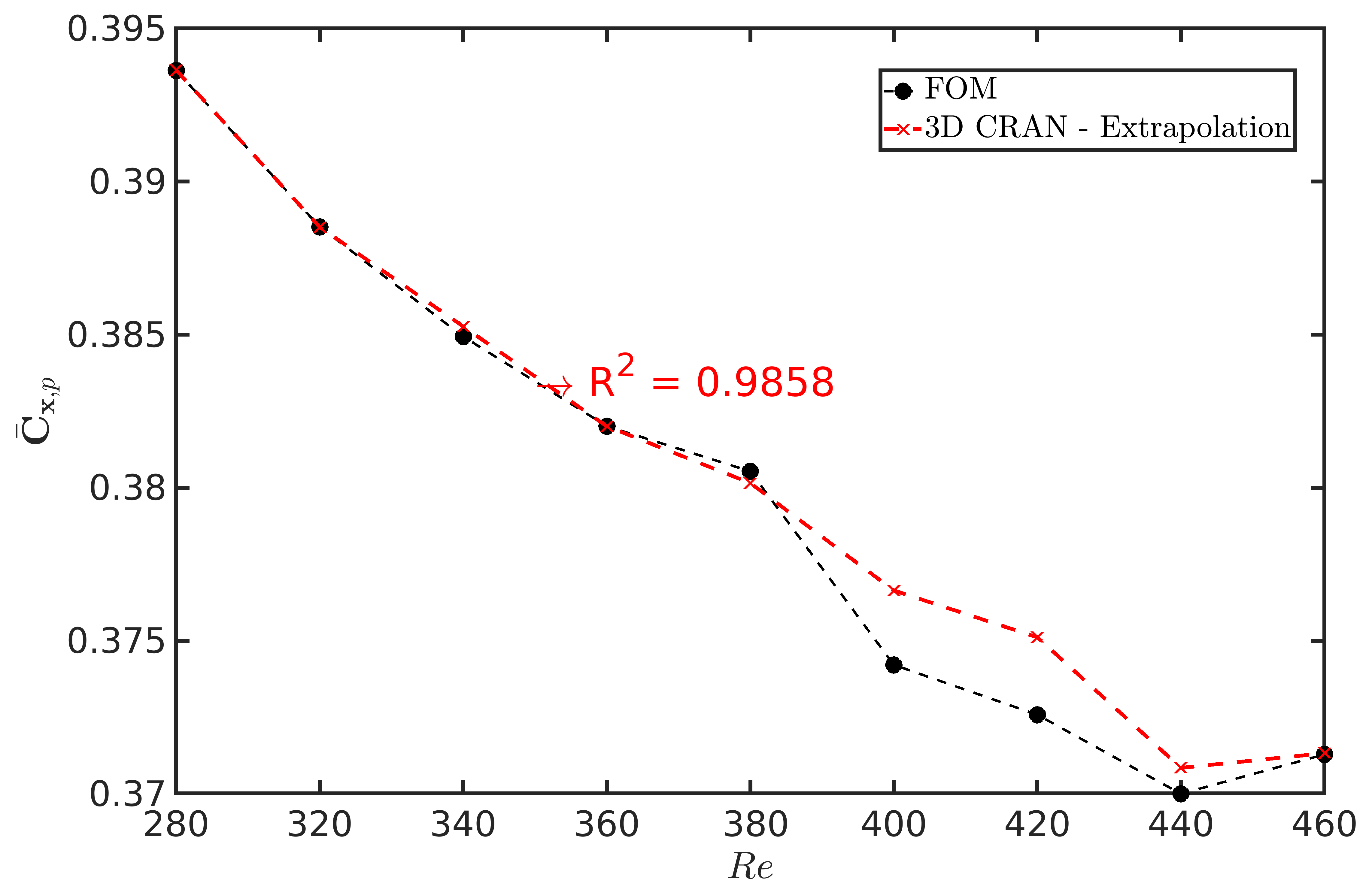}} 
\\
\subfloat[]{\includegraphics[width = 0.48\textwidth]{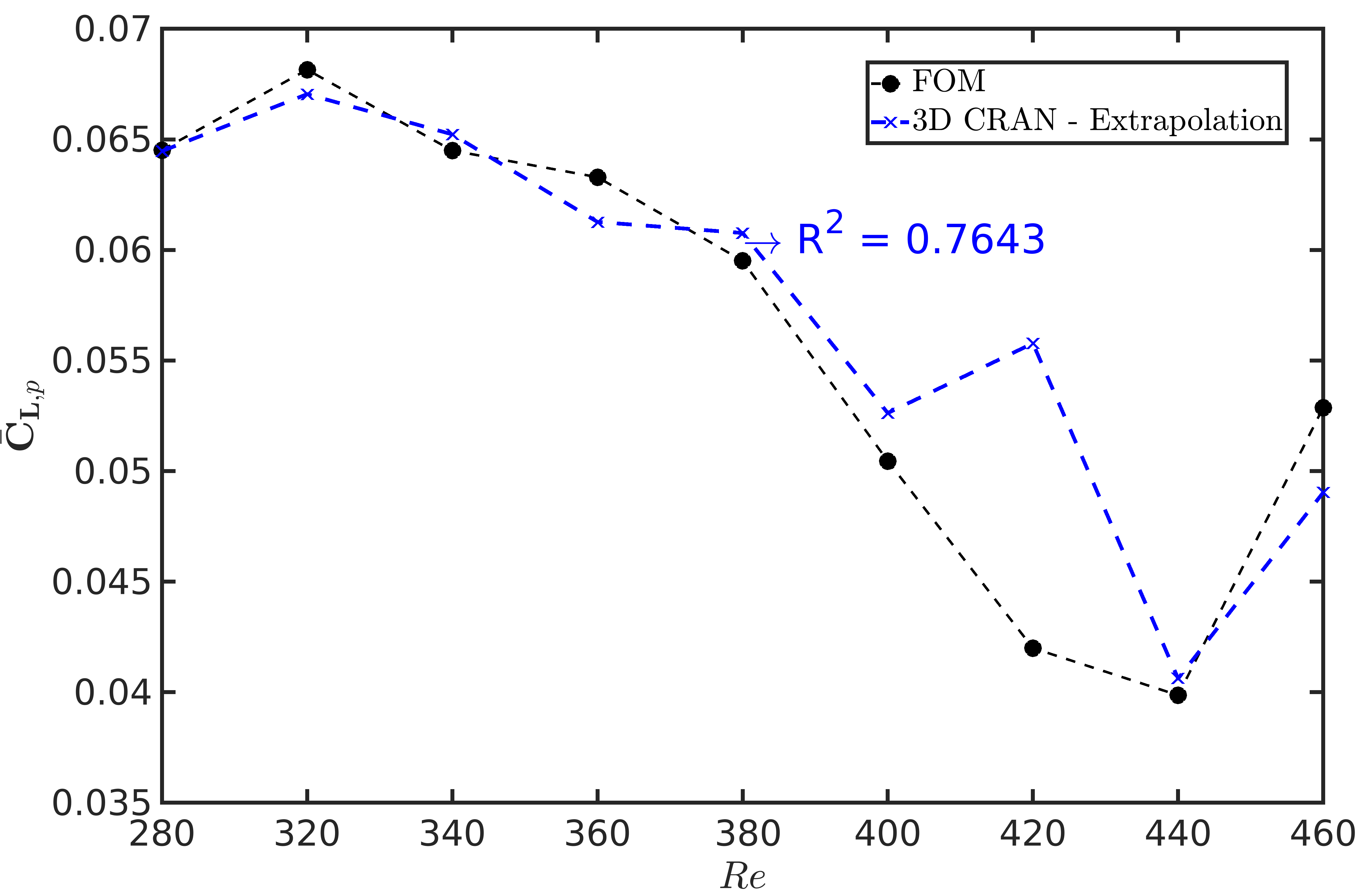}} 
\caption{The variable flow past a sphere: 3D CRAN and FOM comparison of (a) mean drag and (b) mean lift variation over the sphere for different Reynolds numbers. }
\label{forces_behavior_Re}
\end{figure}

Finally, in Table~\ref{tab:cran_pred_multi_Re}, we provide an estimate of the computational costs for the 3D CRAN together with the full-order simulations for flow past a sphere with variable Reynolds number. We recall that the DL-ROM solution offers remarkable speed-ups in online predictions and offline training in this case. Compared to a 32 CPU parallel FOM solver, a single 3D CRAN framework learns the variable $Re$ flow regime 20 times faster via the transfer learning process. At the same time, the online predictions achieved are 1800 times faster than the parallel FOM solver. 
\begin{table}
\centering
\caption{Summary of the offline and online times for 3D CRAN vs. 3D FOM simulations for variable $Re$-based flow.}
\begin{tabular}{p{2.8cm}|p{2.0cm}|p{2.0cm}}
\toprule
\toprule
 &  FOM-HPC &  3D CRAN-PC  \\
\bottomrule
\bottomrule
Processor number & 32 CPUs & 1 GPU\\
Offline time$^{*}$ & $\approx 50 \; \mathrm{h}$  &  $\approx 2.5 \; \mathrm{h}$ \\
Online time$^{**}$  & $\approx 10 \; \mathrm{h}$  & $\approx 20 \; \mathrm{s}$ \\
Offline speed-up   & $1$ & 20 \\
Online speed-up & $1$ &  1800 \\
\bottomrule
\end{tabular}\\ 
$^{*}$ Elapsed time 4000 training steps for 10 $Re$.\\
$^{**}$ Elapsed time 1000 test steps for 10 $Re$.
\label{tab:cran_pred_multi_Re}
\end{table}

\section{Conclusions} \label{conclusions_sph}
We have presented a deep learning framework for the reduced-order modeling of three-dimensional unsteady flow, emphasizing variable $Re$-based flows. The proposed 3D DL-ROM framework relies on the convolutional autoencoder with recurrent neural networks for data-driven predictions. 
While the 3D CNNs provide accurate extraction of the low-dimensional features from full-order flow snapshots, the LSTM-RNN enables the propagation of the features in time. We have successfully demonstrated the inference capability of the proposed 3D CRAN framework by predicting the time series of the unsteady flow fields of three-dimensional flow past a sphere.  
Using coarse-grained learning of $Re$-dependent unsteady flows, a low-dimensional inference of the flow fields with interface load recovery has been discussed. 

We have first analyzed an iterative low interface resolution voxel grid search for the 3D CNNs that preserves the full-order pressure stresses on the fluid-structure interface via snapshot-FTLR. 
We have shown that this snapshot-FTLR method selects a coarse-grain grid for the CRAN architecture by bringing field uniformity and recovering 3D interface information. This reduces the point cloud complexity and the number of nodes in DL space by 3 times compared to CFD space. An end-to-end 3D CRAN is shown to predict the flow dynamics with accurate estimates of flow prediction at a single Reynolds number.
By analyzing an external flow problem past a sphere, we have shown that the 3D CRAN infers and reconstructs the flow fields remarkably for $Re=300$. 
The 3D CRAN extrapolates the field from one input data but requires an expensive offline training cost and hyperparameter search. 
The hyperparameter search has been found to be sensitive to the size of the low-dimensional state of the autoencoder and a detailed study has been performed to tune the network. 

For the first time, we have demonstrated the learning and inference capabilities of the 3D CRAN on a complicated symmetry-breaking flow regime ($280\leq Re\leq 460$) for the flow past a sphere. By leveraging the trained parameters for a single $Re$, we have shown that 3D CRAN can be trained for a variable $Re$ flow regime on a limited data and training time. Using the process of transfer learning, we achieve the offline training speed-up by nearly 20 times compared to the parallel full-order solver.  We find that the predictions perform the best when the field values correspond to $Re \leq 380$, which
characterizes a 3D symmetric shedding of the unsteady vortex patterns. Although the network performs reasonably well for asymmetric flows, the maximum prediction
errors correspond to the transitional flow regime from symmetric to asymmetric vortex shedding. The 3D CRAN offers nearly three order times faster predictions of 3D flow fields with variable Reynolds numbers using a fraction of training time.
It is worth extending the capability of our proposed CRAN framework by incorporating physics-based embedding functions into the autoencoder and considering spatial transformer networks \cite{jaderberg2015} to account for globally invariant features.

\section*{Acknowledgements}
The authors would like to acknowledge the Natural Sciences and Engineering Research Council of Canada (NSERC) for the funding. This research was enabled in part through the computational resources
and services provided by Compute Canada and the Advanced Research Computing facility at the University of British Columbia. 

%
\section*{Data Availability}
The data that support the findings of this study are available on request from the authors.  
\section*{Conflict of interest}
The authors declare that they have no conflict of interest.

\section*{References}
\bibliography{template}   
\bibliographystyle{plain} 

\end{document}